\documentclass[10pt,twocolumn,letterpaper]{article}

\usepackage[letterpaper,top=0.9in,bottom=0.9in,left=0.75in,right=0.75in,columnsep=0.3in]{geometry}
\usepackage[T1]{fontenc}
\usepackage{times}
\usepackage{xspace}
\usepackage[dvipsnames]{xcolor}
\usepackage{graphicx}
\usepackage{amsmath}
\usepackage{amssymb}
\usepackage{booktabs}
\usepackage[numbers,sort&compress]{natbib}
\usepackage[format=plain,labelformat=simple,labelsep=period,font=small]{caption}
\usepackage[font=footnotesize,skip=3pt,subrefformat=parens]{subcaption}
\usepackage[hyphens]{url}
\usepackage[shortlabels,inline]{enumitem}
\setlist[itemize]{noitemsep,leftmargin=*,topsep=0em}
\setlist[enumerate]{noitemsep,leftmargin=*,topsep=0em}
\usepackage{authblk}

\makeatletter
\DeclareRobustCommand\onedot{\futurelet\@let@token\@onedot}
\def\@onedot{\ifx\@let@token.\else.\null\fi\xspace}

\makeatother

\usepackage{cuted} %

\usepackage{currfile} %

\usepackage{caption} %

\usepackage{tikz}
\usetikzlibrary{positioning,arrows.meta,calc}
\definecolor{pdNew}{HTML}{EB6834}     %
\definecolor{pdWan}{HTML}{0F2340}     %
\definecolor{pdOurs}{HTML}{EB6834}    %
\definecolor{pdWalrus}{HTML}{2A78D6}  %
\definecolor{pdUNet}{HTML}{1BAF7A}    %
\definecolor{pdFNO}{HTML}{8A909C}     %
\definecolor{pdDPOT}{HTML}{7B4FA0}    %
\definecolor{pdTruth}{HTML}{1A1F2B}   %
\definecolor{pdCoef}{HTML}{EB6834}    %
\definecolor{pdField}{HTML}{12957A}   %
\definecolor{pdText}{HTML}{7B4FA0}    %
\definecolor{pdNet}{HTML}{8A909C}     %
\definecolor{pdIce}{HTML}{2A78D6}     %



\usepackage{colortbl}
\usepackage{multirow}
\definecolor{pdBand}{gray}{0.925}
\makeatletter
\newcommand{\tabgap}[1]{\rule{0pt}{\dimexpr\ht\@arstrutbox+#1\relax}}
\makeatother

\usepackage{microtype}

\ifdefined\pdfsuppressptexinfo\pdfsuppressptexinfo=-1 \fi

\definecolor{linkblue}{rgb}{0.21,0.49,0.74}
\usepackage[breaklinks,colorlinks,allcolors=linkblue]{hyperref}
\usepackage[capitalize]{cleveref}
\crefname{section}{Sec.}{Secs.}
\Crefname{section}{Section}{Sections}
\crefname{appendix}{App.}{Apps.}
\Crefname{appendix}{Appendix}{Appendices}
\crefname{subappendix}{App.}{Apps.}
\Crefname{subappendix}{Appendix}{Appendices}
\crefname{table}{Tab.}{Tabs.}
\Crefname{table}{Table}{Tables}

\title{\Large\bfseries PhysDial: Adapting Pretrained Video Models for PDE Simulation\\under Changing Physical Conditions}
\author[1]{Yilong Dai}
\author[2]{Yiming Sun}
\author[3]{Shengyu Chen}
\author[1]{Jiaqi Gong}
\author[2]{Xiaowei Jia}
\author[1]{Runlong Yu}
\affil[1]{The University of Alabama}
\affil[2]{Rutgers University}
\affil[3]{University of Pittsburgh}
\date{\small Code: \url{https://github.com/Dyloong1/PhysDial}}

\begin{document}
\maketitle
\begin{abstract}
Simulating how physical systems respond to changing conditions is central to control, design, and scenario studies. To our knowledge, prior PDE foundation models have not been trained or evaluated on changes within trajectories. Our counterfactual test measures how far predictions move toward a changed solver branch from the same state. Models trained on fixed conditions follow changes only partly. A PDE foundation model given a condition input after pretraining barely responds.
PhysDial adapts a pretrained video diffusion transformer to eight two-dimensional PDE families. Coefficients enter with the noise level, condition fields as reference frames, and equation text as a prompt. In Kolmogorov flow, switches within trajectories account for about $5\%$ of the training samples. At matched compute, they halve the error after a change in forcing wavenumber or amplitude. A few trajectories at new amplitudes more than halve extrapolation error. Without physics pretraining, PhysDial matches or exceeds PDE foundation models under training conditions on two turbulent flows. At step 64 on the Kolmogorov test split, PhysDial has $1.07$ times the solver's small-scale energy and Walrus $7.1$ times.
Our experiments suggest that a learned simulator uses a condition only as far as its training data make that condition necessary for prediction. Learning from trajectories that diverge under intervention can help scientific simulators predict what would happen if a system’s forcing or parameters changed.
\end{abstract}

\section{Introduction}
\label{sec:intro}

\suppressfloats[t] %
\begin{figure}[t]
  \centering
  \includegraphics[width=\linewidth]{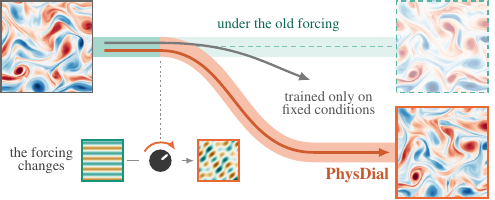}
  \caption{When the forcing of a flow changes, the system leaves the future it was heading for. A simulator trained only on fixed-condition trajectories follows the change only in part. With switch training, PhysDial follows it.}
    \vspace{-0.2cm}
  \label{fig:task}
\end{figure}

Physical systems rarely evolve under constant conditions.
The wind that drives an ocean current varies with the seasons, the heating of a plasma is switched on and off, and the viscosity of a fluid changes with its temperature.
In each case the governing equation stays the same, but its coefficients or forcing change, and the system evolves toward a different future (\cref{fig:task}).
A forecast asks what happens next. A simulator should also answer what would happen if a condition changed now. A numerical solver can compute both futures from the same state. Running it for every change can be costly in control, design, and scenario studies.

Learned simulators are much faster~\cite{li2021fno,lu2021deeponet,gupta2023pdearena,lippe2023pderefiner,dai2026flowrefiner}.
Recent ones are pretrained on many PDE families~\cite{mccabe2024mpp,hao2024dpot,herde2024poseidon,mccabe2026walrus} or built on video generators~\cite{nguyen2025physix,perini2026windinet}.
To our knowledge, none is trained or evaluated with a condition that changes within a trajectory.
They either take no physical condition and infer it from past states~\cite{mccabe2024mpp,hao2024dpot,mccabe2026walrus}, or are conditioned on coefficients, forcing or boundary data that stay fixed within each training trajectory~\cite{takamoto2023cape,zhou2025unisolver,ye2025pdeformer,perini2026windinet}.
We use a counterfactual test to measure how far such models follow a change.
A solver continues the same state under the old and under the new condition, and we locate the prediction of a model between the two futures.
Walrus~\cite{mccabe2026walrus}, fine-tuned on our data with an added condition input, barely moves toward the new future.
Our model, trained only on fixed conditions, moves about $70\%$ of the way after a change of forcing wavenumber ($2\to4$) and $40\%$ after a change of amplitude ($1\to1.4$).

These measurements suggest a hypothesis: a learned simulator uses a condition only as far as its training data make that condition necessary for prediction.
If a condition never changes within a trajectory, the past states already reveal it, so a model can predict well without using its condition input.
Similar shortcuts~\cite{geirhos2020shortcut} have been reported for action-conditioned video world models~\cite{shi2026coco,chen2026worldecho} and for imitation policies that copy past actions~\cite{dehaan2019causalconfusion,wen2020copycat}.
If a condition also takes a single value in all training trajectories, the model never observes its effect.
The hypothesis yields three predictions, which our experiments test.
(H1) A model trained on fixed conditions makes little use of its condition input and follows a change only in part.
(H2) Adding trajectories in which a condition changes makes the model follow the change, and the gain comes from the data rather than from the extra training.
(H3) A few new values reduce the cost of extrapolating a condition held fixed in training.

If the hypothesis holds, training data can make a model follow a change.
Whether the model can use its condition input also depends on the model we start from.
PDE foundation models such as MPP, DPOT and Walrus~\cite{mccabe2024mpp,hao2024dpot,mccabe2026walrus} are designed to infer the physics from past states, so a condition input can be added only after pretraining.
When we added one to Walrus, even training on trajectories with a changing condition did not make it follow the change (\cref{sec:exp-walrus}).
Video generators, in contrast, are pretrained to generate futures that follow a prompt, a reference frame or an action.
They are trained on far more data than any physics corpus offers and are increasingly proposed as world simulators~\cite{brooks2024sora,bruce2024genie,nvidia2025cosmos}.
We therefore start from a video generator and teach it physics, instead of teaching a PDE foundation model to follow conditions.
A single-family U-Net that takes its conditions as inputs from the start also follows changes when trained on such trajectories.
A video generator has a pathway for each physical input and covers multiple families.
Its pretraining improves response to changes (\cref{sec:exp-ablation}).

PhysDial adapts Wan2.1~\cite{wan2025wan}, a text-to-video diffusion transformer with $1.3$B parameters, to autoregressive simulation of eight two-dimensional PDE families (\cref{fig:arch}).
We keep its generation process and route each physical input through the conditioning pathway that plays the same role in video generation.
Coefficients modulate every block together with the noise level.
Condition fields such as a forcing pattern enter as input channels like a reference frame.
A short equation text enters through cross-attention like a prompt.
One model predicts all families autoregressively. It receives the conditions anew in every round, so they act like dials that can be turned during a rollout.
Following the hypothesis, we add two kinds of training data.
Anchor trajectories add a few other values of a condition that is otherwise constant within a family.
Switch trajectories change one condition at a known frame.
In a short final fine-tuning stage, switch training uses about $5\%$ switch samples and supervises only post-change frames.

Our contributions are summarized as follows.
\begin{itemize}
  \item \textbf{A counterfactual test for learned simulators.} A PDE solver gives both futures from the same state when a condition changes. The response $r_k$ measures how far a prediction moves toward the new future. The test exposes weak use of inputs despite high fixed-condition accuracy.
  \item \textbf{Why a simulator can ignore its inputs, and a fix through data.} Under fixed conditions, recent states often reveal the physics. Switch trajectories account for about $5\%$ of training samples. They halve post-change error for forcing wavenumber or amplitude at matched compute and improve response in two architectures with the same losses and backbones. Anchor trajectories more than halve the error when extrapolating the forcing amplitude.
  \item \textbf{From video conditioning to physical conditions.} PhysDial routes coefficients, condition fields, and equation text through the video transformer's existing conditioning pathways. Without physics pretraining, it matches or exceeds two PDE foundation models under the training conditions of two turbulent flows.
\end{itemize}

\section{Related Work}
\label{sec:related}

\paragraph{PDE foundation models and their conditions.}
Building on neural PDE surrogates and benchmarks~\cite{li2021fno,lu2021deeponet,takamoto2022pdebench,gupta2023pdearena,ohana2024well,dai2026physicspreservinglatentcompressionzeroshot}, recent models are pretrained on many PDE families~\cite{mccabe2024mpp,hao2024dpot,herde2024poseidon,shen2024ups,holzschuh2025pdetransformer,mccabe2026walrus}, some as generative models in a latent space~\cite{rozet2025lola,koupai2025echo,chen2025flowmarching,chen2026lgs,couairon2026kastor}.
MPP, DPOT and Walrus take no PDE coefficients and infer them from past states~\cite{mccabe2024mpp,hao2024dpot,mccabe2026walrus}.
Other models are conditioned on explicit coefficients, coefficient fields, boundary data, equation graphs or text~\cite{takamoto2023cape,zhou2025unisolver,ye2024pdeformer,ye2025pdeformer,shen2024ups,zhou2025text2pde,zhu2025equationaware,rozet2025lola,koupai2025echo}, or infer the dynamics from example trajectories or from a code fitted to each environment~\cite{yang2023icon,serrano2025zebra,koupai2025enma,koupai2024geps}.
To our knowledge, these models keep coefficients and forcing parameters fixed within a trajectory.
A forcing may vary in time through fixed parameters~\cite{serrano2025zebra,koupai2024geps}, and some formulations admit time-dependent forcing or boundary data~\cite{zhou2025unisolver,koupai2025echo}.
But none trains or evaluates with a coefficient or forcing parameter that changes partway through a trajectory.
Physics Steering changes the output of Walrus qualitatively by adding concept directions to its activations, without parameter values or solver references~\cite{fear2025physicssteering}.
A model can track a prescribed forcing schedule using its history alone. Testing its response to a new schedule calls for two continuations that share the same history up to the intervention. PhysDial takes explicit conditions that can change during a rollout. Keeping the solver and history fixed isolates the effect of the condition change. The two continuations then provide reference futures for measuring how far the prediction moves in response.

\vspace{-5pt}
\paragraph{Video generators for physics.}
Video generators follow a text prompt, a reference frame or an action and are proposed as world simulators~\cite{brooks2024sora,bruce2024genie,nvidia2025cosmos,wan2025wan}, but they often violate physical laws~\cite{kang2025physicallaw,bansal2025videophy2,yuan2026likephys,cai2026whatif}.
For PDE fields, PhysiX initializes a discrete-token autoregressive model from a video checkpoint and trains it on eight datasets of The Well~\cite{nguyen2025physix}.
Video diffusion has been applied to PDE solving and to specific simulators~\cite{li2025videopde,depellegrini2026lavig,kovacic2026mpmworlds}.
The closest work, WinDiNet~\cite{perini2026windinet}, fine-tunes a video diffusion transformer and its decoder on urban wind flow.
Its conditions are fixed per simulation, and it generates the whole rollout in one denoising pass.
We also fine-tune the decoder. One autoregressive model handles many PDE families and changing conditions.

\paragraph{Changing inputs, shortcuts and extrapolation.}
Learned controllers~\cite{holl2020control,wei2024diffphycon,roysarkar2025dpc,gallagher2026jepc,rautela2026controlfm}, surrogates of flows with ramped boundary conditions~\cite{winkler2026noflunch} and climate emulators~\cite{wattmeyer2023ace,wattmeyer2025ace2,kochkov2024neuralgcm,cachay2024sphericaldyffusion,dheeshjith2025samudra} take inputs that change in time by design, each for a single system.
In action-conditioned video world models, future frames are often predictable from visual inertia, so a model can fit the data without making its dynamics depend on the action~\cite{shi2026coco}.
Such models trained on expert data fail to follow other actions~\cite{chen2026worldecho}.
We study the same shortcut in multi-family PDE simulators, where a solver provides both futures, and we address it through the training data rather than the loss.
Condition values outside the training range have been tested for WinDiNet, ECHO and parameter-conditioned surrogates~\cite{perini2026windinet,koupai2025echo,shokar2025conditioning,vantegelen2026latefusion,zhu2025equationaware,chu2026conditionalgeneralists}, against the error inside the range or a model fine-tuned on a few examples.
We are unaware of prior work that measures extrapolation against a reference of the same architecture and budget trained on the held-out value.

\section{Method}
\label{sec:method}

\begin{figure*}[t]
  \centering
  \includegraphics[width=\linewidth]{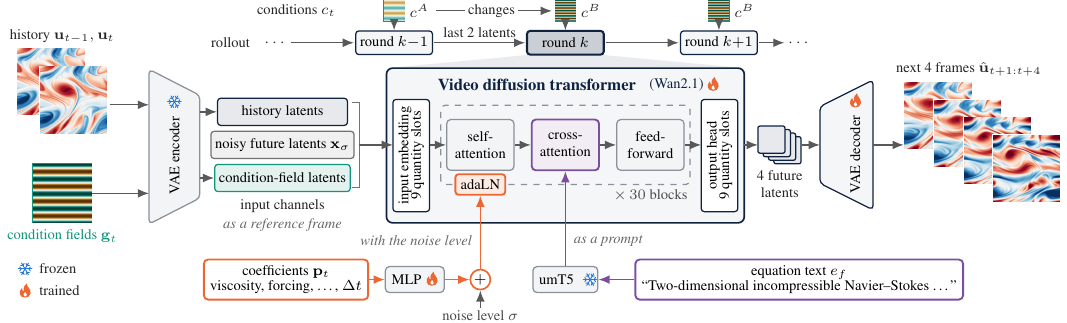}
  \caption{\textbf{Overall framework of PhysDial.} \emph{Top.} In each round, the same network predicts four future states from two past states and the current conditions. The last two predicted latents feed the next round, where the conditions can change. \emph{Bottom.} Coefficients enter with the noise level, condition fields as reference frames, and equation text as a prompt. Snowflakes mark frozen modules; flames mark trained ones.}
  \label{fig:arch}
\end{figure*}

PhysDial reuses a video generator's pathways for specifying which future to generate (\cref{sec:model}). But under fixed conditions, recent states often reveal the physics. A condition that never varies gives the model no examples of what a change would do. We add trajectories with changes within a rollout. From the same past, the supplied condition tells the model which branch to follow (\cref{sec:switch}).

\subsection{Problem setting}
\label{sec:setting}

\paragraph{States and conditions.}
A PDE family $f$ is one equation with its boundary conditions, simulated for many coefficient values and initial states.
The state at frame $t$ is $\mathbf{u}_t \in \mathbb{R}^{C_f \times H_f \times W_f}$, with $C_f$ physical quantities on the solver grid, never downsampled.
The condition at frame $t$ is $c_t = (\mathbf{p}_t, \mathbf{g}_t, e_f)$.
The vector $\mathbf{p}_t$ holds the scalar coefficients, such as viscosity or forcing amplitude, and the frame interval $\Delta t$.
The set $\mathbf{g}_t$ holds spatially varying condition fields, such as a forcing pattern.
The text $e_f$ is one sentence naming the equation, its boundaries and quantities.
We use \emph{condition} for all three and \emph{coefficient} for an entry of $\mathbf{p}_t$.

\paragraph{Autoregressive prediction.}
Given the last two frames and the condition $c$ of the frames to be predicted, the model generates the next four frames jointly,
\begin{equation}
  \hat{\mathbf{u}}_{t+1:t+4} \sim p_\theta\left(\,\cdot \mid \mathbf{u}_{t-1:t},\, c\,\right).
  \label{eq:model}
\end{equation}
We call one such generation a \emph{round} and one predicted frame a \emph{step}.
The last two predicted frames become the history of the next round, so a $16$-step rollout takes four rounds.
The condition is supplied anew in every round. Under fixed conditions it stays the same for the whole trajectory, and under a condition change it differs between rounds.
A step spans $1$ to $16$ stored frames, chosen per family. Its interval is $\Delta t$, and frame indices count steps.

\paragraph{Following a condition change.}
A condition change at frame $t_s$ sets $c_t = c^A$ for $t \le t_s$ and $c_t = c^B$ for $t > t_s$, where $A$ and $B$ differ in one coefficient or in the forcing of the same equation.
The model receives a history generated under $A$ and the condition $c^B$ in every round after $t_s$, so the change takes effect at the first round after $t_s$.
We consider only abrupt changes, the simplest case that a pair of solver branches can test. A continuous change can be approximated by a sequence of small ones.
To measure whether a model follows a change, the solver continues the state at $t_s$ twice.
The continuation under $A$ is the \emph{unchanged branch} $\mathbf{u}^{A}$, and the continuation under $B$ is the \emph{changed branch} $\mathbf{u}^{B}$.
The two branches share their history, so their difference is exactly the effect of the change.
For a prediction $\hat{\mathbf{u}}_k$ at step $k$ after the change, the \emph{response}
\begin{equation}
  r_k = \frac{\langle \hat{\mathbf{u}}_k - \mathbf{u}^{A}_k,\; \mathbf{u}^{B}_k - \mathbf{u}^{A}_k \rangle}{\lVert \mathbf{u}^{B}_k - \mathbf{u}^{A}_k \rVert^2}
  \label{eq:response}
\end{equation}
is the coordinate of the prediction projected onto the line through the two solver states, with $0$ on the unchanged branch and $1$ on the changed branch. It can lie outside $[0,1]$.
We report mean response over evaluation windows and error against~$\mathbf{u}^B$. A prediction can move toward the changed branch and still miss it.

\subsection{From a video generator to a PDE simulator}
\label{sec:model}
A video generator denoises the latents of several frames jointly, guided by its conditioning inputs.
PhysDial keeps this process and changes only its inputs and outputs (\cref{fig:arch}). Physical quantities replace color frames, and physical conditions enter through the existing conditioning pathways.

\paragraph{Physical fields as video latents.}
Every family stays on its solver grid, with $128$ to $512$ points per side, which is far too many tokens for a diffusion transformer.
The video autoencoder downsamples each spatial dimension by a factor of $8$, so with the $2\times2$ patches of the backbone a $512^2$ frame becomes $32^2$ tokens.
We normalize each quantity $i$ of each frame by its spatial mean $\mu_{t,i}$ and its largest absolute deviation $s_{t,i}$ from that mean.
Each normalized quantity is encoded as one image by the frozen encoder $\mathcal{E}$ of the Wan2.1 video autoencoder~\cite{wan2025wan}, without temporal compression.
Quantities are assigned to nine slots shared by all families, and empty slots do not enter the loss.
Normalization removes the amplitude and the mean of each frame, so two constant channels per slot store them relative to the last history frame.
The model generates these channels with the latents and can thus predict a decaying amplitude or a drifting mean.
With the decoder $\mathcal{D}$, a predicted frame is reconstructed as $\hat{\mathbf{u}}_{t,i} = \hat\mu_{t,i} + \hat s_{t,i}\, \mathcal{D}(\hat{\mathbf{z}}_{t,i})$.

\paragraph{Backbone and conditioning.}
We fully fine-tune Wan2.1, a $1.3$B text-to-video diffusion transformer~\mbox{\cite{wan2025wan,peebles2023dit}}.
We~\mbox{expand} the input embedding and output head to nine quantity slots. Each slot starts as a copy of the pretrained layer.
A single-quantity field enters like a video frame latent.
Each physical input takes the matching video pathway (\cref{fig:arch}).
\emph{Coefficients} share the pathway of the noise level.
A vector holds $\mathbf{p}_t$ in fixed slots with a presence mask.
A small MLP maps it to an embedding that is added to the timestep embedding, so it modulates every block through adaptive layer normalization (adaLN)~\cite{peebles2023dit}.
\emph{Condition fields} share the pathway of a reference frame, together with the history.
Each condition field is encoded by the same autoencoder after a fixed scaling that preserves its absolute level.
Its latent is concatenated to the input channels alongside the history latents, as in image-to-video models.
Wan2.1's frozen umT5 encoder~\cite{chung2023unimax} reads the \emph{equation text} $e_f$ as a prompt. The existing cross-attention layers attend to this encoding.
The new input channels and the MLP start at zero. At the first training step, the network therefore computes the same function as the pretrained video model.

\paragraph{Objective and rollout.}
We train with the flow-matching loss of Wan2.1~\cite{lipman2023flowmatching,liu2022rectifiedflow} on the four future frames. History frames are never noised.
At test time we integrate the flow from noise with an Euler sampler.
In a rollout, the last two generated latents become the next history, and frames are decoded only for evaluation.

\paragraph{Decoder and periodic domains.}
The pretrained video decoder adds high-frequency artifacts to fine-scale fields.
As in WinDiNet~\cite{perini2026windinet}, we therefore fine-tune the decoder and keep the encoder frozen, so the trained transformer works with the new decoder without retraining.
Four of our families are periodic in both directions, but the video model is not equivariant to translations of a periodic domain.
We therefore handle periodic axes with circular padding in the autoencoder, rotary position frequencies~\cite{su2024roformer} rounded to harmonics of the periodic token grid, and a shift of each training window by a random number of latent cells.

\subsection{Anchor and switch trajectories}
\label{sec:switch}

We add two kinds of trajectories. Anchors vary a condition that was fixed in the training data. Each anchor simulates the entire rollout at a new value. In switch trajectories, one condition changes mid-rollout. The later frames show how the flow adjusts to that change.

\paragraph{Anchor trajectories.}
Anchor trajectories target a coefficient that has the same value in every training trajectory of a family, such as the forcing amplitude of our Kolmogorov flow.
They have fixed conditions and are simulated at a few other values, here amplitudes $0.7$ and $1.4$ instead of $1$.
The test extrapolates to $0.5$, below both anchors.

\paragraph{Switch training.}
A switch trajectory is simulated under $A$ up to frame $t_s$ and under $B$ afterwards, where $A$ and $B$ are training values of the family that differ in one condition.
Frames before the change were simulated under $c^A$, so they cannot serve as targets under $c^B$.
We therefore take training windows from a switch trajectory only where all four predicted frames lie after $t_s$.
These windows are trained with the condition $c^B$, and frames before the change enter only as history.
The first such window has its whole history under $A$ and its targets under $B$, exactly as at test time. Later windows, up to $16$ steps after the change, cover the transient.
We simulate switch trajectories for Kolmogorov flow only.
They replace about a fifth of its samples, or about $5\%$ of all training samples.
Other families are trained as before.

\paragraph{Training setup.}
PhysDial is fine-tuned from the pretrained video model on the eight families of \cref{sec:exp-setup} in two stages totalling $80$ GPU-hours on H200 GPUs. The second stage adds the anchor trajectories and periodic handling.
Switch training is a third stage of $20$ GPU-hours. \Cref{sec:supp-impl} gives the full training schedule.

\section{Experiments}
\label{sec:exp}

\subsection{Setup}
\label{sec:exp-setup}
\label{sec:protocol}

\begin{table*}[t]
  \caption{\textbf{Relative $L_2$ error under fixed conditions} at the end of rounds 1, 4 and 16 (steps 4, 16 and 64). A row with a family name is its test split, and the rows below are held-out values inside (in) or outside (out) the training range. Bold and underlined mark the lowest and second lowest error. Walrus is pretrained on The Well and not compared there (grey). Round 16 is from separate $64$-step rollouts.}
  \label{tab:main}
  \scriptsize
  \renewcommand{\arraystretch}{0.9}
  \setlength{\tabcolsep}{0pt}
  \newcommand{\wa}{62pt}
  \newcommand{\wb}{65pt}
  \newcommand{\wk}{13pt}
  \newcommand{\wn}{16pt}
  \newcommand{\wN}{19pt}
  \newcolumntype{A}{W{l}{\wa}}
  \newcolumntype{a}{W{l}{\wb}}
  \newcolumntype{K}{W{l}{\wk}}
  \newcolumntype{n}{W{c}{\wn}}
  \newcolumntype{s}{>{\columncolor{pdBand}}W{c}{\wn}}
  \newcolumntype{N}{W{c}{\wN}}
  \newcolumntype{S}{>{\columncolor{pdBand}}W{c}{\wN}}
  \newcommand{\rd}[1]{\multicolumn{1}{c}{#1}}
  \newsavebox{\tabwell}
  \newsavebox{\tabflows}
  \sbox{\tabwell}{%
  \begin{tabular}[t]{@{}AK>{\columncolor{pdBand}}W{c}{14pt}>{\columncolor{pdBand}}W{c}{14pt}W{c}{14pt}W{c}{14pt}>{\columncolor{pdBand}}W{c}{14pt}snW{c}{20pt}s>{\columncolor{pdBand}}W{c}{20pt}@{}}
    \multicolumn{2}{@{}l}{\textbf{The Well}} & \multicolumn{2}{c}{PhysDial} & \multicolumn{2}{c}{Walrus} & \multicolumn{2}{c}{DPOT-H} & \multicolumn{2}{c}{U-Net} & \multicolumn{2}{c}{FNO} \\
    \cmidrule(lr){3-4} \cmidrule(lr){5-6} \cmidrule(lr){7-8} \cmidrule(lr){9-10} \cmidrule(lr){11-12}
    \multicolumn{2}{@{}l}{Family and split \hfill round\hspace{3pt}} & \rd{1} & \rd{4} & \rd{1} & \rd{4} & \rd{1} & \rd{4} & \rd{1} & \rd{4} & \rd{1} & \rd{4} \\
    \midrule
        \emph{Gray--Scott} & test & .191 & .380 & \textcolor{black!50}{.075} & \textcolor{black!50}{.331} & .179 & \phantom{0}.392 & \phantom{0}\textbf{.053} & \phantom{00}\textbf{.314} & \phantom{0}\underline{.090} & \phantom{00}\underline{.340} \\
    \hspace{0.6em}maze & in & \textbf{.133} & \textbf{.321} & \textcolor{black!50}{.028} & \textcolor{black!50}{.107} & \underline{.186} & \phantom{0}\underline{.418} & \phantom{0}.881 & \phantom{00}.953 & \phantom{0}.324 & \phantom{00}.482 \\
    \hspace{0.6em}bubbles & out & \underline{.287} & \underline{.639} & \textcolor{black!50}{.024} & \textcolor{black!50}{.132} & \textbf{.195} & \phantom{0}\textbf{.492} & \phantom{0}.533 & \phantom{00}.691 & \phantom{0}.726 & \phantom{0}2.21\phantom{0} \\
    \tabgap{1.33pt}\emph{Rayleigh--B\'enard} & test & \underline{.070} & .279 & \textcolor{black!50}{.034} & \textcolor{black!50}{.194} & .105 & \phantom{0}.392 & \phantom{0}\textbf{.028} & \phantom{00}\textbf{.119} & \phantom{0}.070 & \phantom{00}\underline{.268} \\
    \hspace{0.6em}Pr $= 2$ & in & .073 & .271 & \textcolor{black!50}{.036} & \textcolor{black!50}{.199} & .095 & \phantom{0}.366 & \phantom{0}\textbf{.037} & \phantom{00}\textbf{.165} & \phantom{0}\underline{.068} & \phantom{00}\underline{.260} \\
    \hspace{0.6em}Ra $= 10^{10}$ & out & .265 & .605 & \textcolor{black!50}{.181} & \textcolor{black!50}{.510} & \underline{.228} & \phantom{0}\underline{.588} & \phantom{0}\textbf{.163} & \phantom{00}\textbf{.468} & \phantom{0}.235 & \phantom{00}.598 \\
    \tabgap{1.33pt}\emph{Shear flow} & test & \textbf{.154} & \textbf{.400} & \textcolor{black!50}{.075} & \textcolor{black!50}{.291} & .268 & \underline{1.95}\phantom{0} & \phantom{0}.305 & \phantom{0}3.53\phantom{0} & \phantom{0}\underline{.246} & \phantom{0}2.03\phantom{0} \\
    \hspace{0.6em}Sc $= 2$ & in & \textbf{.215} & \textbf{.380} & \textcolor{black!50}{.111} & \textcolor{black!50}{.272} & \underline{.437} & \phantom{0}\underline{.890} & \phantom{0}.679 & \phantom{0}5.82\phantom{0} & \phantom{0}.522 & \phantom{0}3.61\phantom{0} \\
    \hspace{0.6em}Re $= 5{\times}10^5$ & out & \textbf{.280} & \textbf{.580} & \textcolor{black!50}{.159} & \textcolor{black!50}{.432} & \underline{.685} & \underline{1.27}\phantom{0} & 1.69\phantom{0} & 15.5\phantom{00} & 5.34\phantom{0} & 89.1\phantom{00} \\
    \tabgap{1.33pt}\emph{Radiative layer} & test & .213 & .379 & \textcolor{black!50}{.132} & \textcolor{black!50}{.347} & \underline{.183} & \phantom{0}\underline{.375} & \phantom{0}\textbf{.142} & \phantom{00}\textbf{.353} & \phantom{0}.231 & \phantom{00}.424 \\
    \hspace{0.6em}$t_{\mathrm{cool}} = 0.32$ & in & .182 & \textbf{.339} & \textcolor{black!50}{.100} & \textcolor{black!50}{.296} & \textbf{.147} & \phantom{0}\underline{.340} & \phantom{0}\underline{.170} & \phantom{00}.378 & \phantom{0}.189 & \phantom{00}.391 \\
    \hspace{0.6em}$t_{\mathrm{cool}} = 3.16$ & out & \underline{.112} & \textbf{.299} & \textcolor{black!50}{.048} & \textcolor{black!50}{.237} & \textbf{.088} & \phantom{0}\underline{.307} & \phantom{0}.308 & \phantom{00}.628 & \phantom{0}.174 & \phantom{00}.388 \\
    \tabgap{1.33pt}\emph{Euler} & test & \underline{.103} & \underline{.251} & \textcolor{black!50}{.044} & \textcolor{black!50}{.138} & .105 & \phantom{0}.276 & \phantom{0}\textbf{.064} & \phantom{00}\textbf{.216} & \phantom{0}.120 & \phantom{00}.276 \\
    \hspace{0.6em}$\gamma = 1.13$ & out & \underline{.138} & \textbf{.264} & \textcolor{black!50}{.065} & \textcolor{black!50}{.156} & .190 & \phantom{0}.428 & \phantom{0}\textbf{.123} & \phantom{00}.366 & \phantom{0}.158 & \phantom{00}\underline{.331} \\
    \hspace{0.6em}$\gamma = 1.76$ & out & \underline{.110} & \textbf{.241} & \textcolor{black!50}{.049} & \textcolor{black!50}{.147} & .135 & \phantom{0}.317 & \phantom{0}\textbf{.093} & \phantom{00}.299 & \phantom{0}.137 & \phantom{00}\underline{.284} \\
    \tabgap{1.33pt}\emph{Acoustic scattering} & test & .083 & .151 & \textcolor{black!50}{.043} & \textcolor{black!50}{.084} & .114 & \phantom{0}.299 & \phantom{0}\textbf{.017} & \phantom{00}\textbf{.053} & \phantom{0}\underline{.036} & \phantom{00}\underline{.086} \\
  \end{tabular}}
  \sbox{\tabflows}{%
  \begin{tabular}[t]{@{}aKSSSNNNSSS@{}}
    \multicolumn{2}{@{}l}{\textbf{Simulated turbulent flows}} & \multicolumn{3}{c}{PhysDial} & \multicolumn{3}{c}{Walrus} & \multicolumn{3}{c}{DPOT-H} \\
    \cmidrule(lr){3-5} \cmidrule(lr){6-8} \cmidrule(lr){9-11}
    \multicolumn{2}{@{}l}{Family and split \hfill round\hspace{3pt}} & \rd{1} & \rd{4} & \rd{16} & \rd{1} & \rd{4} & \rd{16} & \rd{1} & \rd{4} & \rd{16} \\
    \midrule
        \emph{Kolmogorov flow} & test & \textbf{.023} & \textbf{.080} & \textbf{.429} & \underline{.028} & \underline{.089} & \phantom{0}\underline{.592} & .111 & .338 & 1.03 \\
    \hspace{0.6em}$\nu$ inside & in & \textbf{.029} & \underline{.103} & \textbf{.499} & \underline{.029} & \textbf{.085} & \phantom{0}\underline{.551} & .110 & .314 & 1.06 \\
    \hspace{0.6em}dense $\nu$, inside & in & \textbf{.020} & \textbf{.076} & \textbf{.360} & \underline{.023} & \underline{.078} & \phantom{0}\underline{.503} & .111 & .349 & 1.01 \\
    \hspace{0.6em}$k_f = 3$ & in & \textbf{.030} & \underline{.111} & \textbf{.469} & \underline{.032} & \textbf{.092} & \phantom{0}\underline{.591} & .113 & .348 & 1.09 \\
    \hspace{0.6em}multi-mode forcing & in & \textbf{.040} & \underline{.123} & \textbf{.610} & \underline{.042} & \textbf{.119} & \phantom{0}\underline{.759} & .130 & .437 & 1.18 \\
    \hspace{0.6em}$\nu$ below & out & \textbf{.046} & \underline{.180} & \underline{.649} & \underline{.084} & \textbf{.146} & \phantom{0}\textbf{.602} & .161 & .399 & 1.12 \\
    \hspace{0.6em}$\nu$ above & out & \underline{.018} & \underline{.078} & \textbf{.453} & \textbf{.014} & \textbf{.075} & \phantom{0}\underline{.492} & .073 & .249 & 1.04 \\
    \hspace{0.6em}dense $\nu$, outside & out & \textbf{.024} & \underline{.101} & \textbf{.439} & \underline{.031} & \textbf{.089} & \phantom{0}\underline{.515} & .112 & .307 & 1.02 \\
    \hspace{0.6em}$k_f = 6$ & out & \textbf{.039} & \underline{.150} & \textbf{.726} & \underline{.045} & \textbf{.139} & \phantom{0}\underline{.914} & .133 & .390 & 1.38 \\
    \hspace{0.6em}forcing band $6$--$7$ & out & \underline{.054} & \underline{.271} & \textbf{.982} & \textbf{.046} & \textbf{.173} & \underline{1.03}\phantom{0} & .134 & .463 & 1.45 \\
    \hspace{0.6em}amplitude $0.5$ & out & \underline{.022} & \underline{.104} & \textbf{.479} & \textbf{.019} & \textbf{.067} & \phantom{0}\underline{.506} & .115 & .351 & 1.11 \\
    \hspace{0.6em}$k_f$ and amplitude & out & \underline{.033} & \underline{.137} & \textbf{.601} & \textbf{.026} & \textbf{.092} & \phantom{0}\underline{.768} & .128 & .370 & 1.22 \\
    \hspace{0.6em}drag $0.05$ & out & \underline{.050} & \underline{.244} & \underline{.728} & \textbf{.038} & \textbf{.121} & \phantom{0}\textbf{.710} & .135 & .416 & 1.27 \\
    \hspace{0.6em}drag $0.2$ & out & \underline{.028} & \underline{.144} & \underline{.760} & \textbf{.018} & \textbf{.070} & \phantom{0}\textbf{.576} & .091 & .253 & 1.06 \\
    \tabgap{1.85pt}\emph{Decaying turbulence} & test & \textbf{.032} & \textbf{.074} & \textbf{.370} & \underline{.041} & \underline{.086} & \phantom{0}\underline{.464} & .137 & .353 & 1.19 \\
    \hspace{0.6em}$\nu$ inside & in & \textbf{.022} & \textbf{.057} & \textbf{.317} & \underline{.026} & \underline{.070} & \phantom{0}\underline{.394} & .113 & .301 & 1.22 \\
    \hspace{0.6em}$\nu$ outside & out & \textbf{.022} & \textbf{.076} & \underline{.408} & \underline{.039} & \underline{.090} & \phantom{0}\textbf{.396} & .115 & .329 & 1.19 \\
  \end{tabular}}
  \begin{minipage}{\linewidth}
  \hrule height \heavyrulewidth
  \vspace{\belowrulesep}
  \noindent\usebox{\tabwell}\hfill\usebox{\tabflows}\par
  \nointerlineskip
  \nointerlineskip
  \vspace{\aboverulesep}
  \hrule height \heavyrulewidth
  \end{minipage}
\end{table*}

\paragraph{Data.}
We use eight two-dimensional PDE families on regular grids (\cref{tab:main}).
Six come from The Well~\cite{ohana2024well}.
Walrus is pretrained on this benchmark and may have seen its evaluation trajectories, so we do not compare it there.
We therefore simulate two further flows with a pseudo-spectral solver~\cite{dai2026tide}.
Kolmogorov flow and decaying turbulence are the forced and the freely decaying case of two-dimensional turbulence.
No model has seen them in pretraining.
Every family has a \emph{test} split with the training conditions and unseen trajectories.
All but acoustic scattering also have \emph{held-out} splits with condition values inside or outside the training range. \Cref{sec:supp-data} lists these splits.
Ten groups of $32$ Kolmogorov trajectories switch once, five to values inside the training range and five outside. Five more groups test repeated changes over $64$ steps.

\paragraph{Models and baselines.}
\emph{PhysDial} is the model before switch training, and we use it wherever conditions stay fixed.
For condition changes we compare \emph{switch-trained PhysDial} with its \emph{control}, PhysDial fine-tuned for the same compute on fixed conditions only.
A \emph{controlled study} repeats this pair with two seeds. It and the anchor study (\cref{sec:exp-extrap}) use PhysDial trained without periodic handling.
\emph{Walrus}~\cite{mccabe2026walrus}, a $1.3$B PDE foundation model that predicts from six past frames without a condition input, is fine-tuned with its own recipe on our training data.
\emph{DPOT-H}~\cite{hao2024dpot} ($1$B) is fine-tuned on the same families and also takes no conditions.
The \emph{specialists}, a conditioned U-Net~\cite{gupta2023pdearena} and an FNO~\cite{li2021fno}, are trained from scratch on one family each.
A specialist trains for $4$ GPU-hours on one family, whereas the physics pretraining of Walrus far exceeds all of our fine-tuning.

\begin{figure}[t]
  \centering
  \includegraphics[width=\linewidth]{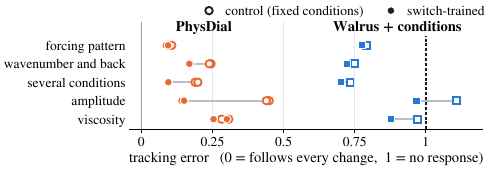}
  \caption{\textbf{Tracking error under repeated changes} ($64$-step Kolmogorov rollouts). The wavenumber changes and returns to its initial value; the mixed sequence changes several conditions in turn. Changes occur every $16$ steps except amplitude ($32$ steps). PhysDial has two seeds and Walrus one run.}
  \label{fig:tracking}
\end{figure}

\paragraph{Metrics and protocol.}
Models roll out $16$ steps (four rounds) on trajectories never used in training, and we report the relative $L_2$ error at step 4 and at step 16, our main horizon.
On the two simulated flows we also roll out $64$ steps. With our frame strides, the trajectories of half of the families from The Well are too short for this.
Generative models draw one sample per window, and predictions for one window under different conditions share the sampling noise.
For each split, we compute the geometric mean of per-window error ratios. We call a model \emph{more accurate} if the mean's $95\%$ bootstrap interval excludes $1$.
If the interval contains $1$ and the mean is within $\pm8\%$ of $1$, the split is a tie.
In all other cases we state only which model has the lower error.
The difference between two seeds exceeds the $8\%$ band on a single split only.
Results come from a single run unless two seeds are stated, and the two seeds repeat only the last training stage.
For a condition change we report the response $r_{16}$ (\cref{eq:response}) and the error with respect to the changed branch.
For repeated changes, let $D_k$ and $\hat D_k$ be the solver and model differences in a flow statistic between changing and fixed conditions, averaged over windows. We use kinetic energy for amplitude changes, enstrophy for viscosity changes, and the flow's projection onto the current forcing otherwise. The \emph{tracking error} is $\mathrm{rms}_k(\hat D_k-D_k)/\mathrm{rms}_k(D_k)$ (\cref{fig:tracking}).
It is $0$ when the model follows every change and $1$ when it does not respond.
To separate the cost of leaving the training range from the difficulty of a held-out value, we train an \emph{in-range reference model}.
It has the same architecture and budget, and its training data also contain other trajectories at that value.
The \emph{extrapolation cost} is the error of a model divided by that of its reference on the same windows, so $1$ means no cost.
We use it in the anchor study (\cref{sec:exp-extrap}), and different models are compared by absolute error.

\subsection{Do models use their conditions? (H1)}
\label{sec:exp-h1}

A model trained on fixed conditions relies mainly on its history.
A model trained without the coefficient inputs has a step-16 error within $5\%$ of the full model on $21$ of the $33$ splits. The coefficients help mainly where they set slow statistics, such as a viscosity outside the training range.
A linear probe recovers viscosity and forcing wavenumber from two past frames ($R^2 = 0.93$ and $0.98$).
Condition fields matter where the history cannot reveal them.
Without them, a model has $2.3$ times the error on random multi-mode forcing but only $1.1$--$1.2$ times on single-mode forcing, where the pattern is visible in the history.

A condition change exposes this reliance (dashed lines in \cref{fig:change} and open markers in \cref{fig:switch-study}).
Trained on fixed conditions only, the control of the controlled study moves $70\%$ of the way to the changed branch after a change of forcing wavenumber and about $40\%$ after a change of amplitude.
Among the changes inside the training range that clearly alter the solution, it follows almost fully only the change of forcing pattern ($r_{16} = 0.94$).
The history does not reveal a multi-mode forcing pattern, so fixed-condition data already make this condition necessary, as the hypothesis predicts.

\suppressfloats[t]

\subsection{Does switch training improve response? (H2)}
\label{sec:exp-changes}

\begin{figure*}[t]
  \centering
  \includegraphics[width=\linewidth]{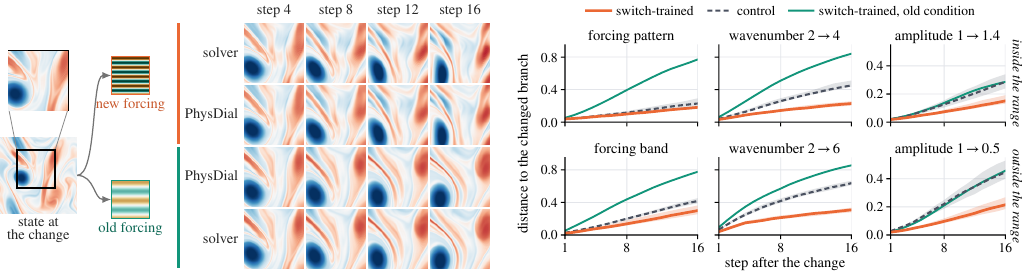}
  \caption{\textbf{Following a condition change} (Kolmogorov flow). \emph{Left.} One window in which the forcing wavenumber changes from $2$ to $4$. Solver and switch-trained PhysDial are shown under the new and under the old forcing in the marked square, where the two solver branches differ most. \emph{Right.} Distance to the changed branch after six kinds of change (median and quartiles over $32$ windows).}
  \label{fig:change}
\end{figure*}

\begin{figure}[t]
  \centering
  \includegraphics[width=\linewidth]{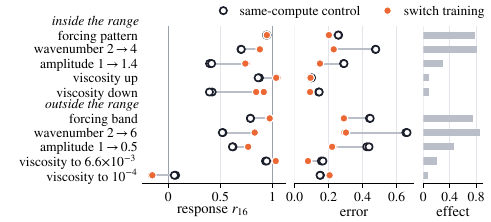}
  \caption{\textbf{Switch training against a same-compute control} (two seeds, Kolmogorov flow, one change, step 16). Errors refer to the changed branch. The effect is the relative difference of the solver branches. Where it is small, the response is only indicative.}
  \label{fig:switch-study}
\end{figure}

In the controlled study on Kolmogorov flow, the control receives the same $20$ GPU-hours on trajectories with fixed conditions, which isolates the effect of the data (\cref{fig:switch-study}).
Inside the training range, switch training halves the error after a change of forcing wavenumber or amplitude.
The~\mbox{control} already follows the forcing pattern. Switch training still cuts the error by about a fifth.
The response rises from $0.70$ to $0.88$ for the wavenumber and from about $0.4$ to $0.75$ for the amplitude.
Changes to values outside the training range show similar gains, and the two seeds differ by at most $0.04$ in every error ratio.
The exception is a viscosity drop to $10^{-4}$, far below the training range, after which the switch-trained model is worse than its control (error ratio $1.35$--$1.38$).

Switch training of PhysDial itself, a single run, reproduces these results.
Over the ten groups, the median error after the change falls to $0.61$ times that of the control (range $0.48$--$1.65$).
The median response rises from $0.62$ to $0.83$, far beyond the spread between seeds.
On the same windows without a change, the error is almost unchanged (ratio $0.96$).
\Cref{fig:change} shows the same behavior at every step and in one window, selected by a fixed rule.
Given the new condition, the switch-trained model stays close to the changed branch.
Given the old condition and the same sampling noise, it stays on the unchanged branch, and its distance grows as the two branches separate.
After a change of forcing pattern or wavenumber, its step-16 error is $0.18$--$0.31$ with the new condition and $0.77$--$0.86$ with the old one.
The two predictions share the history and the sampling noise, so their difference comes from the condition input.
After an amplitude increase the median response is $0.65$, lower than the $0.78$--$0.95$ after a change of forcing pattern or wavenumber.
Switch training also helps when a condition changes several times in one rollout.
In the controlled study, the tracking error for two successive amplitude changes falls from $0.44$--$0.45$ to $0.14$--$0.15$ (\cref{fig:tracking}).
Later segments respond equally well.

\paragraph{Cost of switch training.}
Under fixed conditions, the step-16 error of switch-trained PhysDial is $0.99$ times that of its control (median over the $33$ splits).
The control is never more than $5\%$ worse than PhysDial.
Switch training raises the error on two forcing-extrapolation splits of Kolmogorov flow ($1.34$ and $1.25$ times the error of PhysDial).
\Cref{tab:main} therefore reports PhysDial before switch training.

\subsection{Do anchors reduce extrapolation cost? (H3)}
\label{sec:exp-extrap}

\begin{table}[t]
  \caption{\textbf{Extrapolation cost with and without anchors} (step 16). Without anchors, amplitude and $\beta$ never vary.}
  \label{tab:anchors}
  \centering
  \scriptsize
  \setlength{\tabcolsep}{4pt}
  \begin{tabular}{@{}lcc@{}}
    \toprule
    Held-out condition (Kolmogorov flow) & no anchors & anchors \\
    \midrule
    \multicolumn{3}{@{}l}{\emph{Anchors at amplitudes $0.7$ and $1.4$} (two seeds)} \\
    amplitude $0.5$ & $2.73$--$2.74$ & $1.27$--$1.28$ \\
    amplitude $0.5$ and wavenumber $3$ or $6$ & $2.58$ & $1.33$--$1.34$ \\
    wavenumber $6$ (amplitude $1$) & $1.07$--$1.08$ & $1.15$--$1.16$ \\
    \midrule
    \multicolumn{3}{@{}l}{\emph{Anchors at $\beta = \pm2.4$}} \\
    $\beta = 4$, PhysDial (two seeds) & $5.92$--$5.94$ & $1.61$--$1.66$ \\
    $\beta = 4$, Walrus (one run) & $4.35$ & $1.26$ \\
    \bottomrule
  \end{tabular}
\end{table}

Extrapolation along a condition that never varies in training is costly, and a few trajectories at other values of that condition reduce the cost (\cref{tab:anchors}).
Without anchors, the forcing amplitude of Kolmogorov flow is $1$ in every training trajectory, so its input is constant and the weights that read it receive no gradient.
At the held-out amplitude $0.5$, the model without anchors has a cost of $2.7$.
Anchors at amplitudes $0.7$ and $1.4$ lower it to $1.27$--$1.28$, below the bound of $1.36$ that we fixed before training.
The error of the reference falls by less than $10\%$, so the gain comes mainly from the tested model.
As a side effect of the anchors, the error on the wavenumber-extrapolation split rises by $7\%$ and the cost from $1.07$--$1.08$ to $1.15$--$1.16$.
A second condition that never varies is the coefficient $\beta$ of a $\beta$-plane term added to Kolmogorov flow (zero in all other data).
Anchors at $\pm2.4$ lower the cost at $\beta = 4$ from $5.9$ to $1.6$--$1.7$.
This misses the bound of $1.3$ that we fixed before training.
The same anchors lower the cost of Walrus from $4.35$ to $1.26$, so the effect holds for a second architecture.
Anchors teach the effect of amplitude with no corresponding gain in following amplitude changes.

\begin{figure*}[t]
  \centering
  \includegraphics[width=\linewidth]{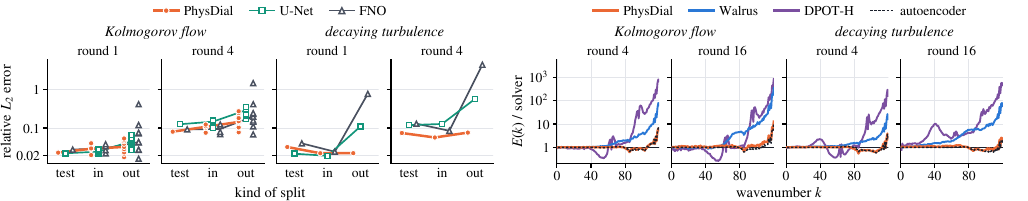}
  \caption{\textbf{Simulated turbulent flows} (round $r$ ends at step $4r$). \emph{Left.} Error of PhysDial and specialists (markers are splits, lines medians). \emph{Right.} Energy spectrum over that of the solver on the test split ($28$ windows). The autoencoder curve is the reconstructed solver frame.}
  \label{fig:flows}
\end{figure*}

\subsection{Do the same data help other models?}
\label{sec:exp-walrus}

We add the conditions to pretrained Walrus as constant input fields and fine-tune it on the same trajectories as PhysDial.
Trained on fixed conditions, Walrus barely uses them. Under repeated changes its tracking error is $0.7$--$1.1$ (\cref{fig:tracking}), where $1$ means no response.
With switch training, the tracking error stays at $0.7$--$1.0$, against at most $0.30$ in our controlled study.
After a single change inside the training range, its error is $0.93$--$1.02$ times that of its control.
Its response is $0.11$--$0.12$ after a change of the forcing and $0.03$ after a change of amplitude, whereas our switch-trained model reaches $0.88$--$0.95$ and $0.75$.
Its share of switch trajectories is no smaller than ours, so a lack of such data is unlikely to be the reason.
A possible cause is its per-sample normalization, after which a linear probe can hardly recover the forcing amplitude from the input ($R^2$ falls from $0.63$ to $0.09$).

A model that is conditioned from the start behaves differently.
A single-family U-Net takes its coefficients through FiLM, and it also follows changes on Kolmogorov flow when trained on the same switch trajectories.
After changes inside the training range its error is close to that of PhysDial. After changes of forcing wavenumber or band to values outside it, PhysDial has the lower error ($0.31$ against $0.54$ for the wavenumber, $0.30$ against $0.47$ for the forcing band).
Without switch trajectories, the error of the U-Net after a change of wavenumber is $0.57$ instead of $0.23$.
These results suggest that switch trajectories work when a model learns the physics with its conditions as inputs, as PhysDial and the conditioned U-Net do.
They did not work for Walrus, where a condition input is attached to a model that already infers the physics from its history.

\subsection{Accuracy under fixed conditions}
\label{sec:exp-standard}

\paragraph{The Well.}
On this benchmark (\cref{tab:main}), PhysDial has a lower error than DPOT-H on $13$ of the $16$ splits at step 16.
The U-Net has the lowest error on five of the six test splits.
This advantage disappears on held-out conditions, where PhysDial has the lowest step-16 error on seven of the ten splits. On shear flow both specialists diverge.

\paragraph{Simulated turbulent flows.}
DPOT-H has the highest error of the three large models on every split and horizon.
\mbox{PhysDial} uses no physics pretraining. It matches Walrus at step 4 and leads at step 64. At step 16 the ranking depends on the family.
At step 4, Walrus is more accurate on five of the $17$ splits and PhysDial on four, with a fifth at the edge of significance.
At step 16, PhysDial is more accurate on all three splits of decaying turbulence and ties with Walrus on the Kolmogorov test split (error ratio $0.92$, interval $[0.82, 1.03]$).
Walrus is more accurate on eight of the other $13$ Kolmogorov splits, six outside training. Its largest margins are on the two drag splits.
At step~64, PhysDial leads on $13$ of $17$ splits.

\paragraph{Physical fidelity.}
By step 64 the pointwise errors of both models are large, and the models differ mainly at the smallest scales (\cref{fig:flows}, right).
On the Kolmogorov test split at step 64, Walrus has $7.1$ times the energy of the solver in the upper half of the wavenumbers, and PhysDial $1.07$ times.
On the decaying-turbulence test split the factors are $4.9$ and $1.06$, and on all $17$ splits PhysDial is closer to the solver than Walrus.
At step 16, this energy is well defined on $26$ of all $33$ splits.
Energy is off by over $1.5\times$ on $2$ splits for PhysDial, $20$ for Walrus, $24$ for DPOT-H, $9$ for U-Net, and $11$ for FNO.

\paragraph{Single-family specialists.}
\Cref{fig:flows} (left) compares PhysDial with the specialists on the simulated flows.
Under training conditions, PhysDial has lower step-16 error than both seeds of each specialist. The U-Net has lower error at step 4.
Outside the training range, PhysDial has a lower step-16 error than the U-Net on $9$ of the $10$ splits and than the FNO on $7$.
Over all eight families, the count against the U-Net is $14$ of $16$.
On decaying turbulence outside the training viscosity range, PhysDial has an error of $0.076$. Across both seeds, the U-Net and FNO have errors of at least $0.57$ and $0.59$, respectively.
The FNO has the lower step-16 error on six of the $14$ Kolmogorov splits.

\subsection{Design choices}
\label{sec:exp-ablation}

The other design choices mainly affect accuracy. \Cref{sec:supp-design} reports the full ablations.
Decoder fine-tuning cuts step-1 and step-16 error by $16\%$ and $1\%$. Periodic handling cuts them by $7\%$ and $4\%$ on the four doubly periodic families, using two seeds for this comparison.
To~compare~\mbox{training objectives,} we keep the network fixed and replace flow matching with regression. The step-16 error on Kolmogorov flow is then $1.2$--$1.9$ times as high.
Video pretraining lowers error and improves response. From random initialization, the same network has $1.15$ times the error and responds less to all six forcing and amplitude changes ($r_{16}$ of $0.37$--$0.79$ against $0.52$--$0.92$).
Increasing the backbone size does not help at matched compute. The $14$B video model has $1.66$--$1.85$ times the error of the $1.3$B model.
Removing the equation text at test time raises the median step-16 error by $12\%$, so the trained model uses it.

\section{Limitations}
All eight PDE families are two-dimensional and use regular grids. The switch-training experiments cover abrupt changes in one condition in Kolmogorov flow. Specialists had lower error under the training conditions for most families from The Well, and the video autoencoder limits single-step accuracy. In the two-seed switch study, the error ratios differed by at most $0.04$ between seeds. Most other comparisons use single runs. The amplitude, drag, and $\beta$ splits test extrapolation in the respective condition and frame interval.

\section{Conclusion}
\label{sec:conclusion}
In this paper, we introduced PhysDial for PDE simulation under changing physical conditions. Our experiments suggest that a learned simulator uses a condition only as far as its training data make that condition necessary for prediction.
The two-branch test reveals weak use of inputs even when fixed-condition accuracy is high. Switch trajectories improved the response of PhysDial and a conditioned U-Net in Kolmogorov flow. Without physics pretraining, PhysDial matched or exceeded two PDE foundation models under the training conditions of two turbulent flows. Walrus responded weakly with a condition input added after pretraining. Future training could include gradual and simultaneous changes. Paired futures also let us measure how video world models respond to actions~\cite{shi2026coco,chen2026worldecho}. We hope this work encourages the use of paired futures to train and evaluate other scientific simulators and action-conditioned world models.

{
    \small
    \bibliographystyle{ieeenat_fullname}
    \bibliography{main}
}

\clearpage
\appendix
\crefalias{section}{appendix}\crefalias{subsection}{subappendix}
\section*{Appendix}

This appendix gives the full version of the method (\cref{sec:supp-method}), implementation details and training settings (\cref{sec:supp-impl}), the data (\cref{sec:supp-data}), the evaluation protocol (\cref{sec:supp-protocol}), additional results for each part of the experiments (\cref{sec:supp-results}), the full list of limitations (\cref{sec:supp-limits}) and notes on reproducibility (\cref{sec:supp-repro}).
As in the main text, a result comes from a single run unless two seeds are stated.

\section{Method details}
\label{sec:supp-method}
\paragraph{Autoregressive prediction.}
Given the last two frames and the condition $c$ of the frames to be predicted, the model generates the next four frames jointly,
\begin{equation}
  \hat{\mathbf{u}}_{t+1:t+4} \sim p_\theta\left(\,\cdot \mid \mathbf{u}_{t-1:t},\, c\,\right).
  \label{eq:supp-model}
\end{equation}
We call one such generation a \emph{round} and one predicted frame a \emph{step}.
The last two predicted frames become the history of the next round, so a $16$-step rollout takes four rounds.
The condition is supplied anew in every round. Under fixed conditions it stays the same for the whole trajectory, and under a condition change it differs between rounds.
A step spans $1$ to $16$ stored frames, chosen per family so that one step changes the state by $10$--$20\%$.
The exceptions are two families in which a single stored frame already changes the state by more (\cref{tab:supp-overview}).
Frames are selected, never resampled.
The interval of one step, $\Delta t$, is taken from each trajectory and given to the model as a coefficient (\cref{tab:slots}).
In the families from The Well it is the same for all trajectories of a family.
In the two simulated flows, frames are stored $0.1$ eddy-turnover times apart.
This time is measured for every simulated condition, so $\Delta t$ varies with the viscosity, the forcing and the drag.
From here on, frame indices count steps.

\paragraph{Condition changes.}
A condition change at frame $t_s$ sets $c_t = c^A$ for $t \le t_s$ and $c_t = c^B$ for $t > t_s$, where $A$ and $B$ differ in one coefficient or in the forcing of the same equation.
The new value $B$ may lie inside or outside the training range.
The model receives a history generated under $A$ and the condition $c^B$ in every round after $t_s$, so the change takes effect at the first round after $t_s$.
In our evaluation groups, the change falls on a round boundary, so the last history frame is the last frame under $A$.
We consider only abrupt changes, the simplest case that a pair of solver branches can test. A continuous change can be approximated by a sequence of small ones.
We leave continuous changes to future work.
Within a rollout, an abrupt change is applied once or several times.

\paragraph{Following a change.}
To measure whether a model follows a change, the solver continues the state at $t_s$ twice.
The continuation under $A$ is the \emph{unchanged branch} $\mathbf{u}^{A}$, and the continuation under $B$ is the \emph{changed branch} $\mathbf{u}^{B}$.
The two branches share their history, so their difference is exactly the effect of the change.
For a prediction $\hat{\mathbf{u}}_k$ at step $k$ after the change, the \emph{response}
\begin{equation}
  r_k = \frac{\langle \hat{\mathbf{u}}_k - \mathbf{u}^{A}_k,\; \mathbf{u}^{B}_k - \mathbf{u}^{A}_k \rangle}{\lVert \mathbf{u}^{B}_k - \mathbf{u}^{A}_k \rVert^2}
  \label{eq:supp-response}
\end{equation}
is $0$ on the unchanged branch and $1$ on the changed branch.
A model that ignores its condition input continues the history as if nothing had changed and has $r_k \approx 0$.
A prediction can move in the right direction without being accurate, so we always report $r_k$, averaged over the evaluation windows, together with the error with respect to $\mathbf{u}^B$.
Our model is generative, so each rollout is one sample.
We draw one sample per evaluation window.
All predictions for the same window share its sampling noise, whichever condition the model receives, so that $r_k$ reflects the condition input rather than the sampling noise.

\paragraph{Physical fields as video latents.}
Every family stays on its solver grid, with $128$ to $512$ points per side, which is far too many tokens for a diffusion transformer.
The video autoencoder downsamples each spatial dimension by a factor of $8$, so with the $2\times2$ patches of the backbone a $512^2$ frame becomes $32^2$ tokens instead of $256^2$.
We normalize each quantity $i$ of each frame by its spatial mean $\mu_{t,i}$ and its largest absolute deviation $s_{t,i}$ from that mean, $\tilde{\mathbf{u}}_{t,i} = (\mathbf{u}_{t,i} - \mu_{t,i}) / s_{t,i}$.
When $s_{t,i}$ is below a small threshold, the normalized field is set to zero, so that the numerical noise of a nearly constant field is not amplified (\cref{sec:supp-impl}).
The normalized field is copied to the three color channels and encoded as one image by the encoder $\mathcal{E}$ of the Wan2.1 video autoencoder~\cite{wan2025wan}, which gives $16$ latent channels.
The decoder $\mathcal{D}$ returns three channels, which we average.
Every physical frame is encoded as a separate image, without the temporal compression of four video frames into one latent frame, so each physical frame is one latent frame.
The encoder stays frozen, so the latent space is the one on which the video model was pretrained.
Quantities are assigned to nine slots shared by all families. The slots are vorticity, two velocity components, pressure, density, buoyancy, a passive tracer and two chemical species.
A family fills only its own slots, and empty slots are zero and do not enter the loss.
Normalization removes the amplitude and the mean of each frame, so two constant channels per slot store them relative to the last history frame (\cref{eq:planes}).
The model generates these channels with the latents and can thus predict a decaying amplitude or a drifting mean, as in decaying turbulence.
A predicted frame is reconstructed as $\hat{\mathbf{u}}_{t,i} = \hat\mu_{t,i} + \hat s_{t,i}\, \mathcal{D}(\hat{\mathbf{z}}_{t,i})$.

\paragraph{Backbone.}
The backbone is the Wan2.1 text-to-video diffusion transformer with $1.3$B parameters (30 blocks, width 1536, 12 heads)~\cite{wan2025wan,peebles2023dit}, and we fine-tune all of its weights.
Its patch embedding maps each $2\times2$ block of latent cells of one frame to a token, and the four future frames lie on its time axis.
We expand its input embedding and output head from one video to the nine quantity slots and initialize each slot with a copy of the pretrained layer.
A single-quantity field then enters exactly like the latent of a video frame.
The output weights for the amplitude and mean channels are initialized to zero.

\paragraph{Conditioning.}
Each physical input uses the conditioning pathway that plays the same role in video generation (\cref{fig:arch}).
\emph{Coefficients} share the pathway of the noise level.
A numeric vector $\mathbf{n}$ holds $\mathbf{p}_t$ in fixed slots, which cover the coefficients of all families, with a presence mask.
It also holds the reference amplitude and mean of each quantity (\cref{sec:supp-impl}).
A small MLP maps $\mathbf{n}$ to an embedding that is added to the timestep embedding, so it modulates every block through adaptive layer normalization (adaLN)~\cite{peebles2023dit}.
\emph{Condition fields} share the pathway of a reference frame, together with the history.
Each condition field is encoded by the same autoencoder after a fixed scaling that preserves its absolute level. The forcing is divided by $16$, $\log_{10}$ of the density by $5$ and $\log_{10}$ of the sound speed by $2.5$.
Its latent is concatenated to the input channels alongside the history latents, as in image-to-video models.
A presence channel is concatenated with it, and the history latents are repeated at every output position.
The \emph{equation text} $e_f$ takes the pathway of the prompt. It is encoded by the frozen umT5~\cite{chung2023unimax}, the text encoder of Wan2.1, and attended to by the existing cross-attention layers.
The released $1.3$B model has no image-to-video weights.
All new input channels (history, condition fields and the mask of present slots) and the numeric MLP are therefore initialized to zero.
At the first training step the network then computes the pretrained function.
With the expanded embedding and head, the numeric MLP and the new input channels, the model has $1.43$B parameters in total.
We write $\mathbf{v}_\theta(\mathbf{x}, \sigma;\, \mathbf{h}, \mathbf{n}, \mathbf{g}, e)$ for the network, where $\mathbf{h}$ denotes the clean history latents with their amplitude and mean channels.

\paragraph{Objective and rollout.}
We train with flow matching~\cite{lipman2023flowmatching,liu2022rectifiedflow} in the convention of Wan2.1.
Let $\mathbf{x}_0$ be the clean future latents with their amplitude and mean channels, $\boldsymbol{\epsilon}\sim\mathcal{N}(0,I)$ the noise and $\sigma\in[0,1]$ the noise level.
The network input is $\mathbf{x}_\sigma = (1-\sigma)\,\mathbf{x}_0 + \sigma\,\boldsymbol{\epsilon}$, and the loss is
\begin{equation}
  \mathcal{L} = \mathbb{E}\, \bigl\lVert \mathbf{m} \odot \bigl(\mathbf{v}_\theta(\mathbf{x}_\sigma, \sigma;\, \mathbf{h}, \mathbf{n}, \mathbf{g}, e) - (\boldsymbol{\epsilon} - \mathbf{x}_0)\bigr) \bigr\rVert^2,
  \label{eq:supp-fm}
\end{equation}
where $\mathbf{m}$ masks the empty slots.
The noise level $\sigma$ follows a shifted logit-normal distribution~\cite{esser2024sd3}.
History frames are never noised.
At test time we integrate the learned flow from noise with an Euler sampler on a shifted grid of noise levels.
In a rollout, the last two generated latents and their amplitude and mean channels become the next history, and frames are decoded only for evaluation.

\paragraph{Decoder fine-tuning and periodic domains.}
The pretrained video decoder adds high-frequency artifacts to fine-scale fields, which limits the accuracy of the first predicted frames.
As in WinDiNet~\cite{perini2026windinet}, we therefore fine-tune the decoder and keep the encoder frozen.
The decoder is trained on single-quantity frames of all training families, with reconstruction, gradient, spectral and adversarial losses (\cref{eq:dec}).
A crop of a periodic field is not periodic.
Each training frame therefore keeps its periodic axes whole, circularly shifted by a random offset, and only non-periodic axes are cropped.
Since the encoder and the latent space are unchanged, the trained transformer works with the new decoder without retraining.
Four of our families are periodic in both directions (Kolmogorov flow, decaying turbulence, Gray--Scott and shear flow).
The video model is not equivariant to translations of a periodic domain, because the video autoencoder pads with zeros and the rotary positions of the transformer do not repeat over the domain.
We therefore handle periodic axes with three changes.
First, the encoder and the fine-tuned decoder use circular padding.
Second, we round the frequencies of the rotary position embedding~\cite{su2024roformer} to harmonics of the periodic token grid, so that attention depends on token positions only modulo the domain size.
Third, during training each window is shifted by a random whole number of latent cells along every periodic axis, as in the patch jittering of Walrus~\cite{mccabe2026walrus}.
Non-periodic sides instead receive a zero-initialized embedding of the distance to the boundary.

\paragraph{Switch training.}
A switch trajectory is simulated under $A$ up to frame $t_s$ and under $B$ afterwards, where $A$ and $B$ are training values of the family that differ in one condition.
After the change its condition is $c^B$, while the frame interval stays the same.
Frames before the change were simulated under $c^A$, so they cannot serve as targets under $c^B$.
We therefore take training windows from a switch trajectory only where all four predicted frames lie after $t_s$.
These windows are trained with the condition $c^B$, and frames before the change enter only as history.
The loss on a switch trajectory averages over these windows,
\begin{equation}
  \mathcal{L}_{\mathrm{sw}} = \mathbb{E}_{j \in \{0,\dots,J\}}\, \mathcal{L}\bigl(\mathbf{x}_0^{(j)},\, \mathbf{h}^{(j)},\, c^B\bigr),
  \label{eq:switch}
\end{equation}
where $\mathcal{L}$ is the loss of \cref{eq:supp-fm} for one window, $\mathbf{x}_0^{(j)}$ holds the latents of frames $t_s+j+1, \dots, t_s+j+4$ and $\mathbf{h}^{(j)}$ those of the history frames $t_s+j-1$ and $t_s+j$.
The window $j=0$ has its whole history under $A$ and its targets under $B$, exactly as at test time.
Later windows, up to $J=16$ steps after the change, cover the transient.
We simulate switch trajectories for Kolmogorov flow only.
In the switch training of PhysDial, switch windows replace at most a share of $0.21$ of the samples of this family, and the other families are trained as before.
In the controlled study, switch windows enter without a cap.
Their share is that of the switch trajectories among the trajectories of the family, also $0.21$.
In both settings the mixture of families is unchanged, and switch windows are about $5\%$ of all training samples.
A control with the same compute, trained on fixed conditions only, separates the effect of the data from that of the extra training.
Training uses $576$ switch trajectories, from $72$ pairs of training values with $8$ trajectories each.
The two values of a pair differ in the amplitude (between $1$ and the anchor values $0.7$ and $1.4$, in both directions), the viscosity, the forcing wavenumber or the forcing pattern.
The random seeds of these trajectories differ from those of the evaluation groups.

\section{Implementation details}
\label{sec:supp-impl}

\begin{table}[t]
  \caption{\textbf{Training stages and budgets.} PhysDial is the model after stage 2, and switch-trained PhysDial is the model after stage 3. The controlled study of switch training repeats stage 3 and its control with two seeds, starting from a stage-2 model trained without periodic handling. The models of the anchor study (\cref{tab:anchors}) and their references are also trained without periodic handling. PhysDial has $1.43$B parameters, and all GPU-hours are on NVIDIA H200 GPUs. Before our fine-tuning, Walrus is pretrained for about $400$k training steps on $19$ physics datasets.}
  \label{tab:training}
  \centering
  \scriptsize
  \setlength{\tabcolsep}{3pt}
  \resizebox{\linewidth}{!}{%
  \begin{tabular}{@{}p{3.3cm}p{2.9cm}r@{}}
    \toprule
    Model and stage & Training data & GPU-hours \\
    \midrule
    \multicolumn{3}{@{}l}{\emph{PhysDial}} \\
    1\quad multi-family training & eight families & 60 \\
    2\quad anchors, periodic handling & + amplitude anchors & 20 \\
    3\quad switch training & + switch trajectories & 20 \\
    \phantom{3}\quad control of stage 3 & as stage 2 & 20 \\
    decoder fine-tuning (one GPU) & single-quantity frames & $20$k steps \\
    in-range references & + held-out values & as above \\
    \midrule
    \multicolumn{3}{@{}l}{\emph{Baselines}} \\
    Walrus ($1.3$B), fine-tuning & as stages 1--2 & $68{+}24$ \\
    DPOT-H ($1$B), fine-tuning & eight families, anchors & 80 \\
    U-Net ($\approx 28$M), one GPU & one family & 4 \\
    FNO ($\approx 76$M), one GPU & one family & 4 \\
    \bottomrule
  \end{tabular}}
\end{table}

\paragraph{Training settings.}
\Cref{tab:training} lists the training stages and budgets of all models.
PhysDial is trained with AdamW ($\beta_2 = 0.95$).
After warm-up, the learning rate is constant ($10^{-4}$ in stage 1 and $3{\times}10^{-5}$ afterwards) and decays linearly over the last $15\%$ of the budget of each stage.
Evaluation uses an exponential moving average of the weights.
Each micro-batch comes from one family and one grid size.
It holds a fixed number of latent tokens per GPU (about $16$k), so small and large grids cost about the same per training step.
Training noise levels are logit-normal with shift $3$.
Sampling uses $25$ Euler integration steps and a sampling shift of $8$ for all families.
The controlled study of switch training and the extrapolation costs of \cref{tab:anchors}, where a model is compared with its reference, use the training shift of $3$ instead.
For every stage, an in-range reference model with the same architecture and budget is trained on data that also contain other trajectories at the held-out values.
One full evaluation takes about $2$ GPU-hours, and the switch evaluation takes $1.5$--$2$.

\paragraph{Normalization threshold.}
A frame whose scale $s$ falls below a threshold is encoded as an all-zero frame.
For each family and quantity, the threshold is $10^{-5}$ times the median of the positive per-frame scales over the training frames.
Frames below it are numerically constant, and their rounding noise would otherwise be stretched to $[-1, 1]$.
They occur in Gray--Scott trajectories that decay to a uniform state ($11\%$ of the Gray--Scott frames) and as initial frames at rest in Rayleigh--B\'enard convection, shear flow, the radiative layer and acoustic scattering.

\paragraph{Amplitude and mean channels.}
For every quantity slot, two constant channels store the amplitude and the mean of the frame relative to the last history frame (subscript $\mathrm{ref}$),
\begin{equation}
  a_{t,i} = 4 \log \frac{s_{t,i}}{s_{\mathrm{ref},i}}, \qquad m_{t,i} = 2\,\mathrm{asinh}\frac{\mu_{t,i} - \mu_{\mathrm{ref},i}}{s_{\mathrm{ref},i}},
  \label{eq:planes}
\end{equation}
and the model generates them together with the latents.

\paragraph{Numeric condition vector.}
The vector $\mathbf{n}$ has $51$ entries. They are one value and one presence flag for each of the $12$ coefficient slots, and three entries for each of the nine quantity slots.
The models of the $\beta$ study (\cref{sec:exp-extrap}) have one more coefficient slot, for $\beta$, and $53$ entries.
The slot values (\cref{tab:slots}) are the scalar coefficients of the families and $\log_{10}\Delta t$.
They are standardized with training statistics and set to zero where a family has no such coefficient.
The three entries of a quantity slot are the $\log_{10}$ of the reference amplitude, the $\mathrm{asinh}$ of the reference mean divided by the reference amplitude, and a flag that indicates whether the family fills this slot.
A slot whose value is the same in every training trajectory stays zero after standardization throughout training, so its weights receive no gradient.
An example is the amplitude of Kolmogorov flow without anchors.
Such slots hold the conditions that never vary in \cref{sec:exp-extrap}.

\begin{table}[t]
  \caption{\textbf{Coefficient slots of the numeric condition vector.} A slot stores the logarithm of a coefficient where its name shows one, and the training values are listed without the logarithm. The drag of Kolmogorov flow has no slot. Acoustic scattering has no scalar coefficient, and all of its conditions are condition fields.}
  \label{tab:slots}
  \centering
  \scriptsize
  \setlength{\tabcolsep}{3pt}
  \begin{tabular}{@{}p{2.3cm}p{1.75cm}p{3.75cm}@{}}
    \toprule
    Slot & Family & Training values \\
    \midrule
    $\log_{10}\nu$ & Kolmogorov, decaying & $\nu$ $3.5{\times}10^{-4}$--$1.87{\times}10^{-3}$ \\
    $k_f$ & Kolmogorov & $2$, $4$, and for random forcing the energy-weighted $|k|$, $2$--$5$ \\
    amplitude & Kolmogorov & $1$ (anchors $0.7$, $1.4$) \\
    $F$, $k$ & Gray--Scott & 4 pattern regimes \\
    $\log_{10}\mathrm{Ra}$, $\log_{10}\mathrm{Pr}$ & Rayleigh--B\'enard & $10^6$--$10^9$, $0.1$--$10$ \\
    $\log_{10}\mathrm{Re}$, $\log_{10}\mathrm{Sc}$ & shear flow & Re $10^4$--$10^5$, Sc $0.1$--$10$ \\
    $\log_{10} t_{\mathrm{cool}}$ & radiative layer & $0.03$--$1.78$ (7 values) \\
    $\gamma$ & Euler & $1.22$--$1.60$ (8 values) \\
    $\log_{10}\Delta t$ & all & interval of one step \\
    $\beta$ ($\beta$ study only) & Kolmogorov & $0$ (anchors $\pm2.4$) \\
    \bottomrule
  \end{tabular}
\end{table}

\paragraph{Family mixture.}
Each family is drawn with probability proportional to the square root of its number of training windows.
A family is capped at $10$ expected passes over its training windows per run, and the surplus weight of a capped family goes to the others in proportion to their weights.
Within a family, the grid size is drawn in proportion to its number of trajectories times the tokens per sample.
The trajectory is then drawn uniformly, and the window start uniformly over the eligible frames.
In decaying turbulence, windows start at frame $20$, after the transient from the smooth initial state.
In the radiative layer they start at frame $24$, because earlier frames of different trajectories are too similar (\cref{sec:supp-data}).

\paragraph{Decoder fine-tuning.}
The decoder is fine-tuned on single-quantity frames $\mathbf{u}$ of all training families.
For a reconstruction $\hat{\mathbf{u}}$, the loss is
\begin{equation}
  \begin{split}
  \mathcal{L}_{\mathcal{D}} ={}& \lVert \hat{\mathbf{u}} - \mathbf{u} \rVert_1 + \lambda_{\nabla} \lVert \nabla\hat{\mathbf{u}} - \nabla\mathbf{u} \rVert_1 + \lambda_{\mathrm{adv}} \mathcal{L}_{\mathrm{adv}} \\
  &+ \lambda_{\mathrm{spec}} \bigl\lVert \log E_{\hat{\mathbf{u}}}(\kappa) - \log E_{\mathbf{u}}(\kappa) \bigr\rVert_1,
  \end{split}
  \label{eq:dec}
\end{equation}
where $E(\kappa)$ is the radially averaged power spectrum and $\mathcal{L}_{\mathrm{adv}}$ is the hinge loss of a patch discriminator~\cite{isola2017pix2pix}.
The loss weights are $\lambda_{\nabla}=0.5$, $\lambda_{\mathrm{spec}}=0.1$, $\lambda_{\mathrm{adv}}=0.01$.

\paragraph{Rotary positions on periodic axes.}
On a periodic axis with $N$ tokens, each frequency $\omega_j$ of the rotary position embedding~\cite{su2024roformer} is replaced by the nearest harmonic of the periodic token grid, $2\pi m_j/N$ with integer $m_j$.

\paragraph{Baselines.}
\emph{Walrus}~\cite{mccabe2026walrus} is a $1.3$B PDE foundation model that predicts from six past frames and takes no conditions.
It is pretrained for about $400$k training steps on $19$ physics datasets, $14$ of them two-dimensional with about $4$M training windows each.
Our backbone, in contrast, has seen no physics data before fine-tuning.
We fine-tune Walrus on the same training data as PhysDial with its own recipe (six history frames, per-sample normalization), first on the eight families and then with the anchor trajectories added.
The conditioned variant of \cref{sec:exp-walrus} adds the conditions as constant input fields.
Walrus is trained causally in time and computes a loss at every history position.
For switch trajectories, the restriction of training windows to frames after the change therefore becomes a mask that sets this loss to zero before the change.
\emph{DPOT-H}~\cite{hao2024dpot} ($1$B) is fine-tuned from its released checkpoint on the eight families with the anchor trajectories, for $80$ GPU-hours at a learning rate of $3{\times}10^{-4}$.
It takes no conditions and has no in-range reference model.
The \emph{specialists} are trained from scratch, one model per family for each of the eight families, with $4$ GPU-hours on one GPU each.
The U-Net~\cite{gupta2023pdearena} ($\approx 28$M parameters) has $96$, $192$, $192$ and $384$ channels per level and circular padding on periodic axes.
It takes the coefficients through a Fourier embedding applied with FiLM in every block.
The FNO~\cite{li2021fno} ($\approx 76$M) has $4$ layers of width $96$ with $32$ modes per axis and takes the coefficients as constant input channels.
Both specialists take four history frames, take the condition fields as input channels, and predict the next frame as a residual.
They are trained with AdamW and a cosine schedule, with a learning rate of $3{\times}10^{-4}$ for the U-Net and $10^{-3}$ for the FNO.
Both add Gaussian noise of standard deviation $0.01$ to the history to counter rollout drift, a value chosen on validation splits.
On Kolmogorov flow, the U-Net is also trained with switch trajectories.
On Kolmogorov flow and decaying turbulence, both specialists are trained with two seeds.
The budgets differ in both directions.
A specialist trains for $4$ GPU-hours on its single family, PhysDial shares its budget among all families, and the physics pretraining of Walrus far exceeds all of our fine-tuning.
Poseidon-B is not among the baselines, because fine-tuning it on the solver grids of the eight families was beyond the budget of this study.

\section{Data}
\label{sec:supp-data}
\paragraph{Scope and sources.}
\Cref{tab:supp-overview} lists the eight families.
All are two-dimensional evolution equations on regular grids and cover incompressible and compressible flow, convection, pattern formation and wave propagation.
Four domains are doubly periodic, and the others have walls or open boundaries on some sides.
Three-dimensional, irregular-geometry and steady problems are outside our scope.
Six families come from The Well~\cite{ohana2024well}.
Kolmogorov flow and decaying turbulence are simulated for this work with a pseudo-spectral solver (\cref{tab:supp-solver}).
For the simulator, all fields stay on the solver grid and are never padded, cropped or coarsened.
Walrus has not seen the two simulated flows in pretraining.
For the six families from The Well, the evaluation trajectories may be part of its pretraining data, so the results of Walrus on these families are reported without drawing conclusions from them.

\begin{table*}[t]
  \caption{\textbf{The eight PDE families.} Step is the number of stored frames per step. It is chosen so that one step changes the state by $10$--$20\%$, except in the radiative layer and acoustic scattering, where a single stored frame already changes it by $20$--$40\%$. The last column states whether the family is part of the pretraining data of Walrus.}
  \label{tab:supp-overview}
  \centering
  \scriptsize
  \setlength{\tabcolsep}{3pt}
  \begin{tabular}{@{}p{2.05cm}p{1.35cm}p{2.9cm}p{2.5cm}p{5.6cm}cc@{}}
    \toprule
    Family & Source & Grid, boundaries & Quantities & Conditions varied in training & Step & Walrus \\
    \midrule
    Kolmogorov flow & simulated & $256^2$ or $512^2$ (by $\nu$), doubly periodic & vorticity, with the forcing as condition field & $\nu$ $3.5{\times}10^{-4}$--$1.87{\times}10^{-3}$, $k_f \in \{2, 4\}$ or a random multi-mode forcing in the band $2$--$5$, and amplitude $1$ (anchors $0.7$, $1.4$) & 2 & no \\
    Decaying turbulence & simulated & $256^2$ or $512^2$, doubly periodic & vorticity & $\nu$ $3.5{\times}10^{-4}$--$1.87{\times}10^{-3}$ ($29$ values) & 1 & no \\
    \midrule
    Gray--Scott & The Well & $128^2$, doubly periodic & two species & $4$ of $6$ $(F, k)$ pattern regimes & 16 & yes \\
    Rayleigh--B\'enard & The Well & $512{\times}128$, periodic in $x$, walls in $z$ & buoyancy, pressure, velocity & Ra $10^6$--$10^9$, Pr $0.1$--$10$ & 1 & yes \\
    Shear flow & The Well & $256{\times}512$, doubly periodic & tracer, pressure, velocity & Re $10^4$--$10^5$, Sc $0.1$--$10$ & 8 & yes \\
    Radiative layer & The Well & $128{\times}384$, periodic in $x$, open in $y$ & density, pressure, velocity & cooling time $0.03$--$1.78$ ($7$ of $9$ values) & 1 & partly \\
    Euler, multi-quadrant & The Well & $512^2$, open & density, pressure, velocity & $\gamma$ $1.22$--$1.60$ ($8$ values) & 2 & yes \\
    Acoustic scattering & The Well & $256^2$, walls on two sides, open on two & pressure and velocity, with density and sound speed as condition fields & the medium fields only (no scalar condition) & 1 & yes \\
    \bottomrule
  \end{tabular}
\end{table*}

\begin{table}[t]
  \caption{\textbf{Equations and solver of the two simulated families.}}
  \label{tab:supp-solver}
  \centering
  \scriptsize
  \setlength{\tabcolsep}{3pt}
  \begin{tabular}{@{}p{1.6cm}p{6.4cm}@{}}
    \toprule
    Family & Equation and solver \\
    \midrule
    Kolmogorov flow & $\partial_t\omega + \mathbf{v}\cdot\nabla\omega + \beta v_y = \nu\Delta\omega - \alpha\omega + F$, with forcing $F = -A k_f \cos(k_f(y+\phi))$ or a random combination of four modes in a band of wavenumbers, and drag $\alpha = 0.1$. The $\beta$-plane term, with $v_y$ the velocity along $y$, is zero except in the $\beta$ study of \cref{sec:exp-extrap}. Pseudo-spectral solver with $2/3$ dealiasing, integrating-factor RK4 and a CFL time step. Random band-limited initial vorticity. After spin-up, $300$ frames are stored, $0.1$ eddy-turnover times apart. \\
    Decaying turbulence & The same equation with $F = 0$ and $\alpha = 0$, and the same solver. Random initial vorticity with $\omega_{\mathrm{rms}} = 5$. \\
    \bottomrule
  \end{tabular}
\end{table}

\paragraph{Trajectories.}
\Cref{tab:supp-counts} gives the number of trajectories per family.
In the two simulated families, the trajectories of each training condition are split into training, validation and test sets ($12 / 2 / 2$ on the viscosity axes).
Every trajectory has its own random seed and spin-up.
For every held-out value, $12$ to $24$ trajectories are simulated.
Of these, $4$ are kept for evaluation ($12$ for the forcing band $6$--$7$), and the others are added only to the training data of the in-range reference model.
The reference model is therefore evaluated on the same trajectories as the tested model.
Several precautions keep a model from recognizing an evaluation trajectory.
Shear flow and Rayleigh--B\'enard convection reuse the same initial states under different conditions.
A fixed part of the initial states is therefore reserved for evaluation and never enters training under any condition.
In the radiative layer, windows start at frame $24$, since trajectories from the same initial state are too similar in earlier frames.
One acoustic-scattering trajectory on which the solver diverged is left out.

\begin{table}[t]
  \caption{\textbf{Trajectories per family.}}
  \label{tab:supp-counts}
  \centering
  \scriptsize
  \setlength{\tabcolsep}{3pt}
  \begin{tabular}{@{}p{2.2cm}p{5.8cm}@{}}
    \toprule
    Family & Trajectories \\
    \midrule
    Kolmogorov flow & main sweep with $8$ conditions $\times$ ($144 / 6 / 12$ training / validation / test). Dense $\nu$ axis with $30 \times (12 / 2 / 2)$. Random multi-mode forcing with $324$ training. Anchors with $384$ ($24$ for each of two amplitudes, two $k_f$ and four $\nu$) \\
    Decaying turbulence & $348 / 58 / 58$ \\
    Gray--Scott & $640$ training \\
    Rayleigh--B\'enard & $729$ training, $276$ test (split by initial state) \\
    Shear flow & $435$ training, $146$ test (split by initial state) \\
    Radiative layer & $56$ training \\
    Euler, multi-quadrant & $304$ training \\
    Acoustic scattering & $2797$ training \\
    \bottomrule
  \end{tabular}
\end{table}

\paragraph{Evaluation splits.}
Every family has a \emph{test} split with the training conditions and unseen trajectories.
\Cref{tab:supp-splits} lists the held-out splits.
A held-out split inside the training range (interpolation) has an unseen value within that range, and a split outside it (extrapolation) has a value beyond it.
The other Kolmogorov splits vary the forcing wavenumber, the forcing pattern, the amplitude or the drag, one of them at a time except for the joint split of wavenumber and amplitude.
Through the random multi-mode forcing, the forcing slot $k_f$ takes continuous values between $2$ and $5$.
The single-mode wavenumber $6$ therefore lies just outside the training range, and the forcing band $6$--$7$ is the main extrapolation split in this direction.

\begin{table*}[t]
  \caption{\textbf{Held-out evaluation splits.} Every family also has a test split with the training conditions. Acoustic scattering has only the test split.}
  \label{tab:supp-splits}
  \centering
  \scriptsize
  \setlength{\tabcolsep}{4pt}
  \begin{tabular}{@{}p{2.6cm}p{2.9cm}p{10.6cm}@{}}
    \toprule
    Family & Split & Held-out condition \\
    \midrule
    Kolmogorov flow & $\nu$ inside & $\nu = 8.1{\times}10^{-4}$ \\
     & $\nu$ below, $\nu$ above & $\nu = 1.0$, $1.5$, $2.3{\times}10^{-4}$ and $\nu = 2.8$, $4.3$, $6.6{\times}10^{-3}$, $10^{-2}$ \\
     & dense $\nu$, inside and outside & every fifth value of a dense axis inside the range ($7$ values, $4.0{\times}10^{-4}$ to $1.6{\times}10^{-3}$) and all $63$ values outside it ($1.0$--$3.4{\times}10^{-4}$ and $2.0{\times}10^{-3}$--$10^{-2}$, up to $3.5$ and $5.3$ times beyond the range) \\
     & $k_f = 3$, $k_f = 6$ & single-mode forcing at these wavenumbers \\
     & multi-mode forcing & unseen random forcing patterns in the training band $2$--$5$ \\
     & forcing band $6$--$7$ & random multi-mode forcing in this band \\
     & amplitude $0.5$ & $A = 0.5$ (training $1$, anchors $0.7$ and $1.4$) \\
     & $k_f$ and amplitude & $(k_f, A) = (3, 0.5)$ and $(6, 0.5)$ \\
     & drag $0.05$, drag $0.2$ & $\alpha = 0.05$ and $0.2$ (training $0.1$). The model has no input for the drag, which reaches it only through $\Delta t$ \\
    Decaying turbulence & $\nu$ inside, $\nu$ outside & every fifth value inside the range ($7$ values) and $65$ values outside it, $28$ down to $10^{-4}$ and $37$ up to $10^{-2}$ \\
    \midrule
    Gray--Scott & maze, bubbles & the maze pattern regime (inside the range) and the bubble regime (largest $F$) \\
    Rayleigh--B\'enard & Pr $= 2$, Ra $= 10^{10}$ & \\
    Shear flow & Sc $= 2$, Re $= 5{\times}10^5$ & \\
    Radiative layer & $t_{\mathrm{cool}} = 0.32$, $3.16$ & $18$ windows each \\
    Euler, multi-quadrant & $\gamma = 1.13$, $\gamma = 1.76$ & \\
    \bottomrule
  \end{tabular}
\end{table*}

\paragraph{Switch evaluation groups.}
On Kolmogorov flow, ten evaluation groups of $32$ trajectories each have a single condition change after stored frame $10$.
Five groups change to a value inside the training range. They are forcing wavenumber $2\to4$, single-mode to multi-mode forcing, amplitude $1\to1.4$, and an increase and a decrease of viscosity.
Five change to a value outside it. They are wavenumber $2\to6$, a forcing band outside the training band, amplitude $1\to0.5$, and viscosity up to $6.6\times10^{-3}$ and down to $10^{-4}$.
Five further groups change conditions several times within one $64$-step rollout.
They change the forcing pattern repeatedly, the forcing wavenumber and back, several conditions in turn, the amplitude, and the viscosity.
In these groups, the conditions change every $16$ steps, four times per rollout, starting with the first predicted step.
The amplitude changes every $32$ steps, twice per rollout.

\paragraph{Quality checks.}
Every family passes two checks.
First, with the frozen encoder and the fine-tuned decoder, the video autoencoder reconstructs every evaluation split and quantity with a relative $L_2$ error of at most $0.05$ and a ratio of gradient energies in $[0.94, 1.01]$ ($24$ frames per split). The drag splits of Kolmogorov flow were not checked.
The largest error is that of Rayleigh--B\'enard convection ($0.050$).
With the pretrained decoder, Gray--Scott and Rayleigh--B\'enard convection slightly exceed $0.05$ ($0.051$--$0.059$).
Second, one step changes the state by $10$--$20\%$, except in the two families noted in \cref{tab:supp-overview}.

\section{Evaluation protocol}
\label{sec:supp-protocol}
\paragraph{Windows and metrics.}
Most evaluation splits have $32$ windows, and the held-out splits of the radiative layer have $18$.
Each window comes from a trajectory never used in training.
Where a split has too few trajectories, one trajectory provides several non-overlapping windows.
All models use the same windows and start predicting from the same frame.
Walrus takes six history frames, and our model takes the last two of them.
Models roll out $16$ steps without further true frames.
We report the per-frame relative $L_2$ error $\lVert \hat{\mathbf{u}}_k - \mathbf{u}_k \rVert / \lVert \mathbf{u}_k \rVert$, averaged over the windows of a split, at the end of the first round ($k=4$) and of the fourth round ($k=16$, our main horizon).
On the two turbulent flows we also roll out $64$ steps and report the error at the end of round 16.
With the steps of \cref{tab:supp-overview}, the trajectories of half of the families from The Well are too short for such a rollout.
Some design choices act mainly on the first predicted frame, and errors at $k=1$ are always marked as such.
Generative models draw one sample per window, with the sampling noise fixed per window.
Errors are never averaged across families or splits, since one step corresponds to a different amount of change in each family.
Two further reference values help to interpret the errors.
The persistence baseline copies the last history frame.
The video autoencoder reconstructs the true frame, and no model in this latent space can have a lower error than this reconstruction.

\paragraph{Comparing two models.}
We compare models split by split.
For each split we report the geometric mean of the per-window error ratio, with its $95\%$ interval from a bootstrap over trajectories ($2000$ resamples).
We also report the share of windows on which the first model has the lower error and the $p$-value of a Wilcoxon signed-rank test.
We call a model \emph{more accurate} on a split only when the interval of the ratio does not contain $1$.
When it contains $1$ and the ratio is within $\pm8\%$, the split is treated as a tie.
The band follows from the variation between seeds (\cref{tab:supp-seeds}).
Between the two seeds of the anchor study, no Kolmogorov split changes by more than $7.8\%$.
Between two seeds of PhysDial, only the test split of shear flow changes by more than $8\%$ (by $28\%$ at step 16).
Between two runs of Walrus, no split does.
In all other cases we state only which model has the lower error.
The same holds for the specialists, which have no per-window records.

\paragraph{Selection, seeds and single runs.}
The windows of every split are divided into two halves by a fixed hash of the trajectory, or of the initial state for shear flow and Rayleigh--B\'enard convection.
Choices that depend on results, such as the sampling shift, were made on one half.
We report all windows.
The anchor costs, the step-64 counts and the direction of the comparisons with Walrus agree between the two halves, while the number of splits with a significant difference changes by up to two.
``Two seeds'' denotes two seeds of the last training stage, both started from the same checkpoint.
They are not an independent repetition of the whole training.
A result comes from a single run unless two seeds are stated.

\paragraph{Responses to a change.}
For a single change we report, at step 16, the response $r_{16}$ of \cref{eq:supp-response} and the error with respect to the changed branch.
This error is given both when the model receives $c^B$ and when it receives $c^A$.
The two runs use the same sampling noise, so they differ only through the condition input.
The effect of a change is the relative difference of the two solver branches, $\lVert\mathbf{u}^B_k - \mathbf{u}^A_k\rVert / \lVert\mathbf{u}^B_k\rVert$.
The response $r_k$ is not reported when the effect is below $0.05$, and we treat it as indicative between $0.05$ and $0.1$.
For several changes within one rollout we report a tracking error on one statistic $X$ of the flow.
This statistic is the kinetic energy for changes of amplitude, the enstrophy for changes of viscosity, and the projection of the flow on the current forcing otherwise.
Let $D_k$ be the difference of $X$ between the changed and the unchanged solver branch at step $k$.
Let $\hat D_k$ be the difference between the predictions that receive the changing conditions and those that receive the initial condition, both averaged over the windows.
The tracking error is $\mathrm{rms}_k(\hat D_k - D_k) / \mathrm{rms}_k(D_k)$.
It is $0$ when the model follows every change and $1$ when it does not respond, as for any model that uses only the history.

\paragraph{Extrapolation cost.}
For a held-out value we train an in-range reference model with the same architecture and compute budget, whose training data also contain other trajectories at the held-out value.
On the same windows, we define
\begin{equation}
  \mathrm{cost} = \frac{\mathrm{err}(\text{model})}{\mathrm{err}(\text{in-range reference})}
  \label{eq:cost}
\end{equation}
A cost of $1$ means that leaving the training range costs nothing.
The reference error measures how hard the held-out condition is in itself.
An absolute error alone mixes these two factors.
We use the cost in the anchor study and compare different models by their absolute errors.

\section{Additional results}
\label{sec:supp-results}

\subsection{Use of the condition inputs (H1)}
All models compared in this subsection are trained for the $60$ GPU-hours of the first stage.

\paragraph{Without coefficient inputs.}
We train the same network without the coefficient slots, keeping the equation text and the condition fields as inputs.
On $21$ splits, the step-16 error of this model is $0.97$--$1.05$ times that of the model with the slots.
These splits cover Kolmogorov flow and decaying turbulence inside the training range, Rayleigh--B\'enard convection, the radiative layer, Euler flow and acoustic scattering.
The slots help where a coefficient sets slow statistics.
Without them, the model has $1.13$--$1.14$ times the error on the high-viscosity extrapolation and the held-out forcing band of Kolmogorov flow.
It has $1.13$ times the error on the extrapolation split of decaying turbulence and $1.36$ times on the test split of shear flow.
At values far outside the training range, the slots can mislead the model.
Without them, the error is lower on the amplitude and joint extrapolation splits of Kolmogorov flow (error ratio $0.68$--$0.78$) and on a held-out pattern of Gray--Scott ($0.66$).

\paragraph{Recovering the condition from the history.}
A linear probe on physical statistics of two history frames recovers $\log\nu$ with $R^2 = 0.93$, the forcing wavenumber with $0.98$ and the $\beta$-plane coefficient with $0.88$.
Six history frames give no improvement over two.

\paragraph{Without condition fields.}
A model trained without condition fields has $2.26$ times the error on random multi-mode forcing and $1.93$ times on the held-out forcing band.
It has $1.61$ times the error on the single-mode wavenumber extrapolation and $1.76$ times on acoustic scattering, where the condition field is the medium.
On single-mode forcing inside the training range, it has only $1.11$--$1.23$ times the error, since two frames already show a single-mode pattern.

\subsection{Condition changes (H2)}

\begin{table}[t]
  \caption{\textbf{Switch training on Kolmogorov flow, controlled study} (two seeds each of switch training and its control, one change, step 16, $32$ windows per group, sampling shift $3$). The effect is the relative difference of the two solver branches. The error is the relative $L_2$ error with respect to the changed branch when the model receives the new condition. Ranges span the two seeds. The ratio is the error of the switch-trained model divided by that of its same-compute control, per seed.}
  \label{tab:switch-kolmo}
  \centering
  \scriptsize
  \setlength{\tabcolsep}{3pt}
  \resizebox{\linewidth}{!}{%
  \begin{tabular}{@{}lcccccc@{}}
    \toprule
    & & \multicolumn{2}{c}{switch training} & \multicolumn{2}{c}{control} & ratio \\
    \cmidrule(lr){3-4} \cmidrule(lr){5-6}
    Change ($A \to B$) & effect & error & $r_{16}$ & error & $r_{16}$ & seed 0 / 1 \\
    \midrule
    \multicolumn{7}{@{}l}{\emph{inside the training range}} \\
    wavenumber $2\to4$ & 0.81 & 0.23 & 0.88 & 0.48 & 0.70 & 0.48 / 0.48 \\
    forcing pattern & 0.78 & 0.20 & 0.95 & 0.26 & 0.94 & 0.77 / 0.79 \\
    amplitude $1\to1.4$ & 0.30 & 0.14--0.15 & 0.74--0.75 & 0.29 & 0.40--0.42 & 0.50 / 0.51 \\
    viscosity up & 0.09 & 0.09 & 1.04--1.05 & 0.10 & 0.86--0.87 & 0.93 / 0.96 \\
    viscosity down & 0.08 & 0.09 & 0.84--0.92 & 0.14 & 0.40--0.42 & 0.63 / 0.63 \\
    \midrule
    \multicolumn{7}{@{}l}{\emph{outside the training range}} \\
    wavenumber $2\to6$ & 0.85 & 0.29--0.30 & 0.83--0.84 & 0.65--0.66 & 0.52 & 0.45 / 0.46 \\
    forcing band & 0.76 & 0.29 & 0.97 & 0.44 & 0.79 & 0.66 / 0.65 \\
    amplitude $1\to0.5$ & 0.46 & 0.22--0.23 & 0.77 & 0.42--0.44 & 0.62 & 0.54 / 0.50 \\
    viscosity to $6.6{\times}10^{-3}$ & 0.21 & 0.08 & 1.03--1.04 & 0.16--0.17 & 0.93--0.94 & 0.51 / 0.47 \\
    viscosity to $10^{-4}$ & 0.07 & 0.21 & $-0.15$ & 0.15 & 0.06--0.07 & 1.35 / 1.38 \\
    \bottomrule
  \end{tabular}}
\end{table}

\paragraph{Controlled study.}
\Cref{tab:switch-kolmo} gives the numbers behind \cref{fig:switch-study}.
Switch training about halves the error after a change of forcing wavenumber or amplitude.
For the forcing pattern, which the control already follows, the error falls by about a fifth.
The two seeds differ by at most $0.04$ in every ratio.
When the viscosity drops to $10^{-4}$, far outside the training range, the switch-trained model moves away from the changed branch ($r_{16} < 0$) and has a higher error than the control.
Three of the four viscosity changes alter the solution by less than $0.1$ by step 16, so their responses are only indicative.

\begin{table}[t]
  \caption{\textbf{Tracking error} under several condition changes in one $64$-step rollout of Kolmogorov flow. These are the numbers behind \cref{fig:tracking}. The tracking error is $0$ when a model follows every change and $1$ when it does not respond. The control is trained on fixed conditions for the same compute, and the switch-trained model is trained with switch trajectories. PhysDial entries give seed 0 / seed 1. For the viscosity, the difference between switch training and its control is not significant. A model without a condition input has a tracking error of $1$ by definition.}
  \label{tab:supp-te}
  \centering
  \scriptsize
  \setlength{\tabcolsep}{3pt}
  \resizebox{\linewidth}{!}{%
  \begin{tabular}{@{}lcccc@{}}
    \toprule
    & \multicolumn{2}{c}{PhysDial (controlled study)} & \multicolumn{2}{c}{Walrus + condition input (one run)} \\
    \cmidrule(lr){2-3} \cmidrule(lr){4-5}
    Repeated changes of & control & switch-trained & control & switch-trained \\
    \midrule
    forcing pattern & 0.11 / 0.09 & 0.09 / 0.09 & 0.79 & 0.77 \\
    forcing wavenumber and back & 0.25 / 0.24 & 0.17 / 0.17 & 0.75 & 0.72 \\
    several conditions in turn & 0.19 / 0.20 & 0.09 / 0.10 & 0.73 & 0.70 \\
    amplitude & 0.45 / 0.44 & 0.14 / 0.15 & 1.11 & 0.97 \\
    viscosity & 0.31 / 0.28 & 0.25 / 0.30 & 0.97 & 0.88 \\
    \bottomrule
  \end{tabular}}
\end{table}

\paragraph{Several changes in one rollout.}
\Cref{tab:supp-te} gives the tracking error for the five groups with repeated changes.
In the controlled study, the gain for the amplitude is far larger than the bootstrap intervals of the tracking error in both seeds.
The response in later segments is as large as in the first, so the model continues to follow the changes after the first one.
Even without switch training, our model follows changes of the forcing much better than Walrus with an added condition input.
It also partly follows changes of amplitude and viscosity.

\begin{figure*}[t]
  \centering
  \includegraphics[width=\linewidth]{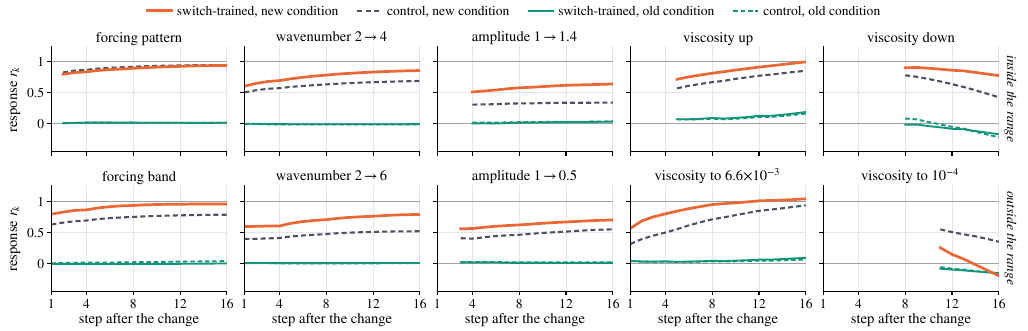}
  \caption{\textbf{Response at every step after a change} (Kolmogorov flow, switch-trained PhysDial and its same-compute control, mean over $32$ windows). At $0$ the prediction is on the unchanged branch, and at $1$ it is on the changed branch. Steps at which the two solver branches differ by less than $0.05$ are not drawn, since the response there is dominated by noise.}
  \label{fig:supp-response}
\end{figure*}

\paragraph{Switch training of PhysDial.}
For PhysDial itself, we compare the switch-trained model with its same-compute control on the ten single-change groups at step 16.
The error after the change, divided by that of the control, has a median of $0.61$ over the groups (range $0.48$--$1.65$).
The largest value occurs for the viscosity drop to $10^{-4}$.
The median response rises from $0.62$ to $0.83$.
The ratio of the error under the new condition to that under the old condition falls from $0.80$ to $0.45$.
Receiving the new condition thus lowers the error by a fifth for the control and by more than half for the switch-trained model.
On the same histories without a change, the error ratio of the two models is $0.96$.
\Cref{fig:supp-response} shows the response at every step.
With the old condition as input, both models stay on the unchanged branch.
With the new condition, the switch-trained model responds more strongly than its control at every step in eight groups.
The two responses are equal for the forcing pattern inside the training range, which the control already follows.
After the viscosity drop to $10^{-4}$, the switch-trained model moves in the wrong direction.

\begin{table}[t]
  \caption{\textbf{Switch-trained PhysDial with the new and with the old condition} (Kolmogorov flow, step 16, medians over the $32$ windows of a group). Windows is the number of windows that pass the selection rule. The effect is the relative difference of the two solver branches. The error is the relative $L_2$ error with respect to the changed branch. With the old condition as input, the model stays on the unchanged branch ($r_{16} \approx 0$). Where the effect is below $0.1$, the response is only indicative.}
  \label{tab:supp-cases}
  \centering
  \scriptsize
  \setlength{\tabcolsep}{3pt}
  \resizebox{\linewidth}{!}{%
  \begin{tabular}{@{}lcccc@{}}
    \toprule
    & & & error & $r_{16}$ \\
    Change & windows & effect & new / old & new / old \\
    \midrule
    pattern, one $\to$ several modes & 32 / 32 & 0.78 & 0.18 / 0.77 & 0.94 / 0.02 \\
    wavenumber $2\to4$ & 32 / 32 & 0.82 & 0.23 / 0.84 & 0.86 / $-0.01$ \\
    forcing band (outside) & 32 / 32 & 0.76 & 0.30 / 0.78 & 0.95 / $-0.01$ \\
    wavenumber $2\to6$ (outside) & 32 / 32 & 0.86 & 0.31 / 0.86 & 0.78 / 0.00 \\
    amplitude $1\to0.5$ (outside) & 17 / 32 & 0.46 & 0.21 / 0.46 & 0.71 / 0.00 \\
    amplitude $1\to1.4$ & 6 / 32 & 0.29 & 0.15 / 0.29 & 0.65 / 0.03 \\
    viscosity up & 0 / 32 & 0.09 & 0.08 / 0.11 & 0.99 / 0.15 \\
    viscosity down & 0 / 32 & 0.08 & 0.09 / 0.12 & 0.80 / $-0.14$ \\
    viscosity to $6.6{\times}10^{-3}$ (outside) & 4 / 32 & 0.21 & 0.07 / 0.21 & 1.03 / 0.07 \\
    viscosity to $10^{-4}$ (outside) & 0 / 32 & 0.06 & 0.18 / 0.10 & $-0.15$ / $-0.14$ \\
    \bottomrule
  \end{tabular}}
\end{table}

\begin{figure*}[t]
  \centering
  \includegraphics[page=1]{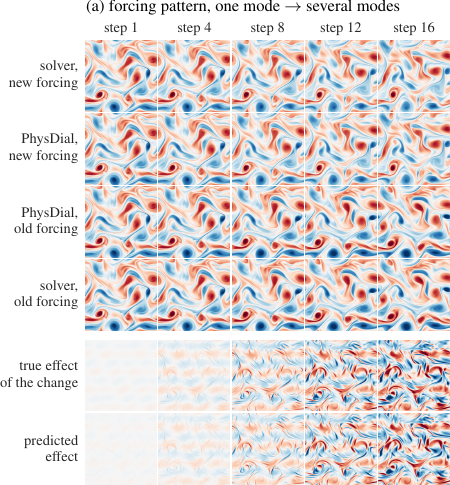}\hfill
  \includegraphics[page=2]{figs/cases.pdf}
  \caption{\textbf{Two windows under a change of forcing} (switch-trained PhysDial, Kolmogorov flow). The bottom rows compare the true effect of the change (the difference of the two solver branches) with the predicted effect (the difference between the prediction under the new forcing and the unchanged branch). (a) The window of \cref{fig:task,fig:supp-zoom}. At step 16 the effect is $0.99$. With the new and the old forcing, the error is $0.18$ and $0.99$ and the response is $0.95$ and $0.01$. (b) The window of \cref{fig:change}, with effect $0.86$, errors $0.21$ and $0.87$, and responses $0.89$ and $0.00$.}
  \label{fig:supp-cases}
\end{figure*}

\begin{figure}[t]
  \centering
  \includegraphics[width=\linewidth]{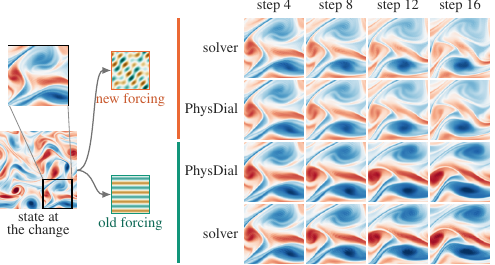}
  \caption{\textbf{A change of the forcing pattern in close-up} (switch-trained PhysDial, Kolmogorov flow, window (a) of \cref{fig:supp-cases}). The forcing changes from one mode to several. The layout is that of \cref{fig:change} (left), and the tiles show the marked square.}
  \label{fig:supp-zoom}
\end{figure}

\begin{figure}[t]
  \centering
  \includegraphics[width=\linewidth]{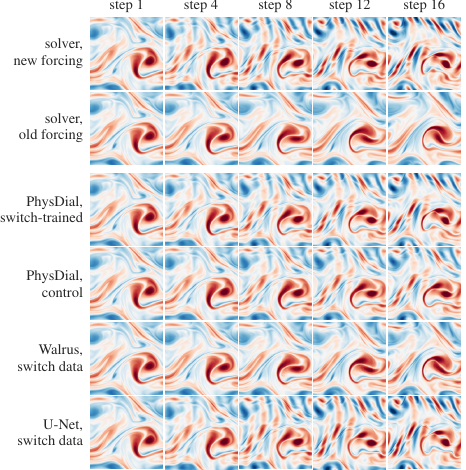}
  \caption{\textbf{Four models after a change of forcing wavenumber from $2$ to $6$}, a value outside the training range (one window). Every model receives the new condition. At step 16, the response and the error with respect to the changed branch are $0.80$ and $0.34$ for switch-trained PhysDial and $0.56$ and $0.66$ for its control. They are $0.13$ and $0.90$ for Walrus with a condition input and $0.82$ and $0.53$ for the U-Net specialist, both trained with switch trajectories.}
  \label{fig:supp-compare}
\end{figure}

\begin{figure*}[t]
  \centering
  \includegraphics[width=\linewidth]{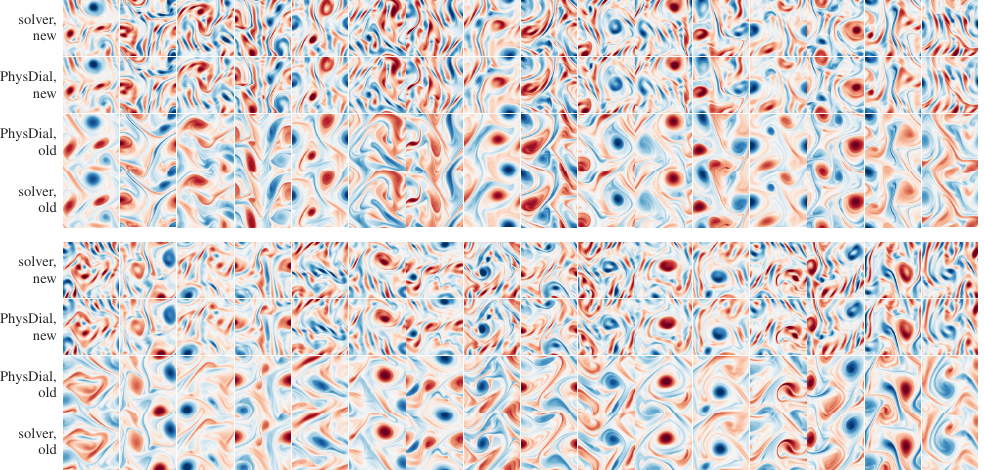}
  \caption{\textbf{All $32$ windows of one group} at step 16 (Kolmogorov flow, forcing wavenumber $2\to6$, one column per window). Given the new condition, switch-trained PhysDial shows the fine bands of the changed branch. Given the old condition, it shows the large vortices of the unchanged branch.}
  \label{fig:supp-mosaic}
\end{figure*}

\paragraph{Cases and their selection.}
The windows shown in \cref{fig:change,fig:supp-cases,fig:supp-compare} are selected by a fixed rule.
A window qualifies when it meets three criteria at step 16.
First, the effect of the change is at least $0.25$.
Second, the prediction with the new condition is clearly closer to the changed branch, with an error of at most $0.6$ times its distance to the unchanged branch.
Third, the prediction with the old condition stays closer to the unchanged branch than to the changed one, which is equivalent to a response below $0.5$.
Among the qualifying windows, we show those with the highest score.
The score is the effect times the difference between two quantities under the new condition. These are the distance of the prediction to the unchanged branch and its error with respect to the changed branch.
Each tile of \cref{fig:change} shows a square whose side is half that of the domain.
It is the square in which the two solver branches differ most, summed over the four displayed steps.
\Cref{fig:supp-zoom} shows window (a) of \cref{fig:supp-cases} in the same layout, with a smaller square.
\Cref{tab:supp-cases} gives the number of qualifying windows and the medians over all windows of each group.
After a change of the forcing pattern or wavenumber, every window qualifies, both inside and outside the training range.
The error is $0.2$--$0.3$ with the new condition and about $0.8$ with the old one.
\Cref{fig:supp-mosaic} shows all $32$ windows of one such group.
After a decrease of the amplitude, about half of the windows qualify, and after an increase only six.
After a change of viscosity at most four windows qualify, since three of the four changes alter the solution by less than $0.1$ by step 16.
The model therefore does not follow every kind of change equally well.

\paragraph{Cost under fixed conditions.}
Under fixed conditions, the third training stage changes the error little, with or without switch trajectories.
Over the $33$ splits of the standard evaluation, the error of switch-trained PhysDial divided by that of its control has a median of $0.99$ at both step 4 and step 16.
The error of the control divided by that of PhysDial has a median of $0.99$ at step 4 and $1.00$ at step 16.
On no split is the error of the control more than $5\%$ higher.
The changes are concentrated on the extrapolation splits of Kolmogorov flow and go in both directions (\cref{tab:supp-cost}).
On the two forcing extrapolation splits, the higher errors are far outside the variation between seeds (\cref{tab:supp-seeds}).
They recur in the controlled study, with error ratios of $1.36$ and $1.22$ against the control in both seeds (decoded with the pretrained decoder).
We therefore treat them as a real cost.

\begin{table}[t]
  \caption{\textbf{Splits most affected by switch training} under fixed conditions (error of switch-trained PhysDial divided by that of PhysDial). The test split of shear flow also varies by $28\%$ between seeds, so we do not treat it as evidence.}
  \label{tab:supp-cost}
  \centering
  \scriptsize
  \setlength{\tabcolsep}{4pt}
  \resizebox{\linewidth}{!}{%
  \begin{tabular}{@{}lc@{}}
    \toprule
    Split (step 16) & switch-trained / PhysDial \\
    \midrule
    Kolmogorov flow, drag $0.05$ & 0.74 \\
    Kolmogorov flow, forcing band $6$--$7$ & 1.34 \\
    Kolmogorov flow, $k_f = 6$ & 1.25 \\
    Shear flow, test & 1.20 \\
    \bottomrule
  \end{tabular}}
\end{table}

\paragraph{Pretraining and following a change.}
Pretraining also helps the model follow changes.
\Cref{tab:supp-init} compares the model after the first stage with the same network trained from random initialization on the same data and with the same compute.
Neither has seen anchor or switch trajectories.
On all six groups with a change of forcing or amplitude, the randomly initialized network responds less and has the higher error, at both step 4 and step 16 (sign test, $p \approx 0.03$).
Its error under fixed conditions is also $15\%$ higher, so part of the difference may reflect a generally weaker model.
However, after a change of the forcing pattern or wavenumber, it already follows the change only about half as far as the pretrained model at step 4 (median $r_4$ of $0.21$--$0.56$ against $0.40$--$0.86$).
Accumulated error cannot explain a difference of this size after four steps.
We make no claim about the mechanism.

\begin{table}[t]
  \caption{\textbf{Video pretraining and following a change} (Kolmogorov flow, one change, step 16). The model after the first stage is compared with the same network trained from random initialization. Neither received switch training. Errors are with respect to the changed branch.}
  \label{tab:supp-init}
  \centering
  \scriptsize
  \setlength{\tabcolsep}{4pt}
  \resizebox{\linewidth}{!}{%
  \begin{tabular}{@{}lcc@{}}
    \toprule
    & $r_{16}$ & error \\
    Change & pretrained / random & pretrained / random \\
    \midrule
    amplitude down (outside) & 0.84 / 0.73 & 0.17 / 0.31 \\
    amplitude up & 0.76 / 0.60 & 0.15 / 0.26 \\
    wavenumber & 0.69 / 0.53 & 0.50 / 0.58 \\
    wavenumber (outside) & 0.52 / 0.37 & 0.69 / 0.76 \\
    forcing pattern & 0.92 / 0.79 & 0.29 / 0.39 \\
    forcing pattern (outside) & 0.77 / 0.58 & 0.49 / 0.58 \\
    \bottomrule
  \end{tabular}}
\end{table}

\paragraph{The same data on Walrus.}
Trained on fixed conditions, Walrus with an added condition input has a tracking error between $0.73$ and $1.11$ (\cref{tab:supp-te}).
After a change of amplitude, its prediction is slightly further from the changed branch than without the new condition.
With switch trajectories, its error after a single change inside the training range is $0.93$--$1.02$ times that of its control.
Its response is $0.11$--$0.12$ for changes of the forcing, $0.03$ for the amplitude, and $0.64$--$0.76$ for the viscosity.
Its share of switch trajectories is no smaller than ours, so a lack of data is not the obvious explanation.
A linear probe points to an alternative.
After the per-sample normalization of Walrus, the forcing amplitude can hardly be recovered from the input ($R^2$ falls from $0.63$ to $0.09$).
This hypothesis is supported by the probe but was not ablated.

\begin{table}[t]
  \caption{\textbf{U-Net specialist trained without and with switch trajectories} (Kolmogorov flow, $4$ GPU-hours each, one change, step 16, on the $32$ windows per group of \cref{tab:switch-kolmo}). Errors are with respect to the changed branch. Three viscosity changes (up, down and to $10^{-4}$) have an effect below $0.1$, so their responses are only indicative.}
  \label{tab:supp-unet-switch}
  \centering
  \scriptsize
  \setlength{\tabcolsep}{4pt}
  \resizebox{\linewidth}{!}{%
  \begin{tabular}{@{}lcc@{}}
    \toprule
    & $r_{16}$ & error \\
    Change ($A \to B$) & without / with & without / with \\
    \midrule
    \multicolumn{3}{@{}l}{\emph{inside the training range}} \\
    wavenumber $2\to4$ & 0.70 / 0.91 & 0.57 / 0.23 \\
    forcing pattern & 0.88 / 0.90 & 0.23 / 0.21 \\
    amplitude $1\to1.4$ & 0.59 / 0.86 & 0.21 / 0.14 \\
    viscosity up & 1.16 / 1.13 & 0.12 / 0.11 \\
    viscosity down & $-0.67$ / $-0.78$ & 0.33 / 0.33 \\
    \midrule
    \multicolumn{3}{@{}l}{\emph{outside the training range}} \\
    wavenumber $2\to6$ & 0.52 / 0.76 & 0.95 / 0.54 \\
    forcing band & 0.63 / 0.76 & 0.88 / 0.47 \\
    amplitude $1\to0.5$ & 0.82 / 0.87 & 0.32 / 0.20 \\
    viscosity to $6.6{\times}10^{-3}$ & 0.72 / 0.50 & 0.23 / 0.18 \\
    viscosity to $10^{-4}$ & $-1.41$ / $-1.87$ & 0.38 / 0.44 \\
    \bottomrule
  \end{tabular}}
\end{table}

\paragraph{The same data on a U-Net specialist.}
Trained on Kolmogorov flow with the same switch trajectories, a U-Net follows changes inside the training range about as well as PhysDial.
After changes of amplitude, wavenumber and forcing pattern, its errors are $0.14$--$0.23$, against $0.16$--$0.24$ for PhysDial.
After changes to values outside the range, PhysDial has the lower error, with $0.31$ against $0.54$ for the wavenumber, $0.30$ against $0.47$ for the forcing band, and $0.20$ against $0.44$ for a decrease of viscosity.
After a decrease of viscosity, the U-Net moves in the wrong direction, with $r_{16}$ of $-0.8$ inside the training range and $-1.9$ outside it.
PhysDial has $r_{16}$ of $0.8$ inside the range and $-0.2$ outside it.
It does not follow the drop to $10^{-4}$ either, but it deviates far less.
Both changes alter the solution by less than $0.1$, so these responses are only indicative.

For the U-Net, too, the gain comes from the data.
\Cref{tab:supp-unet-switch} compares it with the same U-Net trained for the same $4$ GPU-hours without switch trajectories.
After a change of amplitude, wavenumber or forcing pattern, the switch trajectories raise the response and lower the error on all six groups.
The error after a change of wavenumber falls from $0.57$ to $0.23$ inside the training range and from $0.95$ to $0.54$ outside it.
For the viscosity, the picture is mixed.
After a decrease, both U-Nets move in the wrong direction, and the one trained with switch trajectories moves further.
After an increase beyond the range, the response falls from $0.72$ to $0.50$, while the error also falls.
After the drop to $10^{-4}$, the error rises from $0.38$ to $0.44$, as it does for PhysDial.
The U-Net takes its coefficients through FiLM in every block, and PhysDial takes them through adaLN.
Both are conditioned from the start of training, unlike Walrus, whose condition input is attached after pretraining.

\subsection{Anchor trajectories (H3)}
All PhysDial models of this study and their in-range reference models are trained without periodic handling and sampled with a shift of $3$.
The costs along the amplitude and the wavenumber are decoded with the pretrained decoder and those along $\beta$ with the fine-tuned decoder. The two models of a ratio always share the decoder.

\paragraph{Amplitude anchors.}
The anchor trajectories are simulated at amplitudes $0.7$ and $1.4$ for the two forcing wavenumbers and four viscosities of the main sweep, with $24$ trajectories each and $384$ in total.
Before training we expected an extrapolation cost of about $2.3$--$2.7$ at amplitude $0.5$ without anchors and fixed a bound of at most $1.36$ with them.
The measured cost without anchors is $2.73$--$2.74$, slightly above the expected range, and the bound holds in both seeds (\cref{tab:anchors}).
With anchors, the error is $0.42$ times that of the model without them on the amplitude split and $0.49$--$0.50$ times on the joint split.
The reference gains much less. Its error with anchors is $0.905$ times that without them on the amplitude split and $0.955$ times on the joint split (one seed).
On six of the other seven families, the geometric mean of the error ratios lies between $0.98$ and $1.01$.
On shear flow it is $1.06$--$1.07$, mainly because of the test split ($1.19$ and $1.09$ in the two seeds), whose error also varies by $28\%$ between seeds of PhysDial (\cref{tab:supp-seeds}).
On the wavenumber extrapolation split, the cost rises from $1.07$--$1.08$ to $1.15$--$1.16$.
The error itself rises by $7\%$ in both seeds.
The amplitude also changes the eddy-turnover time, and with it the frame interval.
Without anchors, the held-out amplitude $0.5$ moves the $\Delta t$ input outside its training range for $k_f = 2$ by $0.16$--$0.19$ decades.
This is comparable to the training span of $0.21$ decades.
With anchors it remains outside by $0.07$--$0.10$ decades.
For $k_f = 4$ it stays inside the range.
A similar shift at $k_f = 2$ accompanies the drag splits and the held-out $\beta$, so part of these costs may come from the $\Delta t$ input, which our runs cannot separate from the condition itself.
In one test on the first-stage model, doubling the $\Delta t$ input did not raise the error on the amplitude split (ratio $0.88$).

\paragraph{Anchors along $\beta$.}
The coefficient $\beta$ of the $\beta$-plane term (\cref{tab:supp-solver}) is a second condition that takes a single value in training. It is zero in all trajectories except the anchors.
The models of this study receive it through an additional coefficient slot.
The anchors are $192$ trajectories at each of $\beta = \pm2.4$, $24$ for each of the two forcing wavenumbers and four viscosities of the main sweep.
The held-out value $\beta = 4$ is simulated at the same wavenumbers and at five viscosities, with $16$ trajectories each.
Of these, $4$ are evaluated and $12$ are added to the training data of the reference model.
For the held-out value $\beta = 4$, we predicted before training a cost of at least $2$ without anchors and at most $1.3$ with anchors at $\beta = \pm2.4$.
The first prediction holds. Without anchors, the cost is $5.92$--$5.94$ in both seeds.
The second does not hold. With anchors, the cost falls to $1.61$--$1.66$.
The remaining cost lies at forcing wavenumber $2$ ($2.28$--$2.33$), whereas at wavenumber $4$ the cost is $1.08$--$1.12$.
One possible explanation, which was not tested, is that the test value lies far beyond the anchors.
On Walrus, the same anchors lower the error at $\beta = 4$ from $0.861$ to $0.250$ and the cost from $4.35$ to $1.26$ (one run).
Its error on the other $14$ Kolmogorov splits rises by a median of $5.6\%$, more than the variation between runs.
The other seven families are unaffected.

\paragraph{Anchors and switch trajectories.}
The model without anchors responds more to a change of amplitude than the model with them.
Anchors make the amplitude matter for prediction, and the model with anchors then appears to infer it from the energy of its history.
Making the model use its condition input after a change is the role of switch training.

\subsection{Accuracy under fixed conditions}

\begin{table*}[t]
  \caption{\textbf{Errors on every split of the two simulated turbulent flows} (relative $L_2$, mean over the windows of a split). Bold marks the lowest error per split and step. PhysDial is the model before switch training. U-Net and FNO are seed 0.}
  \label{tab:supp-flows}
  \centering
  \scriptsize
  \setlength{\tabcolsep}{5pt}
  \begin{tabular}{@{}llcccccccccc@{}}
    \toprule
    & & \multicolumn{2}{c}{PhysDial} & \multicolumn{2}{c}{Walrus} & \multicolumn{2}{c}{DPOT-H} & \multicolumn{2}{c}{U-Net} & \multicolumn{2}{c}{FNO} \\
    \cmidrule(lr){3-4} \cmidrule(lr){5-6} \cmidrule(lr){7-8} \cmidrule(lr){9-10} \cmidrule(lr){11-12}
    Family & Split & step 4 & step 16 & step 4 & step 16 & step 4 & step 16 & step 4 & step 16 & step 4 & step 16 \\
    \midrule
        Kolmogorov flow & test & 0.0231 & \textbf{0.0804} & 0.0278 & 0.0894 & 0.1112 & 0.3380 & \textbf{0.0222} & 0.1253 & 0.0277 & 0.0923 \\
     & $\nu$ inside & 0.0285 & 0.1033 & 0.0291 & 0.0852 & 0.1101 & 0.3140 & \textbf{0.0252} & 0.1464 & 0.0267 & \textbf{0.0795} \\
     & dense $\nu$, inside & \textbf{0.0200} & 0.0756 & 0.0230 & 0.0783 & 0.1114 & 0.3487 & 0.0238 & 0.1455 & 0.0221 & \textbf{0.0695} \\
     & $k_f = 3$ & 0.0298 & 0.1114 & 0.0322 & \textbf{0.0916} & 0.1125 & 0.3481 & \textbf{0.0275} & 0.1581 & 0.0338 & 0.1235 \\
     & multi-mode forcing & 0.0403 & 0.1230 & 0.0420 & 0.1193 & 0.1300 & 0.4374 & \textbf{0.0218} & \textbf{0.1012} & 0.0381 & 0.1127 \\
     & $\nu$ below & \textbf{0.0457} & 0.1804 & 0.0844 & \textbf{0.1457} & 0.1607 & 0.3991 & 0.0659 & 0.3483 & 0.4134 & 1.4396 \\
     & $\nu$ above & 0.0176 & 0.0776 & \textbf{0.0144} & 0.0753 & 0.0732 & 0.2487 & 0.0498 & 0.2916 & 0.0161 & \textbf{0.0682} \\
     & dense $\nu$, outside & \textbf{0.0235} & 0.1014 & 0.0309 & \textbf{0.0887} & 0.1124 & 0.3067 & 0.0507 & 0.2890 & 0.1193 & 0.4111 \\
     & $k_f = 6$ & \textbf{0.0393} & 0.1502 & 0.0448 & \textbf{0.1386} & 0.1330 & 0.3904 & 0.0549 & 0.3115 & 0.0726 & 0.2316 \\
     & forcing band $6$--$7$ & 0.0536 & 0.2713 & 0.0457 & \textbf{0.1726} & 0.1336 & 0.4628 & \textbf{0.0454} & 0.2923 & 0.0759 & 0.2413 \\
     & amplitude $0.5$ & 0.0216 & 0.1037 & \textbf{0.0188} & \textbf{0.0674} & 0.1154 & 0.3513 & 0.0364 & 0.2224 & 0.0276 & 0.1107 \\
     & $k_f$ and amplitude & 0.0325 & 0.1374 & \textbf{0.0255} & \textbf{0.0919} & 0.1281 & 0.3697 & 0.0370 & 0.2112 & 0.0433 & 0.1973 \\
     & drag $0.05$ & 0.0497 & 0.2436 & \textbf{0.0382} & \textbf{0.1208} & 0.1353 & 0.4162 & 0.0395 & 0.2303 & 0.0423 & 0.1564 \\
     & drag $0.2$ & 0.0284 & 0.1438 & \textbf{0.0184} & \textbf{0.0701} & 0.0906 & 0.2533 & 0.0269 & 0.1691 & 0.0271 & 0.1451 \\
    \midrule
    Decaying turbulence & test & 0.0323 & \textbf{0.0743} & 0.0410 & 0.0862 & 0.1369 & 0.3531 & \textbf{0.0217} & 0.1191 & 0.0401 & 0.1296 \\
     & $\nu$ inside & 0.0223 & \textbf{0.0569} & 0.0264 & 0.0701 & 0.1131 & 0.3007 & \textbf{0.0191} & 0.1274 & 0.0244 & 0.0847 \\
     & $\nu$ outside & \textbf{0.0224} & \textbf{0.0761} & 0.0392 & 0.0902 & 0.1149 & 0.3287 & 0.1104 & 0.5682 & 0.7644 & 4.3235 \\
    \bottomrule
  \end{tabular}
\end{table*}

\begin{table*}[t]
  \caption{\textbf{Errors on every split of the six families from The Well.} Walrus may have seen these evaluation trajectories in pretraining, so its numbers (grey) are not compared. Bold marks the lowest error among the other models.}
  \label{tab:supp-well}
  \centering
  \scriptsize
  \setlength{\tabcolsep}{5pt}
  \begin{tabular}{@{}llcccccccccc@{}}
    \toprule
    & & \multicolumn{2}{c}{PhysDial} & \multicolumn{2}{c}{Walrus} & \multicolumn{2}{c}{DPOT-H} & \multicolumn{2}{c}{U-Net} & \multicolumn{2}{c}{FNO} \\
    \cmidrule(lr){3-4} \cmidrule(lr){5-6} \cmidrule(lr){7-8} \cmidrule(lr){9-10} \cmidrule(lr){11-12}
    Family & Split & step 4 & step 16 & step 4 & step 16 & step 4 & step 16 & step 4 & step 16 & step 4 & step 16 \\
    \midrule
        Gray--Scott & test & 0.1911 & 0.3801 & \textcolor{black!50}{0.0754} & \textcolor{black!50}{0.3307} & 0.1788 & 0.3917 & \textbf{0.0526} & \textbf{0.3138} & 0.0901 & 0.3399 \\
     & maze & \textbf{0.1331} & \textbf{0.3213} & \textcolor{black!50}{0.0278} & \textcolor{black!50}{0.1072} & 0.1860 & 0.4179 & 0.8814 & 0.9527 & 0.3236 & 0.4818 \\
     & bubbles & 0.2868 & 0.6389 & \textcolor{black!50}{0.0238} & \textcolor{black!50}{0.1322} & \textbf{0.1950} & \textbf{0.4917} & 0.5334 & 0.6909 & 0.7263 & 2.2096 \\
    \midrule
    Rayleigh--B\'enard & test & 0.0697 & 0.2791 & \textcolor{black!50}{0.0336} & \textcolor{black!50}{0.1943} & 0.1052 & 0.3918 & \textbf{0.0277} & \textbf{0.1194} & 0.0698 & 0.2676 \\
     & Pr $= 2$ & 0.0729 & 0.2712 & \textcolor{black!50}{0.0355} & \textcolor{black!50}{0.1994} & 0.0951 & 0.3662 & \textbf{0.0368} & \textbf{0.1648} & 0.0677 & 0.2595 \\
     & Ra $= 10^{10}$ & 0.2651 & 0.6046 & \textcolor{black!50}{0.1812} & \textcolor{black!50}{0.5096} & 0.2279 & 0.5878 & \textbf{0.1632} & \textbf{0.4682} & 0.2351 & 0.5980 \\
    \midrule
    Shear flow & test & \textbf{0.1541} & \textbf{0.4004} & \textcolor{black!50}{0.0747} & \textcolor{black!50}{0.2913} & 0.2679 & 1.9462 & 0.3051 & 3.5343 & 0.2463 & 2.0280 \\
     & Sc $= 2$ & \textbf{0.2146} & \textbf{0.3797} & \textcolor{black!50}{0.1113} & \textcolor{black!50}{0.2715} & 0.4368 & 0.8904 & 0.6793 & 5.8175 & 0.5218 & 3.6091 \\
     & Re $= 5{\times}10^5$ & \textbf{0.2799} & \textbf{0.5800} & \textcolor{black!50}{0.1588} & \textcolor{black!50}{0.4315} & 0.6846 & 1.2674 & 1.6937 & 15.5036 & 5.3448 & 89.1193 \\
    \midrule
    Radiative layer & test & 0.2132 & 0.3787 & \textcolor{black!50}{0.1322} & \textcolor{black!50}{0.3468} & 0.1834 & 0.3745 & \textbf{0.1415} & \textbf{0.3529} & 0.2312 & 0.4244 \\
     & $t_{\mathrm{cool}} = 0.32$ & 0.1820 & \textbf{0.3385} & \textcolor{black!50}{0.1001} & \textcolor{black!50}{0.2964} & \textbf{0.1468} & 0.3404 & 0.1698 & 0.3778 & 0.1894 & 0.3913 \\
     & $t_{\mathrm{cool}} = 3.16$ & 0.1120 & \textbf{0.2988} & \textcolor{black!50}{0.0484} & \textcolor{black!50}{0.2370} & \textbf{0.0883} & 0.3069 & 0.3083 & 0.6275 & 0.1744 & 0.3883 \\
    \midrule
    Euler & test & 0.1034 & 0.2511 & \textcolor{black!50}{0.0439} & \textcolor{black!50}{0.1378} & 0.1050 & 0.2763 & \textbf{0.0641} & \textbf{0.2164} & 0.1196 & 0.2764 \\
     & $\gamma = 1.13$ & 0.1382 & \textbf{0.2644} & \textcolor{black!50}{0.0650} & \textcolor{black!50}{0.1561} & 0.1898 & 0.4278 & \textbf{0.1226} & 0.3663 & 0.1575 & 0.3306 \\
     & $\gamma = 1.76$ & 0.1097 & \textbf{0.2409} & \textcolor{black!50}{0.0489} & \textcolor{black!50}{0.1473} & 0.1352 & 0.3174 & \textbf{0.0925} & 0.2991 & 0.1371 & 0.2843 \\
    \midrule
    Acoustic scattering & test & 0.0827 & 0.1511 & \textcolor{black!50}{0.0430} & \textcolor{black!50}{0.0841} & 0.1137 & 0.2989 & \textbf{0.0171} & \textbf{0.0525} & 0.0361 & 0.0859 \\
    \bottomrule
  \end{tabular}
\end{table*}

\paragraph{Errors on every split.}
\Cref{tab:supp-flows,tab:supp-well} give the errors of the five models on all $33$ splits at steps 4 and 16.
\Cref{tab:supp-stats4,tab:supp-stats16} give the per-window statistics for PhysDial, Walrus and DPOT-H, and \cref{tab:supp-wins} counts the splits on which PhysDial and Walrus differ significantly.
At step 16, PhysDial is more accurate on all splits of decaying turbulence.
On Kolmogorov flow, the test split is a tie.
Walrus is more accurate on eight Kolmogorov splits, six outside the training range and the two interpolation splits in $\nu$ and $k_f$ (ratios of $1.21$--$1.22$).

\begin{table}[t]
  \caption{\textbf{Splits with a significant difference between PhysDial and Walrus.} Number of splits on which PhysDial / Walrus is better, that is, on which the $95\%$ interval of the error ratio lies below / above $1$. At step 4, the count for PhysDial on Kolmogorov flow includes one borderline split, dense $\nu$, inside. Its interval is $[0.842, 0.998]$ ($p = 0.050$), and its upper end reaches $1$ for $10$ of $50$ other seeds of the bootstrap.}
  \label{tab:supp-wins}
  \centering
  \scriptsize
  \setlength{\tabcolsep}{5pt}
  \begin{tabular}{@{}lcc@{}}
    \toprule
    Family (splits) & step 4 & step 16 \\
    \midrule
    Kolmogorov flow (14) & 2 / 5 & 0 / 8 \\
    Decaying turbulence (3) & 3 / 0 & 3 / 0 \\
    \bottomrule
  \end{tabular}
\end{table}

\begin{table*}[t]
  \caption{\textbf{Statistics per window at step 4.} The ratio is the geometric mean of the per-window error ratio, with its $95\%$ interval from a bootstrap over trajectories. Wins is the share of windows (in $\%$) on which PhysDial has the lower error, and $p$ comes from a Wilcoxon signed-rank test. On the test split of Gray--Scott, windows whose true field is nearly constant have no defined error and are left out here. Its means therefore differ by about $8\%$ from those of \cref{tab:supp-well}, in the same direction for all three models.}
  \label{tab:supp-stats4}
  \centering
  \scriptsize
  \setlength{\tabcolsep}{3.2pt}
  \resizebox{\linewidth}{!}{%
  \begin{tabular}{@{}llccccccccccc@{}}
    \toprule
    & & \multicolumn{3}{c}{error, mean $\pm$ standard deviation} & \multicolumn{4}{c}{PhysDial / Walrus} & \multicolumn{4}{c}{PhysDial / DPOT-H} \\
    \cmidrule(lr){3-5} \cmidrule(lr){6-9} \cmidrule(lr){10-13}
    Family & Split & PhysDial & Walrus & DPOT-H & ratio & $95\%$ interval & wins & $p$ & ratio & $95\%$ interval & wins & $p$ \\
    \midrule
        Kolmogorov flow & test & 0.0231 $\pm$ 0.0099 & 0.0278 $\pm$ 0.0176 & 0.1112 $\pm$ 0.0311 & 0.91 & [0.82, 1.02] & 53 & 0.239 & 0.20 & [0.18, 0.22] & 100 & $4.7{\times}10^{-10}$ \\
     & $\nu$ inside & 0.0285 $\pm$ 0.0085 & 0.0291 $\pm$ 0.0100 & 0.1101 $\pm$ 0.0177 & 1.00 & [0.96, 1.04] & 50 & 0.875 & 0.25 & [0.22, 0.28] & 100 & $4.7{\times}10^{-10}$ \\
     & dense $\nu$, inside & 0.0200 $\pm$ 0.0068 & 0.0230 $\pm$ 0.0110 & 0.1114 $\pm$ 0.0343 & 0.92 & [0.84, 1.00] & 62 & 0.050 & 0.18 & [0.16, 0.19] & 100 & $4.7{\times}10^{-10}$ \\
     & $k_f = 3$ & 0.0298 $\pm$ 0.0116 & 0.0322 $\pm$ 0.0170 & 0.1125 $\pm$ 0.0285 & 0.99 & [0.92, 1.07] & 59 & 0.548 & 0.26 & [0.24, 0.27] & 100 & $4.7{\times}10^{-10}$ \\
     & multi-mode forcing & 0.0403 $\pm$ 0.0140 & 0.0420 $\pm$ 0.0188 & 0.1300 $\pm$ 0.0250 & 1.00 & [0.94, 1.06] & 50 & 0.934 & 0.30 & [0.27, 0.32] & 100 & $4.7{\times}10^{-10}$ \\
     & $\nu$ below & 0.0457 $\pm$ 0.0179 & 0.0844 $\pm$ 0.0331 & 0.1607 $\pm$ 0.0269 & 0.54 & [0.52, 0.57] & 100 & $4.7{\times}10^{-10}$ & 0.27 & [0.22, 0.31] & 100 & $4.7{\times}10^{-10}$ \\
     & $\nu$ above & 0.0176 $\pm$ 0.0049 & 0.0144 $\pm$ 0.0052 & 0.0732 $\pm$ 0.0077 & 1.24 & [1.11, 1.39] & 19 & 0.001 & 0.23 & [0.21, 0.25] & 100 & $4.7{\times}10^{-10}$ \\
     & dense $\nu$, outside & 0.0235 $\pm$ 0.0143 & 0.0309 $\pm$ 0.0260 & 0.1124 $\pm$ 0.0492 & 0.92 & [0.79, 1.07] & 53 & 0.270 & 0.20 & [0.18, 0.22] & 100 & $4.7{\times}10^{-10}$ \\
     & $k_f = 6$ & 0.0393 $\pm$ 0.0126 & 0.0448 $\pm$ 0.0177 & 0.1330 $\pm$ 0.0209 & 0.91 & [0.74, 1.13] & 41 & 0.512 & 0.29 & [0.26, 0.31] & 100 & $4.7{\times}10^{-10}$ \\
     & forcing band $6$--$7$ & 0.0536 $\pm$ 0.0209 & 0.0457 $\pm$ 0.0099 & 0.1336 $\pm$ 0.0208 & 1.10 & [0.84, 1.41] & 47 & 0.102 & 0.37 & [0.33, 0.42] & 100 & $4.7{\times}10^{-10}$ \\
     & amplitude $0.5$ & 0.0216 $\pm$ 0.0054 & 0.0188 $\pm$ 0.0099 & 0.1154 $\pm$ 0.0242 & 1.27 & [1.09, 1.48] & 19 & 0.004 & 0.19 & [0.17, 0.20] & 100 & $4.7{\times}10^{-10}$ \\
     & $k_f$ and amplitude & 0.0325 $\pm$ 0.0155 & 0.0255 $\pm$ 0.0137 & 0.1281 $\pm$ 0.0255 & 1.30 & [1.13, 1.50] & 19 & $7.7{\times}10^{-4}$ & 0.23 & [0.20, 0.26] & 100 & $4.7{\times}10^{-10}$ \\
     & drag $0.05$ & 0.0497 $\pm$ 0.0207 & 0.0382 $\pm$ 0.0221 & 0.1353 $\pm$ 0.0443 & 1.42 & [1.26, 1.59] & 19 & $8.7{\times}10^{-6}$ & 0.36 & [0.33, 0.39] & 100 & $4.7{\times}10^{-10}$ \\
     & drag $0.2$ & 0.0284 $\pm$ 0.0060 & 0.0184 $\pm$ 0.0104 & 0.0906 $\pm$ 0.0160 & 1.75 & [1.45, 2.07] & 19 & $2.7{\times}10^{-6}$ & 0.31 & [0.29, 0.33] & 100 & $4.7{\times}10^{-10}$ \\
    \midrule
    Decaying turbulence & test & 0.0323 $\pm$ 0.0136 & 0.0410 $\pm$ 0.0193 & 0.1369 $\pm$ 0.0275 & 0.82 & [0.73, 0.90] & 75 & $4.0{\times}10^{-4}$ & 0.22 & [0.20, 0.24] & 100 & $4.7{\times}10^{-10}$ \\
     & $\nu$ inside & 0.0223 $\pm$ 0.0089 & 0.0264 $\pm$ 0.0135 & 0.1131 $\pm$ 0.0202 & 0.88 & [0.78, 0.97] & 72 & 0.011 & 0.19 & [0.17, 0.21] & 100 & $4.7{\times}10^{-10}$ \\
     & $\nu$ outside & 0.0224 $\pm$ 0.0130 & 0.0392 $\pm$ 0.0390 & 0.1149 $\pm$ 0.0425 & 0.79 & [0.66, 0.93] & 50 & 0.239 & 0.18 & [0.17, 0.20] & 100 & $4.7{\times}10^{-10}$ \\
    \midrule
    Gray--Scott & test & 0.1762 $\pm$ 0.1842 & 0.0695 $\pm$ 0.0890 & 0.1649 $\pm$ 0.1680 & 4.20 & [3.36, 5.40] & 0 & $1.5{\times}10^{-8}$ & 1.02 & [0.96, 1.08] & 48 & 0.594 \\
     & maze & 0.1331 $\pm$ 0.0053 & 0.0278 $\pm$ 0.0090 & 0.1860 $\pm$ 0.0051 & 5.01 & [4.52, 5.55] & 0 & $4.7{\times}10^{-10}$ & 0.72 & [0.71, 0.72] & 100 & $4.7{\times}10^{-10}$ \\
     & bubbles & 0.2868 $\pm$ 0.0838 & 0.0238 $\pm$ 0.0115 & 0.1950 $\pm$ 0.0094 & 13.52 & [10.41, 18.19] & 0 & $4.7{\times}10^{-10}$ & 1.43 & [1.31, 1.56] & 0 & $4.7{\times}10^{-10}$ \\
    \midrule
    Rayleigh--B\'enard & test & 0.0697 $\pm$ 0.0656 & 0.0336 $\pm$ 0.0321 & 0.1052 $\pm$ 0.0928 & 2.00 & [1.83, 2.18] & 3 & $9.3{\times}10^{-10}$ & 0.54 & [0.40, 0.70] & 84 & $9.5{\times}10^{-7}$ \\
     & Pr $= 2$ & 0.0729 $\pm$ 0.0716 & 0.0355 $\pm$ 0.0371 & 0.0951 $\pm$ 0.0516 & 2.09 & [1.85, 2.30] & 3 & $9.6{\times}10^{-8}$ & 0.57 & [0.47, 0.69] & 81 & $1.1{\times}10^{-6}$ \\
     & Ra $= 10^{10}$ & 0.2651 $\pm$ 0.0388 & 0.1812 $\pm$ 0.0313 & 0.2279 $\pm$ 0.0330 & 1.47 & [1.43, 1.51] & 0 & $4.7{\times}10^{-10}$ & 1.16 & [1.14, 1.18] & 0 & $4.7{\times}10^{-10}$ \\
    \midrule
    Shear flow & test & 0.1541 $\pm$ 0.1136 & 0.0747 $\pm$ 0.0595 & 0.2679 $\pm$ 0.3145 & 2.02 & [1.78, 2.33] & 0 & $4.7{\times}10^{-10}$ & 0.56 & [0.45, 0.68] & 78 & $9.9{\times}10^{-6}$ \\
     & Sc $= 2$ & 0.2146 $\pm$ 0.1266 & 0.1113 $\pm$ 0.0815 & 0.4368 $\pm$ 0.6121 & 2.02 & [1.74, 2.41] & 0 & $2.4{\times}10^{-7}$ & 0.60 & [0.45, 0.78] & 74 & 0.003 \\
     & Re $= 5{\times}10^5$ & 0.2799 $\pm$ 0.1318 & 0.1588 $\pm$ 0.1012 & 0.6846 $\pm$ 0.8675 & 1.89 & [1.64, 2.19] & 0 & $4.7{\times}10^{-10}$ & 0.58 & [0.44, 0.75] & 72 & $1.1{\times}10^{-4}$ \\
    \midrule
    Radiative layer & test & 0.2132 $\pm$ 0.0817 & 0.1322 $\pm$ 0.0788 & 0.1834 $\pm$ 0.0774 & 1.79 & [1.58, 2.02] & 0 & $9.5{\times}10^{-7}$ & 1.18 & [1.10, 1.26] & 19 & $7.2{\times}10^{-4}$ \\
     & $t_{\mathrm{cool}} = 0.32$ & 0.1820 $\pm$ 0.0295 & 0.1001 $\pm$ 0.0264 & 0.1468 $\pm$ 0.0328 & 1.85 & [1.75, 1.94] & 0 & $7.6{\times}10^{-6}$ & 1.25 & [1.18, 1.35] & 11 & $1.9{\times}10^{-4}$ \\
     & $t_{\mathrm{cool}} = 3.16$ & 0.1120 $\pm$ 0.0166 & 0.0484 $\pm$ 0.0099 & 0.0883 $\pm$ 0.0141 & 2.33 & [2.10, 2.54] & 0 & $7.6{\times}10^{-6}$ & 1.27 & [1.18, 1.34] & 6 & $1.5{\times}10^{-5}$ \\
    \midrule
    Euler & test & 0.1034 $\pm$ 0.0371 & 0.0439 $\pm$ 0.0186 & 0.1050 $\pm$ 0.0323 & 2.38 & [2.22, 2.54] & 0 & $4.7{\times}10^{-10}$ & 0.96 & [0.90, 1.02] & 56 & 0.369 \\
     & $\gamma = 1.13$ & 0.1382 $\pm$ 0.1025 & 0.0650 $\pm$ 0.0560 & 0.1898 $\pm$ 0.1339 & 2.24 & [1.92, 2.54] & 0 & $1.9{\times}10^{-6}$ & 0.70 & [0.48, 0.94] & 80 & 0.003 \\
     & $\gamma = 1.76$ & 0.1097 $\pm$ 0.0607 & 0.0489 $\pm$ 0.0305 & 0.1352 $\pm$ 0.0732 & 2.30 & [2.12, 2.53] & 0 & $1.9{\times}10^{-6}$ & 0.81 & [0.75, 0.87] & 85 & $2.7{\times}10^{-5}$ \\
    \midrule
    Acoustic scattering & test & 0.0827 $\pm$ 0.0310 & 0.0430 $\pm$ 0.0202 & 0.1137 $\pm$ 0.0572 & 2.08 & [1.90, 2.30] & 0 & $4.7{\times}10^{-10}$ & 0.76 & [0.69, 0.83] & 94 & $2.6{\times}10^{-8}$ \\
    \bottomrule
  \end{tabular}}
\end{table*}

\begin{table*}[t]
  \caption{\textbf{Statistics per window at step 16.} Columns as in \cref{tab:supp-stats4}.}
  \label{tab:supp-stats16}
  \centering
  \scriptsize
  \setlength{\tabcolsep}{3.2pt}
  \resizebox{\linewidth}{!}{%
  \begin{tabular}{@{}llccccccccccc@{}}
    \toprule
    & & \multicolumn{3}{c}{error, mean $\pm$ standard deviation} & \multicolumn{4}{c}{PhysDial / Walrus} & \multicolumn{4}{c}{PhysDial / DPOT-H} \\
    \cmidrule(lr){3-5} \cmidrule(lr){6-9} \cmidrule(lr){10-13}
    Family & Split & PhysDial & Walrus & DPOT-H & ratio & $95\%$ interval & wins & $p$ & ratio & $95\%$ interval & wins & $p$ \\
    \midrule
        Kolmogorov flow & test & 0.0804 $\pm$ 0.0409 & 0.0894 $\pm$ 0.0540 & 0.3380 $\pm$ 0.1078 & 0.92 & [0.82, 1.03] & 69 & 0.032 & 0.22 & [0.20, 0.25] & 100 & $4.7{\times}10^{-10}$ \\
     & $\nu$ inside & 0.1033 $\pm$ 0.0310 & 0.0852 $\pm$ 0.0298 & 0.3140 $\pm$ 0.0740 & 1.22 & [1.09, 1.38] & 22 & $5.4{\times}10^{-5}$ & 0.32 & [0.30, 0.35] & 100 & $4.7{\times}10^{-10}$ \\
     & dense $\nu$, inside & 0.0756 $\pm$ 0.0303 & 0.0783 $\pm$ 0.0387 & 0.3487 $\pm$ 0.1198 & 0.99 & [0.87, 1.13] & 50 & 0.832 & 0.21 & [0.19, 0.24] & 100 & $4.7{\times}10^{-10}$ \\
     & $k_f = 3$ & 0.1114 $\pm$ 0.0512 & 0.0916 $\pm$ 0.0438 & 0.3481 $\pm$ 0.1016 & 1.21 & [1.08, 1.39] & 34 & 0.008 & 0.30 & [0.26, 0.36] & 100 & $4.7{\times}10^{-10}$ \\
     & multi-mode forcing & 0.1230 $\pm$ 0.0364 & 0.1193 $\pm$ 0.0328 & 0.4374 $\pm$ 0.0827 & 1.03 & [0.95, 1.11] & 44 & 0.410 & 0.28 & [0.25, 0.30] & 100 & $4.7{\times}10^{-10}$ \\
     & $\nu$ below & 0.1804 $\pm$ 0.0642 & 0.1457 $\pm$ 0.0377 & 0.3991 $\pm$ 0.0776 & 1.20 & [1.12, 1.29] & 16 & $6.0{\times}10^{-6}$ & 0.43 & [0.38, 0.50] & 100 & $4.7{\times}10^{-10}$ \\
     & $\nu$ above & 0.0776 $\pm$ 0.0284 & 0.0753 $\pm$ 0.0274 & 0.2487 $\pm$ 0.0698 & 1.02 & [0.93, 1.12] & 44 & 0.625 & 0.30 & [0.27, 0.34] & 100 & $4.7{\times}10^{-10}$ \\
     & dense $\nu$, outside & 0.1014 $\pm$ 0.0642 & 0.0887 $\pm$ 0.0439 & 0.3067 $\pm$ 0.1059 & 1.09 & [0.97, 1.21] & 41 & 0.144 & 0.29 & [0.25, 0.34] & 100 & $4.7{\times}10^{-10}$ \\
     & $k_f = 6$ & 0.1502 $\pm$ 0.0327 & 0.1386 $\pm$ 0.0353 & 0.3904 $\pm$ 0.0558 & 1.09 & [0.92, 1.28] & 38 & 0.246 & 0.38 & [0.35, 0.41] & 100 & $4.7{\times}10^{-10}$ \\
     & forcing band $6$--$7$ & 0.2713 $\pm$ 0.1074 & 0.1726 $\pm$ 0.0405 & 0.4628 $\pm$ 0.0552 & 1.48 & [1.12, 1.90] & 41 & 0.003 & 0.54 & [0.46, 0.62] & 100 & $4.7{\times}10^{-10}$ \\
     & amplitude $0.5$ & 0.1037 $\pm$ 0.0335 & 0.0674 $\pm$ 0.0241 & 0.3513 $\pm$ 0.0715 & 1.56 & [1.36, 1.78] & 16 & $2.3{\times}10^{-6}$ & 0.29 & [0.26, 0.32] & 100 & $4.7{\times}10^{-10}$ \\
     & $k_f$ and amplitude & 0.1374 $\pm$ 0.0709 & 0.0919 $\pm$ 0.0315 & 0.3697 $\pm$ 0.0769 & 1.35 & [1.19, 1.52] & 31 & $1.6{\times}10^{-4}$ & 0.32 & [0.28, 0.37] & 100 & $4.7{\times}10^{-10}$ \\
     & drag $0.05$ & 0.2436 $\pm$ 0.0772 & 0.1208 $\pm$ 0.0546 & 0.4162 $\pm$ 0.1108 & 2.09 & [1.92, 2.28] & 0 & $4.7{\times}10^{-10}$ & 0.58 & [0.53, 0.62] & 100 & $4.7{\times}10^{-10}$ \\
     & drag $0.2$ & 0.1438 $\pm$ 0.0354 & 0.0701 $\pm$ 0.0214 & 0.2533 $\pm$ 0.0370 & 2.09 & [1.86, 2.33] & 0 & $4.7{\times}10^{-10}$ & 0.56 & [0.52, 0.61] & 100 & $4.7{\times}10^{-10}$ \\
    \midrule
    Decaying turbulence & test & 0.0743 $\pm$ 0.0254 & 0.0862 $\pm$ 0.0258 & 0.3531 $\pm$ 0.0834 & 0.86 & [0.80, 0.92] & 75 & $3.1{\times}10^{-4}$ & 0.21 & [0.19, 0.23] & 100 & $4.7{\times}10^{-10}$ \\
     & $\nu$ inside & 0.0569 $\pm$ 0.0154 & 0.0701 $\pm$ 0.0212 & 0.3007 $\pm$ 0.0434 & 0.82 & [0.75, 0.89] & 75 & $4.9{\times}10^{-5}$ & 0.18 & [0.17, 0.20] & 100 & $4.7{\times}10^{-10}$ \\
     & $\nu$ outside & 0.0761 $\pm$ 0.0364 & 0.0902 $\pm$ 0.0471 & 0.3287 $\pm$ 0.1131 & 0.86 & [0.81, 0.91] & 81 & $4.9{\times}10^{-5}$ & 0.22 & [0.20, 0.24] & 100 & $4.7{\times}10^{-10}$ \\
    \midrule
    Gray--Scott & test & 0.3504 $\pm$ 0.3291 & 0.3048 $\pm$ 0.3576 & 0.3612 $\pm$ 0.3223 & 1.58 & [0.88, 2.47] & 18 & $4.7{\times}10^{-4}$ & 0.64 & [0.31, 0.96] & 68 & 0.012 \\
     & maze & 0.3213 $\pm$ 0.0106 & 0.1072 $\pm$ 0.0233 & 0.4179 $\pm$ 0.0117 & 3.06 & [2.84, 3.29] & 0 & $4.7{\times}10^{-10}$ & 0.77 & [0.76, 0.78] & 100 & $4.7{\times}10^{-10}$ \\
     & bubbles & 0.6389 $\pm$ 0.1995 & 0.1322 $\pm$ 0.0669 & 0.4917 $\pm$ 0.0308 & 5.69 & [4.21, 8.10] & 0 & $4.7{\times}10^{-10}$ & 1.26 & [1.16, 1.36] & 0 & $4.7{\times}10^{-10}$ \\
    \midrule
    Rayleigh--B\'enard & test & 0.2791 $\pm$ 0.1837 & 0.1943 $\pm$ 0.1090 & 0.3918 $\pm$ 0.1514 & 1.27 & [1.11, 1.44] & 22 & 0.002 & 0.58 & [0.46, 0.71] & 88 & $4.2{\times}10^{-7}$ \\
     & Pr $= 2$ & 0.2712 $\pm$ 0.2114 & 0.1994 $\pm$ 0.1434 & 0.3662 $\pm$ 0.1595 & 1.23 & [1.06, 1.40] & 22 & $3.1{\times}10^{-4}$ & 0.56 & [0.47, 0.67] & 84 & $8.1{\times}10^{-7}$ \\
     & Ra $= 10^{10}$ & 0.6046 $\pm$ 0.0941 & 0.5096 $\pm$ 0.1064 & 0.5878 $\pm$ 0.0990 & 1.20 & [1.16, 1.24] & 3 & $4.7{\times}10^{-9}$ & 1.03 & [0.99, 1.07] & 38 & 0.119 \\
    \midrule
    Shear flow & test & 0.4004 $\pm$ 0.5864 & 0.2912 $\pm$ 0.5182 & 1.9462 $\pm$ 7.4147 & 1.81 & [1.56, 2.11] & 3 & $8.8{\times}10^{-9}$ & 0.50 & [0.38, 0.67] & 88 & $2.8{\times}10^{-5}$ \\
     & Sc $= 2$ & 0.3797 $\pm$ 0.2415 & 0.2715 $\pm$ 0.2409 & 0.8904 $\pm$ 1.3941 & 1.62 & [1.35, 1.93] & 13 & $4.9{\times}10^{-5}$ & 0.56 & [0.38, 0.81] & 74 & 0.007 \\
     & Re $= 5{\times}10^5$ & 0.5800 $\pm$ 0.3277 & 0.4315 $\pm$ 0.3070 & 1.2674 $\pm$ 1.8157 & 1.50 & [1.30, 1.74] & 16 & $1.3{\times}10^{-5}$ & 0.72 & [0.49, 1.00] & 50 & 0.350 \\
    \midrule
    Radiative layer & test & 0.3787 $\pm$ 0.1343 & 0.3441 $\pm$ 0.1270 & 0.3745 $\pm$ 0.1066 & 1.11 & [1.05, 1.19] & 33 & 0.029 & 0.99 & [0.91, 1.09] & 52 & 0.785 \\
     & $t_{\mathrm{cool}} = 0.32$ & 0.3385 $\pm$ 0.0561 & 0.2963 $\pm$ 0.0393 & 0.3404 $\pm$ 0.0465 & 1.14 & [1.07, 1.19] & 11 & 0.001 & 0.99 & [0.90, 1.09] & 61 & 0.671 \\
     & $t_{\mathrm{cool}} = 3.16$ & 0.2988 $\pm$ 0.0725 & 0.2370 $\pm$ 0.0549 & 0.3069 $\pm$ 0.0679 & 1.26 & [1.20, 1.32] & 6 & $3.8{\times}10^{-5}$ & 0.97 & [0.86, 1.07] & 50 & 0.702 \\
    \midrule
    Euler & test & 0.2511 $\pm$ 0.0846 & 0.1378 $\pm$ 0.0518 & 0.2763 $\pm$ 0.0795 & 1.87 & [1.75, 2.01] & 0 & $4.7{\times}10^{-10}$ & 0.90 & [0.84, 0.97] & 66 & 0.014 \\
     & $\gamma = 1.13$ & 0.2644 $\pm$ 0.1435 & 0.1561 $\pm$ 0.1078 & 0.4278 $\pm$ 0.3757 & 1.79 & [1.60, 2.02] & 0 & $1.9{\times}10^{-6}$ & 0.66 & [0.43, 0.93] & 60 & 0.064 \\
     & $\gamma = 1.76$ & 0.2409 $\pm$ 0.0979 & 0.1473 $\pm$ 0.0717 & 0.3174 $\pm$ 0.1185 & 1.70 & [1.54, 1.90] & 0 & $1.9{\times}10^{-6}$ & 0.75 & [0.68, 0.82] & 95 & $3.8{\times}10^{-6}$ \\
    \midrule
    Acoustic scattering & test & 0.1511 $\pm$ 0.0529 & 0.0841 $\pm$ 0.0379 & 0.2989 $\pm$ 0.2194 & 1.89 & [1.75, 2.02] & 0 & $4.7{\times}10^{-10}$ & 0.55 & [0.48, 0.60] & 100 & $4.7{\times}10^{-10}$ \\
    \bottomrule
  \end{tabular}}
\end{table*}

\paragraph{Specialists.}
Outside the training range, PhysDial has a lower error than the specialists on most splits.
It has a lower error than the U-Net on $14$ of the $16$ splits at step 16 and on $10$ at step 4, and than the FNO on $12$ at both steps.
Under training conditions, the specialists are stronger at step 4 than at step 16.
The U-Net has a lower error than PhysDial on seven of the eight test splits at step 4 and on five at step 16.
On the test splits of Kolmogorov flow and decaying turbulence, PhysDial has the lower error at step 16 against both seeds of each specialist.
Against the U-Net, its errors are $0.080$ against $0.125$ and $0.130$ on Kolmogorov flow and $0.074$ against $0.119$ and $0.529$ on decaying turbulence.
The two seeds of the FNO reach $0.092$ and $0.093$ on Kolmogorov flow and $0.130$ and $0.124$ on decaying turbulence.
Outside the viscosity range of decaying turbulence, the two seeds reach $0.57$ and $1.67$ for the U-Net and $4.3$ and $0.59$ for the FNO, against $0.076$ for PhysDial.
The second seed of the U-Net is unstable on decaying turbulence. With the same number of training steps, its final training loss is $2.2$ times higher (\cref{tab:supp-unetseeds} gives both seeds).
Inside the training range, PhysDial does not have the lower error everywhere.
There, the FNO has the lower error on three held-out splits of Kolmogorov flow. They are $\nu$ inside ($0.080$ against $0.103$), dense $\nu$, inside ($0.070$ against $0.076$) and multi-mode forcing ($0.113$ against $0.123$).
On the six families from The Well, the U-Net has the lower error on five test splits.
It is ahead most clearly on Rayleigh--B\'enard convection ($0.12$ against $0.28$), acoustic scattering ($0.053$ against $0.151$) and Gray--Scott ($0.31$ against $0.38$).
The FNO has the lower error on three test splits (Gray--Scott, Rayleigh--B\'enard convection and acoustic scattering).
PhysDial has the lower error on shear flow, where both specialists diverge with errors of $2$ to $89$.
It also has the lower error on the extrapolation split of the radiative layer and on the two held-out patterns of Gray--Scott.

\begin{table}[t]
  \caption{\textbf{U-Net specialist, two seeds.} Error at step 16 and ratio of seed 1 to seed 0.}
  \label{tab:supp-unetseeds}
  \centering
  \scriptsize
  \setlength{\tabcolsep}{5pt}
  \begin{tabular}{@{}llccc@{}}
    \toprule
    Family & Split & seed 0 & seed 1 & ratio \\
    \midrule
        Kolmogorov flow & test & 0.1253 & 0.1298 & 1.04 \\
     & $\nu$ inside & 0.1464 & 0.0982 & 0.67 \\
     & dense $\nu$, inside & 0.1455 & 0.1450 & 1.00 \\
     & $k_f = 3$ & 0.1581 & 0.1598 & 1.01 \\
     & multi-mode forcing & 0.1012 & 0.1066 & 1.05 \\
     & $\nu$ below & 0.3483 & 0.3519 & 1.01 \\
     & $\nu$ above & 0.2916 & 0.2701 & 0.93 \\
     & dense $\nu$, outside & 0.2890 & 0.2904 & 1.00 \\
     & $k_f = 6$ & 0.3115 & 0.2880 & 0.92 \\
     & forcing band $6$--$7$ & 0.2923 & 0.4265 & 1.46 \\
     & amplitude $0.5$ & 0.2224 & 0.2316 & 1.04 \\
     & $k_f$ and amplitude & 0.2112 & 0.1866 & 0.88 \\
     & drag $0.05$ & 0.2303 & 0.2266 & 0.98 \\
     & drag $0.2$ & 0.1691 & 0.1706 & 1.01 \\
    \midrule
    Decaying turbulence & test & 0.1191 & 0.5286 & 4.44 \\
     & $\nu$ inside & 0.1274 & 0.4624 & 3.63 \\
     & $\nu$ outside & 0.5682 & 1.6724 & 2.94 \\
    \bottomrule
  \end{tabular}
\end{table}

\subsection{Long rollouts and physical fidelity}

\begin{table}[t]
  \caption{\textbf{Long rollouts.} Error of PhysDial (ours) and Walrus at steps 16, 32 and 64 of one $64$-step rollout ($32$ windows per split). The ratio is PhysDial / Walrus at step 64, and the DPOT column is the error of DPOT-H at step 64.}
  \label{tab:supp-long}
  \centering
  \scriptsize
  \setlength{\tabcolsep}{2.6pt}
  \resizebox{\linewidth}{!}{%
  \begin{tabular}{@{}llcccccccc@{}}
    \toprule
    & & \multicolumn{2}{c}{step 16} & \multicolumn{2}{c}{step 32} & \multicolumn{4}{c}{step 64} \\
    \cmidrule(lr){3-4} \cmidrule(lr){5-6} \cmidrule(lr){7-10}
    Family & Split & ours & Wal. & ours & Wal. & ours & Wal. & ratio & DPOT \\
    \midrule
        Kolmogorov flow & test & 0.079 & 0.085 & 0.187 & 0.217 & 0.429 & 0.592 & 0.72 & 1.033 \\
     & $\nu$ inside & 0.108 & 0.083 & 0.229 & 0.202 & 0.499 & 0.551 & 0.91 & 1.064 \\
     & dense $\nu$, inside & 0.080 & 0.085 & 0.178 & 0.217 & 0.360 & 0.503 & 0.72 & 1.009 \\
     & $k_f = 3$ & 0.110 & 0.094 & 0.238 & 0.239 & 0.469 & 0.591 & 0.79 & 1.089 \\
     & multi-mode forcing & 0.122 & 0.118 & 0.285 & 0.314 & 0.610 & 0.759 & 0.80 & 1.175 \\
     & $\nu$ below & 0.167 & 0.151 & 0.349 & 0.290 & 0.649 & 0.602 & 1.08 & 1.117 \\
     & $\nu$ above & 0.077 & 0.073 & 0.180 & 0.176 & 0.453 & 0.492 & 0.92 & 1.044 \\
     & dense $\nu$, outside & 0.105 & 0.103 & 0.216 & 0.230 & 0.439 & 0.515 & 0.85 & 1.024 \\
     & $k_f = 6$ & 0.149 & 0.132 & 0.354 & 0.367 & 0.726 & 0.914 & 0.79 & 1.379 \\
     & forcing band $6$--$7$ & 0.267 & 0.170 & 0.559 & 0.485 & 0.982 & 1.028 & 0.95 & 1.452 \\
     & amplitude $0.5$ & 0.100 & 0.066 & 0.235 & 0.182 & 0.479 & 0.506 & 0.95 & 1.113 \\
     & $k_f$ and amplitude & 0.140 & 0.091 & 0.302 & 0.274 & 0.601 & 0.768 & 0.78 & 1.218 \\
     & drag $0.05$ & 0.238 & 0.121 & 0.455 & 0.322 & 0.728 & 0.710 & 1.02 & 1.268 \\
     & drag $0.2$ & 0.145 & 0.068 & 0.366 & 0.209 & 0.760 & 0.576 & 1.32 & 1.056 \\
    \midrule
    Decaying turbulence & test & 0.074 & 0.085 & 0.156 & 0.180 & 0.370 & 0.464 & 0.80 & 1.185 \\
     & $\nu$ inside & 0.058 & 0.073 & 0.127 & 0.154 & 0.317 & 0.394 & 0.80 & 1.216 \\
     & $\nu$ outside & 0.072 & 0.088 & 0.180 & 0.170 & 0.408 & 0.396 & 1.03 & 1.187 \\
    \bottomrule
  \end{tabular}}
\end{table}

\begin{figure*}[t]
  \centering
  \includegraphics[width=\linewidth]{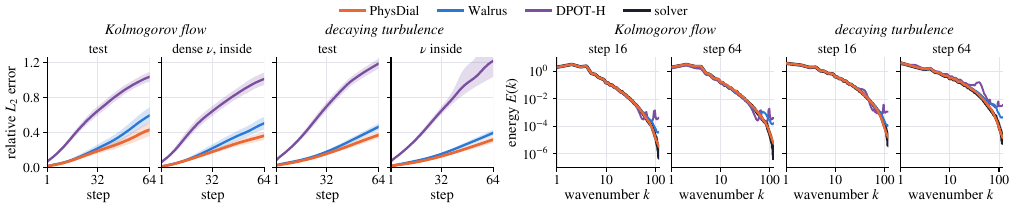}
  \caption{\textbf{Long rollouts, further views}. \emph{Left.} Error at every step on the test split and on one interpolation split of each turbulent flow (mean over $32$ windows, $95\%$ interval). \emph{Right.} Radial energy spectra of the last frame on the test splits ($28$ windows), from which the ratios of \cref{fig:flows} are computed. Shells above wavenumber $120$, in which the solver has almost no energy, are not shown.}
  \label{fig:supp-long-curves}
\end{figure*}

\paragraph{Errors up to step 64.}
Over longer horizons, the comparison between PhysDial and Walrus shifts toward PhysDial (\cref{tab:supp-long}, errors at steps 16, 32 and 64 of one rollout).
On Kolmogorov flow, Walrus has the lower error on $12$ of $14$ splits at step 16, and PhysDial on $11$ at step 64.
On the test splits, the error ratio of PhysDial to Walrus falls from $0.93$ to $0.72$ on Kolmogorov flow and from $0.87$ to $0.80$ on decaying turbulence.
These are ratios of mean errors in the $64$-step rollouts, whose windows differ from those of the $16$-step evaluation.
The values at step 16 therefore differ slightly from the per-window ratios of \cref{tab:supp-stats16} ($0.92$ and $0.86$).
\Cref{fig:supp-long-curves} (left) shows the error at every step on the test split and on one interpolation split of each turbulent flow.
On all four, the error of PhysDial is below that of Walrus from the second step on, and the gap widens with the horizon.
The two seeds of PhysDial trained without periodic handling (sampling shift $3$) show the same shift.
They have the lower error on one of the $14$ Kolmogorov splits at step 16 and on $10$ at step 64.
The geometric mean of their error ratio to Walrus falls from $1.29$--$1.30$ to $0.91$--$0.92$.
On the three splits of decaying turbulence, they have the lower error at both steps (ratios of $0.87$--$0.88$ and $0.81$--$0.82$).
At step 64, most errors are between $0.4$ and $1$, so the predictions are far from the solver trajectory point by point.
What remains to compare at this horizon are the statistics of the field.

\begin{figure*}[t]
  \centering
  \includegraphics[width=\linewidth]{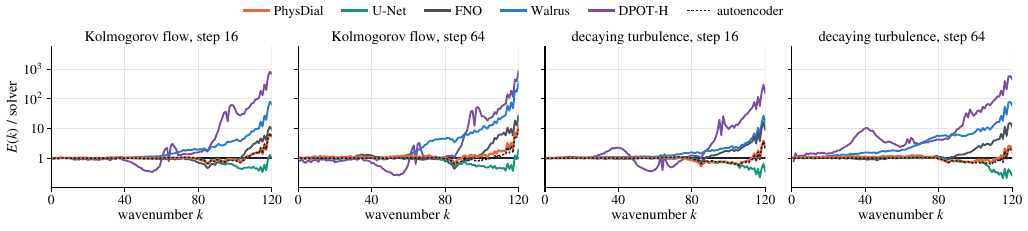}
  \caption{\textbf{Energy spectrum divided by that of the solver, all models} (test splits, last frame, $28$ windows, seed 0 of U-Net and FNO). Autoencoder denotes the solver frame encoded and decoded by the video autoencoder.}
  \label{fig:supp-spectra}
\end{figure*}

\begin{table}[t]
  \caption{\textbf{Physical statistics} at steps 16 and 64 (first $8$ windows of every split). Spectrum is the mean of $|\log(\text{predicted energy} / \text{solver energy})|$ over four wavenumber bands, which is $0$ when the energy is correct at every scale. Energy drift is the relative error of the kinetic energy. Persist.\ is persistence, which copies the last history frame.}
  \label{tab:supp-fidelity}
  \centering
  \scriptsize
  \setlength{\tabcolsep}{3pt}
  \resizebox{\linewidth}{!}{%
  \begin{tabular}{@{}llcccc@{}}
    \toprule
    Family & Quantity (step 16 / 64) & PhysDial & Walrus & DPOT-H & persist. \\
    \midrule
    Kolmogorov & spectrum & 0.055 / 0.14 & 0.44 / 0.85 & 0.95 / 1.12 & 0.078 / 0.14 \\
    Decaying & spectrum & 0.032 / 0.067 & 0.18 / 0.51 & 0.54 / 1.15 & 0.18 / 0.49 \\
    Kolmogorov & energy drift & 0.017 / 0.047 & 0.009 / 0.054 & 0.039 / 0.32 & 0.061 / 0.18 \\
    Decaying & energy drift & 0.012 / 0.046 & 0.004 / 0.033 & 0.037 / 0.20 & 0.004 / 0.014 \\
    \bottomrule
  \end{tabular}}
\end{table}

\paragraph{Spectra and energy.}
On the two turbulent flows, the energy spectrum of PhysDial is the closest to that of the solver (\cref{tab:supp-fidelity}).
Its gap to Walrus and DPOT-H grows from step 16 to step 64.
The same distance can be computed over all windows from the band ratios of \cref{tab:supp-bands-a,tab:supp-bands-b}, averaged over the splits of a family.
At step 16 it is $0.10$ for PhysDial, against $0.36$ for Walrus and $0.84$ for DPOT-H, on Kolmogorov flow, and $0.05$ against $0.17$ and $0.50$ on decaying turbulence.
The drift of the kinetic energy is smaller for Walrus at step 16 and similar at step 64.
The distribution of field values is measured by the Wasserstein-1 distance between predicted and true values, divided by the standard deviation of the true field (first $8$ windows of every split).
It is similar for PhysDial and Walrus at step 16 ($0.011$ against $0.010$ on Kolmogorov flow, $0.007$ against $0.008$ on decaying turbulence) and closer to the solver for PhysDial at step 64 ($0.029$ against $0.043$, and $0.020$ against $0.032$).
In the periodic families, the spatial mean of our prediction drifts slightly, which is a property of the reconstruction from latents.

\paragraph{Energy per scale.}
We divide the radial energy spectrum of the last frame into four wavenumber bands, given as fractions of the largest resolved wavenumber. The bands are $(0, 1/8]$, $(1/8, 1/4]$, $(1/4, 1/2]$ and $(1/2, 1]$.
For each band we report the predicted energy divided by that of the solver, pooled over windows and quantities.
A ratio of $1$ is correct, a larger value means too much energy, and a smaller value too little.
Where the solver has less than $10^{-4}$ of its energy in the highest band, the ratio in this band has a very small denominator and is not reported ($^\dagger$).
This concerns all splits of Euler flow and a few of Kolmogorov and shear flow.
\Cref{tab:supp-bands-a,tab:supp-bands-b} give all splits at step 16, with the reconstruction of the true frame by the autoencoder as a reference, and \cref{tab:supp-bands4,tab:supp-bands64} give steps 4 and 64.
All models are correct at the largest scales, and the differences lie at the smallest.
On the $26$ splits where it is reported, the energy in the highest band is off by more than a factor of $1.5$ on $2$ splits for PhysDial, $20$ for Walrus, $24$ for DPOT-H, $9$ for the U-Net and $11$ for the FNO.
For Walrus, $18$ of the $20$ are an excess of energy.
For DPOT-H, $16$ of the $24$ are an excess and $8$ a deficit (Rayleigh--B\'enard convection, shear flow, the radiative layer and acoustic scattering).
The highest band of PhysDial is mostly within $0.9$--$1.25$, close to the reconstruction by the autoencoder.
That of Walrus is above $1.5$ on $11$ of the $12$ Kolmogorov splits on which the band is reported.
The U-Net specialist is accurate on the test splits of its own family ($1.18$ on Kolmogorov flow, $0.93$ on decaying turbulence) and deviates mostly on held-out splits.
The difference to Walrus grows with the horizon.
On the Kolmogorov test split at step 64, the highest band has $1.07$ times the energy of the solver for PhysDial, against $7.06$ for Walrus and $7.56$ for DPOT-H.
The spectra of \cref{fig:flows,fig:supp-long-curves} resolve these bands shell by shell.
For them, the squared Fourier amplitudes of the last frame are summed over shells of integer wavenumber and over quantities, and averaged over windows.
They use the $28$ of the $32$ windows of the $64$-step rollouts that lie on the $256^2$ grid, since spectra on different grids cannot be averaged shell by shell.
The band ratios at steps 16 and 64 (\cref{tab:supp-bands-a,tab:supp-bands-b,tab:supp-bands64}) instead pool all windows.
Those at step 16 come from the $16$-step evaluation, which uses different windows, so band ratios and spectra differ slightly.
\Cref{tab:supp-bands4,tab:supp-fidelity} use only the first $8$ windows of every split.
\Cref{fig:supp-spectra} adds the two specialists and the autoencoder to these spectra.
At step 16, the spectrum of PhysDial lies on that of the autoencoder at every shell, so its excess in the highest shells is the reconstruction error of the autoencoder.
At step 64, it lies slightly above.
The U-Net loses energy in the highest shells, and the FNO gains some, though less than Walrus.

\begin{table*}[t]
  \caption{\textbf{Energy per scale at step 16}, the two bands of larger scales. Predicted energy divided by that of the solver (all windows, seed 0 of U-Net and FNO). Autoenc.\ is the true frame encoded and decoded by the video autoencoder.}
  \label{tab:supp-bands-a}
  \centering
  \scriptsize
  \setlength{\tabcolsep}{4.2pt}
  \resizebox{\linewidth}{!}{%
  \begin{tabular}{@{}llcccccccccccc@{}}
    \toprule
    & & \multicolumn{6}{c}{largest scales, $(0, 1/8]$} & \multicolumn{6}{c}{$(1/8, 1/4]$} \\
    \cmidrule(lr){3-8} \cmidrule(lr){9-14}
    Family & Split & PhysDial & Walrus & DPOT-H & U-Net & FNO & autoenc. & PhysDial & Walrus & DPOT-H & U-Net & FNO & autoenc. \\
    \midrule
        Kolmogorov flow & test & 1.01 & 1.01 & 1.00 & 1.00 & 1.00 & 1.00 & 1.00 & 0.99 & 0.95 & 1.01 & 0.99 & 1.00 \\
     & $\nu$ inside & 1.01 & 1.00 & 0.98 & 1.00 & 1.00 & 1.00 & 1.02 & 0.95 & 0.87 & 0.91 & 0.99 & 1.00 \\
     & dense $\nu$, inside & 1.00 & 1.00 & 0.98 & 1.00 & 1.00 & 1.00 & 1.00 & 0.98 & 0.96 & 0.99 & 1.00 & 1.00 \\
     & $k_f = 3$ & 1.01 & 1.00 & 0.98 & 1.01 & 0.97 & 1.00 & 1.01 & 1.04 & 0.95 & 0.99 & 0.98 & 1.00 \\
     & multi-mode forcing & 1.00 & 0.99 & 0.94 & 1.00 & 1.00 & 1.00 & 1.03 & 0.99 & 0.99 & 1.00 & 1.01 & 1.00 \\
     & $\nu$ below & 1.03 & 0.99 & 1.00 & 1.00 & 2.65 & 1.00 & 1.05 & 0.79 & 0.87 & 0.78 & 3.35 & 1.00 \\
     & $\nu$ above & 0.96 & 1.02 & 1.01 & 1.11 & 1.05 & 1.00 & 1.06 & 1.73 & 2.25 & 2.53 & 1.43 & 0.99 \\
     & dense $\nu$, outside & 0.98 & 1.01 & 1.02 & 1.02 & 1.50 & 1.00 & 0.96 & 1.03 & 1.14 & 1.18 & 4.43 & 1.00 \\
     & $k_f = 6$ & 1.01 & 0.97 & 1.02 & 1.02 & 1.08 & 1.00 & 1.12 & 1.01 & 1.27 & 1.00 & 0.89 & 1.00 \\
     & forcing band $6$--$7$ & 1.04 & 0.96 & 1.01 & 1.04 & 1.05 & 1.00 & 1.15 & 1.02 & 1.42 & 1.16 & 0.85 & 1.00 \\
     & amplitude $0.5$ & 1.01 & 1.02 & 0.97 & 1.03 & 0.99 & 1.00 & 1.01 & 1.05 & 0.80 & 1.13 & 0.94 & 1.00 \\
     & $k_f$ and amplitude & 0.99 & 1.00 & 1.01 & 1.04 & 0.91 & 1.00 & 1.06 & 1.04 & 0.91 & 1.07 & 0.96 & 1.00 \\
     & drag $0.05$ & 0.95 & 0.98 & 0.97 & 0.95 & 0.94 & 1.00 & 0.94 & 1.00 & 1.06 & 0.93 & 0.96 & 1.00 \\
     & drag $0.2$ & 1.12 & 1.03 & 1.04 & 1.14 & 1.14 & 1.00 & 1.20 & 1.02 & 0.93 & 1.14 & 1.17 & 1.00 \\
    \midrule
    Decaying turbulence & test & 1.01 & 1.00 & 1.00 & 1.01 & 1.00 & 1.00 & 1.02 & 1.03 & 1.08 & 1.00 & 0.98 & 1.00 \\
     & $\nu$ inside & 1.01 & 1.00 & 0.97 & 1.00 & 1.00 & 1.00 & 1.00 & 0.91 & 1.01 & 1.04 & 0.99 & 1.00 \\
     & $\nu$ outside & 1.01 & 1.01 & 1.00 & 1.09 & 9.58 & 1.00 & 1.03 & 0.99 & 1.30 & 1.16 & 16.60 & 1.00 \\
    \midrule
    Gray--Scott & test & 0.99 & 0.98 & 1.01 & 1.01 & 0.99 & 1.00 & 0.96 & 0.97 & 0.99 & 0.99 & 0.99 & 1.00 \\
     & maze & 3.14 & 1.21 & 5.42 & 0.00 & 34.30 & 0.98 & 1.23 & 0.96 & 1.28 & 0.00 & 0.08 & 0.98 \\
     & bubbles & 0.58 & 0.91 & 0.53 & 0.16 & 25.60 & 1.00 & 1.23 & 1.06 & 1.06 & 0.62 & 7.32 & 0.99 \\
    \midrule
    Rayleigh--B\'enard & test & 0.98 & 0.99 & 1.00 & 1.00 & 1.00 & 1.00 & 0.87 & 0.86 & 0.47 & 1.02 & 0.88 & 1.00 \\
     & Pr $= 2$ & 0.98 & 1.00 & 1.03 & 1.02 & 1.00 & 1.00 & 0.80 & 0.87 & 0.56 & 0.88 & 0.88 & 1.00 \\
     & Ra $= 10^{10}$ & 1.00 & 1.02 & 1.01 & 1.01 & 1.00 & 1.00 & 0.65 & 0.68 & 0.54 & 1.02 & 0.94 & 0.95 \\
    \midrule
    Shear flow & test & 0.99 & 0.99 & 0.99 & 0.99 & 0.99 & 1.00 & 0.80 & 0.47 & 0.68 & 0.89 & 0.95 & 1.00 \\
     & Sc $= 2$ & 0.99 & 0.99 & 0.99 & 0.97 & 0.98 & 1.00 & 0.66 & 0.73 & 1.60 & 0.45 & 0.83 & 1.00 \\
     & Re $= 5{\times}10^5$ & 0.98 & 0.97 & 0.97 & 0.99 & 0.97 & 1.00 & 0.75 & 0.32 & 0.64 & 0.46 & 0.13 & 0.99 \\
    \midrule
    Radiative layer & test & 1.02 & 0.94 & 0.98 & 1.02 & 0.86 & 1.00 & 0.79 & 0.75 & 1.00 & 1.01 & 0.80 & 1.00 \\
     & $t_{\mathrm{cool}} = 0.32$ & 1.03 & 0.95 & 1.00 & 1.15 & 0.96 & 1.00 & 0.83 & 0.80 & 0.90 & 1.08 & 0.80 & 1.00 \\
     & $t_{\mathrm{cool}} = 3.16$ & 0.99 & 0.95 & 1.04 & 1.28 & 0.95 & 1.00 & 0.96 & 0.75 & 1.08 & 1.24 & 1.03 & 1.00 \\
    \midrule
    Euler & test & 0.92 & 1.00 & 0.92 & 0.92 & 0.92 & 1.00 & 0.73 & 0.88 & 0.68 & 0.96 & 0.36 & 1.00 \\
     & $\gamma = 1.13$ & 1.02 & 1.02 & 0.78 & 1.01 & 1.04 & 1.00 & 1.10 & 1.06 & 0.61 & 3.21 & 0.34 & 1.00 \\
     & $\gamma = 1.76$ & 1.02 & 1.01 & 0.95 & 0.80 & 1.00 & 1.00 & 1.05 & 0.85 & 0.60 & 1.27 & 0.33 & 1.00 \\
    \midrule
    Acoustic scattering & test & 0.99 & 0.99 & 0.99 & 1.00 & 1.00 & 1.00 & 0.86 & 0.94 & 1.00 & 1.01 & 1.01 & 1.00 \\
    \bottomrule
  \end{tabular}}
\end{table*}

\begin{table*}[t]
  \caption{\textbf{Energy per scale at step 16}, the two bands of smaller scales. Share is the fraction of the solver energy in the smallest scales. At $^\dagger$ this share is below $10^{-4}$ and the ratio in the smallest scales is not reported.}
  \label{tab:supp-bands-b}
  \centering
  \scriptsize
  \setlength{\tabcolsep}{3.4pt}
  \resizebox{\linewidth}{!}{%
  \begin{tabular}{@{}llccccccccccccc@{}}
    \toprule
    & & \multicolumn{6}{c}{$(1/4, 1/2]$} & \multicolumn{6}{c}{smallest scales, $(1/2, 1]$} & \\
    \cmidrule(lr){3-8} \cmidrule(lr){9-14}
    Family & Split & PhysDial & Walrus & DPOT-H & U-Net & FNO & autoenc. & PhysDial & Walrus & DPOT-H & U-Net & FNO & autoenc. & share \\
    \midrule
        Kolmogorov flow & test & 1.02 & 1.04 & 0.85 & 1.03 & 0.97 & 1.00 & 0.94 & 2.20 & 8.58 & 1.18 & 0.81 & 1.00 & $4.6{\times}10^{-4}$ \\
     & $\nu$ inside & 1.03 & 0.95 & 0.71 & 0.77 & 0.95 & 0.99 & 0.99 & 1.63 & 4.29 & 0.54 & 0.77 & 0.99 & $7.0{\times}10^{-4}$ \\
     & dense $\nu$, inside & 0.97 & 1.04 & 0.88 & 1.02 & 0.97 & 1.00 & 0.92 & 4.35 & 23.28 & 1.06 & 0.87 & 0.99 & $1.5{\times}10^{-4}$ \\
     & $k_f = 3$ & 1.01 & 1.14 & 0.79 & 0.98 & 0.95 & 0.99 & 0.96 & 2.75 & 7.71 & 0.95 & 0.75 & 0.99 & $4.9{\times}10^{-4}$ \\
     & multi-mode forcing & 1.02 & 0.97 & 0.78 & 1.00 & 0.95 & 1.00 & 0.91 & 1.57 & 2.04 & 0.86 & 0.69 & 0.98 & $1.4{\times}10^{-3}$ \\
     & $\nu$ below & 0.94 & 0.69 & 0.41 & 0.51 & 3.22 & 0.99 & 0.64 & 2.46 & 4.27 & 0.20 & 9.43 & 0.96 & $1.3{\times}10^{-3}$ \\
     & $\nu$ above$^\dagger$ & 1.15 & 3.14 & 9.17 & 7.06 & 1.61 & 0.99 & 3.50 & 48.31 & 2209 & 46.24 & 2.73 & 3.12 & $1.3{\times}10^{-6}$ \\
     & dense $\nu$, outside$^\dagger$ & 0.82 & 1.02 & 1.43 & 0.92 & 7.58 & 1.00 & 0.49 & 6.75 & 53.89 & 0.84 & 42.43 & 0.97 & $1.0{\times}10^{-4}$ \\
     & $k_f = 6$ & 1.13 & 1.23 & 1.77 & 0.96 & 0.77 & 1.00 & 1.03 & 2.53 & 8.73 & 0.83 & 0.71 & 0.99 & $7.0{\times}10^{-4}$ \\
     & forcing band $6$--$7$ & 1.18 & 1.30 & 2.49 & 1.38 & 0.71 & 1.00 & 1.03 & 3.38 & 13.74 & 1.63 & 0.63 & 0.97 & $4.8{\times}10^{-4}$ \\
     & amplitude $0.5$ & 1.01 & 1.09 & 0.94 & 1.25 & 0.99 & 0.99 & 0.92 & 2.48 & 19.96 & 1.38 & 0.83 & 0.98 & $2.0{\times}10^{-4}$ \\
     & $k_f$ and amplitude & 1.04 & 1.23 & 1.35 & 1.10 & 0.87 & 1.00 & 0.91 & 2.00 & 13.83 & 1.06 & 0.66 & 0.98 & $3.9{\times}10^{-4}$ \\
     & drag $0.05$ & 0.90 & 1.13 & 1.00 & 0.89 & 1.00 & 1.00 & 0.78 & 3.44 & 13.66 & 0.80 & 0.99 & 0.98 & $5.3{\times}10^{-4}$ \\
     & drag $0.2$ & 1.24 & 0.97 & 0.93 & 1.12 & 1.17 & 1.00 & 1.22 & 1.46 & 9.51 & 1.02 & 0.94 & 0.99 & $3.4{\times}10^{-4}$ \\
    \midrule
    Decaying turbulence & test & 1.02 & 1.08 & 1.67 & 1.00 & 0.98 & 1.00 & 0.92 & 1.61 & 3.12 & 0.93 & 1.14 & 0.98 & $1.6{\times}10^{-3}$ \\
     & $\nu$ inside & 0.98 & 0.79 & 1.41 & 1.18 & 0.97 & 1.00 & 0.89 & 0.85 & 4.22 & 2.27 & 0.99 & 0.98 & $8.3{\times}10^{-4}$ \\
     & $\nu$ outside & 0.97 & 0.75 & 1.34 & 1.17 & 416 & 0.99 & 0.75 & 2.03 & 6.30 & 1.21 & 6629 & 0.97 & $7.8{\times}10^{-4}$ \\
    \midrule
    Gray--Scott & test & 0.99 & 0.98 & 1.00 & 0.99 & 0.99 & 1.00 & 0.97 & 1.03 & 1.13 & 1.00 & 1.01 & 0.99 & $3.1{\times}10^{-3}$ \\
     & maze & 1.38 & 1.09 & 2.08 & 0.00 & 0.05 & 0.99 & 2.89 & 1.54 & 5.13 & 0.06 & 0.50 & 1.67 & $1.2{\times}10^{-3}$ \\
     & bubbles & 0.96 & 1.03 & 0.89 & 0.44 & 2.55 & 0.98 & 0.86 & 0.91 & 0.96 & 0.82 & 20.24 & 0.94 & $2.9{\times}10^{-3}$ \\
    \midrule
    Rayleigh--B\'enard & test & 0.69 & 0.66 & 0.13 & 0.99 & 0.53 & 0.99 & 1.14 & 0.84 & 0.12 & 0.69 & 0.09 & 0.94 & $1.5{\times}10^{-3}$ \\
     & Pr $= 2$ & 0.57 & 0.58 & 0.15 & 0.86 & 0.47 & 0.99 & 0.98 & 0.66 & 0.09 & 0.55 & 0.07 & 0.94 & $2.3{\times}10^{-3}$ \\
     & Ra $= 10^{10}$ & 0.45 & 0.31 & 0.09 & 0.67 & 0.32 & 0.96 & 0.89 & 0.50 & 0.01 & 0.11 & 0.02 & 0.77 & $1.9{\times}10^{-2}$ \\
    \midrule
    Shear flow & test$^\dagger$ & 0.73 & 0.35 & 0.29 & 0.67 & 0.80 & 1.01 & 0.89 & 2.33 & 0.32 & 0.34 & 0.42 & 1.07 & $8.6{\times}10^{-5}$ \\
     & Sc $= 2$$^\dagger$ & 0.62 & 1.54 & 1.48 & 0.32 & 0.58 & 1.00 & 0.66 & 13.78 & 1.96 & 0.30 & 0.25 & 1.10 & $2.7{\times}10^{-5}$ \\
     & Re $= 5{\times}10^5$ & 0.90 & 0.47 & 0.21 & 0.21 & 0.02 & 1.01 & 1.14 & 1.83 & 0.06 & 0.05 & 0.00 & 1.00 & $1.0{\times}10^{-3}$ \\
    \midrule
    Radiative layer & test & 0.70 & 0.81 & 0.60 & 1.00 & 0.92 & 1.01 & 0.86 & 2.19 & 0.44 & 0.92 & 1.60 & 1.00 & $2.0{\times}10^{-3}$ \\
     & $t_{\mathrm{cool}} = 0.32$ & 0.66 & 0.80 & 0.59 & 1.09 & 0.70 & 1.00 & 0.76 & 1.94 & 0.36 & 1.12 & 1.01 & 1.01 & $2.3{\times}10^{-3}$ \\
     & $t_{\mathrm{cool}} = 3.16$ & 0.92 & 0.76 & 0.79 & 1.42 & 0.85 & 0.99 & 1.06 & 2.15 & 0.63 & 1.99 & 1.09 & 0.98 & $8.7{\times}10^{-4}$ \\
    \midrule
    Euler & test$^\dagger$ & 0.62 & 1.71 & 0.43 & 0.54 & 0.23 & 0.99 & 0.52 & 7.14 & 2.00 & 1.05 & 0.16 & 0.98 & $2.4{\times}10^{-5}$ \\
     & $\gamma = 1.13$$^\dagger$ & 0.95 & 4.13 & 0.44 & 2.77 & 0.20 & 0.99 & 1.07 & 28.39 & 4.85 & 6.68 & 0.17 & 0.99 & $8.1{\times}10^{-6}$ \\
     & $\gamma = 1.76$$^\dagger$ & 0.81 & 2.34 & 0.51 & 0.78 & 0.29 & 0.99 & 0.73 & 14.45 & 5.30 & 2.47 & 0.30 & 0.98 & $9.5{\times}10^{-6}$ \\
    \midrule
    Acoustic scattering & test & 0.72 & 0.87 & 0.78 & 1.05 & 0.94 & 0.99 & 0.99 & 1.13 & 0.55 & 1.05 & 0.98 & 1.04 & $1.9{\times}10^{-3}$ \\
    \bottomrule
  \end{tabular}}
\end{table*}

\begin{table*}[t]
  \caption{\textbf{Energy per scale at step 4} (first $8$ windows of every split). At $^\dagger$ the solver has less than $10^{-4}$ of its energy in the smallest scales.}
  \label{tab:supp-bands4}
  \centering
  \scriptsize
  \setlength{\tabcolsep}{3.6pt}
  \resizebox{\linewidth}{!}{%
  \begin{tabular}{@{}llcccccccccccc@{}}
    \toprule
    & & \multicolumn{3}{c}{$(0, 1/8]$} & \multicolumn{3}{c}{$(1/8, 1/4]$} & \multicolumn{3}{c}{$(1/4, 1/2]$} & \multicolumn{3}{c}{$(1/2, 1]$} \\
    \cmidrule(lr){3-5} \cmidrule(lr){6-8} \cmidrule(lr){9-11} \cmidrule(lr){12-14}
    Family & Split & PhysDial & Walrus & DPOT-H & PhysDial & Walrus & DPOT-H & PhysDial & Walrus & DPOT-H & PhysDial & Walrus & DPOT-H \\
    \midrule
        Kolmogorov flow & test & 1.00 & 1.00 & 1.00 & 0.99 & 1.01 & 0.96 & 0.99 & 1.00 & 0.75 & 0.88 & 1.25 & 11.46 \\
     & $\nu$ inside & 1.00 & 1.00 & 1.00 & 0.99 & 0.98 & 0.95 & 0.98 & 0.94 & 0.76 & 0.99 & 1.17 & 15.02 \\
     & dense $\nu$, inside & 1.00 & 1.00 & 0.99 & 0.99 & 0.99 & 0.89 & 0.99 & 1.03 & 0.88 & 0.93 & 1.71 & 25.11 \\
     & $k_f = 3$ & 1.00 & 1.00 & 0.99 & 0.98 & 1.01 & 0.88 & 0.98 & 1.05 & 0.55 & 0.91 & 1.75 & 5.16 \\
     & multi-mode forcing & 1.00 & 1.00 & 0.99 & 0.99 & 1.00 & 0.93 & 0.99 & 1.01 & 0.62 & 0.90 & 1.11 & 1.12 \\
     & $\nu$ below & 1.00 & 1.00 & 0.99 & 0.97 & 0.90 & 0.67 & 0.90 & 0.75 & 0.56 & 0.68 & 2.18 & 18.53 \\
     & $\nu$ above$^\dagger$ & 0.99 & 1.00 & 1.00 & 1.02 & 1.20 & 1.50 & 1.15 & 1.94 & 11.13 & 24.23 & 119 & 16388 \\
     & dense $\nu$, outside$^\dagger$ & 0.99 & 1.00 & 1.00 & 1.07 & 1.24 & 1.55 & 1.12 & 1.57 & 9.21 & 5.46 & 28.24 & 3722 \\
     & $k_f = 6$ & 1.01 & 1.00 & 1.01 & 1.02 & 0.98 & 1.09 & 1.02 & 1.00 & 0.92 & 1.00 & 2.36 & 15.51 \\
     & forcing band $6$--$7$ & 1.01 & 0.99 & 1.01 & 1.03 & 1.03 & 1.15 & 1.07 & 1.18 & 1.19 & 1.18 & 3.49 & 28.27 \\
     & amplitude $0.5$$^\dagger$ & 1.00 & 1.00 & 1.01 & 0.99 & 1.03 & 0.89 & 0.99 & 1.21 & 1.02 & 1.10 & 5.93 & 131 \\
     & $k_f$ and amplitude$^\dagger$ & 1.01 & 1.00 & 1.01 & 1.01 & 1.02 & 1.00 & 1.03 & 1.10 & 1.09 & 1.07 & 3.13 & 54.07 \\
     & drag $0.05$ & 0.99 & 1.00 & 1.00 & 0.98 & 0.97 & 0.97 & 0.95 & 1.01 & 1.02 & 0.90 & 3.50 & 57.43 \\
     & drag $0.2$ & 1.03 & 1.00 & 1.01 & 1.03 & 1.01 & 0.99 & 1.02 & 1.01 & 0.93 & 1.05 & 1.52 & 19.27 \\
    \midrule
    Decaying turbulence & test & 1.00 & 1.00 & 1.00 & 1.00 & 1.01 & 1.00 & 1.00 & 1.01 & 1.01 & 0.94 & 1.02 & 1.84 \\
     & $\nu$ inside & 1.00 & 1.00 & 0.99 & 1.00 & 0.97 & 0.97 & 0.99 & 0.89 & 0.88 & 0.94 & 0.88 & 6.36 \\
     & $\nu$ outside$^\dagger$ & 1.00 & 1.00 & 1.00 & 0.99 & 1.05 & 1.12 & 1.00 & 1.15 & 1.88 & 1.15 & 1.85 & 57.70 \\
    \bottomrule
  \end{tabular}}
\end{table*}

\begin{table*}[t]
  \caption{\textbf{Energy per scale at step 64} (all windows). The mark $^\dagger$ is used as in \cref{tab:supp-bands-b}.}
  \label{tab:supp-bands64}
  \centering
  \scriptsize
  \setlength{\tabcolsep}{3.6pt}
  \resizebox{\linewidth}{!}{%
  \begin{tabular}{@{}llcccccccccccc@{}}
    \toprule
    & & \multicolumn{3}{c}{$(0, 1/8]$} & \multicolumn{3}{c}{$(1/8, 1/4]$} & \multicolumn{3}{c}{$(1/4, 1/2]$} & \multicolumn{3}{c}{$(1/2, 1]$} \\
    \cmidrule(lr){3-5} \cmidrule(lr){6-8} \cmidrule(lr){9-11} \cmidrule(lr){12-14}
    Family & Split & PhysDial & Walrus & DPOT-H & PhysDial & Walrus & DPOT-H & PhysDial & Walrus & DPOT-H & PhysDial & Walrus & DPOT-H \\
    \midrule
        Kolmogorov flow & test & 1.04 & 0.99 & 0.87 & 1.06 & 1.02 & 1.01 & 1.06 & 1.26 & 0.84 & 1.07 & 7.06 & 7.56 \\
     & $\nu$ inside & 1.02 & 0.99 & 0.79 & 1.09 & 0.89 & 0.87 & 1.10 & 0.91 & 1.01 & 1.01 & 4.02 & 5.37 \\
     & dense $\nu$, inside & 1.03 & 1.00 & 0.91 & 1.06 & 1.05 & 1.18 & 1.02 & 1.38 & 1.37 & 1.03 & 16.41 & 26.96 \\
     & $k_f = 3$ & 1.02 & 1.00 & 0.84 & 1.04 & 1.17 & 1.09 & 1.02 & 1.52 & 1.30 & 0.97 & 9.07 & 8.24 \\
     & multi-mode forcing & 1.01 & 0.94 & 0.66 & 1.12 & 1.08 & 0.78 & 1.17 & 1.30 & 0.88 & 0.98 & 4.26 & 2.13 \\
     & $\nu$ below & 1.05 & 0.97 & 1.00 & 1.09 & 0.81 & 1.11 & 0.94 & 1.14 & 0.51 & 0.63 & 6.65 & 4.11 \\
     & $\nu$ above$^\dagger$ & 0.91 & 1.09 & 0.82 & 0.96 & 2.94 & 3.12 & 0.91 & 7.08 & 11.86 & 2.02 & 190 & 1294 \\
     & dense $\nu$, outside & 0.96 & 1.05 & 0.99 & 0.94 & 1.51 & 2.90 & 0.77 & 2.06 & 2.78 & 0.50 & 20.99 & 78.48 \\
     & $k_f = 6$ & 0.99 & 0.78 & 1.15 & 1.15 & 1.08 & 3.22 & 1.25 & 1.85 & 12.41 & 1.15 & 9.53 & 65.69 \\
     & forcing band $6$--$7$ & 1.07 & 0.73 & 1.19 & 1.27 & 1.08 & 3.61 & 1.38 & 2.06 & 12.10 & 1.18 & 14.43 & 65.55 \\
     & amplitude $0.5$ & 0.99 & 1.08 & 0.91 & 1.05 & 1.26 & 0.60 & 1.08 & 1.52 & 0.71 & 0.94 & 8.85 & 19.68 \\
     & $k_f$ and amplitude & 0.95 & 0.92 & 1.16 & 1.03 & 1.32 & 1.28 & 0.99 & 2.03 & 1.81 & 0.78 & 7.90 & 17.43 \\
     & drag $0.05$ & 0.88 & 0.93 & 1.03 & 0.83 & 1.05 & 21.06 & 0.76 & 1.49 & 67.73 & 0.70 & 10.88 & 538 \\
     & drag $0.2$ & 1.20 & 1.15 & 1.16 & 1.84 & 1.33 & 1.41 & 2.18 & 1.33 & 1.53 & 2.20 & 4.63 & 12.60 \\
    \midrule
    Decaying turbulence & test & 1.04 & 1.01 & 1.17 & 1.08 & 1.18 & 1.74 & 1.11 & 1.52 & 5.93 & 1.06 & 4.91 & 9.62 \\
     & $\nu$ inside & 1.04 & 1.01 & 1.15 & 1.04 & 0.96 & 7.58 & 1.07 & 1.00 & 36.17 & 1.00 & 2.38 & 177 \\
     & $\nu$ outside & 1.07 & 1.06 & 1.37 & 1.06 & 1.23 & 3.34 & 0.97 & 1.26 & 8.13 & 0.68 & 5.29 & 17.00 \\
    \bottomrule
  \end{tabular}}
\end{table*}

\subsection{Design choices}
\label{sec:supp-design}
\Cref{tab:supp-ablation} lists the tested variants.
Most of them come from a single run.

\begin{table*}[t]
  \caption{\textbf{Design choices}. Ratios are errors relative to the full model unless stated otherwise. Decaying turbulence shares the vorticity slot with Kolmogorov flow, so the equation text is the only input that tells the two apart.}
  \label{tab:supp-ablation}
  \centering
  \scriptsize
  \setlength{\tabcolsep}{4pt}
  \begin{tabular}{@{}p{4.4cm}p{8.1cm}p{4.1cm}@{}}
    \toprule
    Variant & Result & Interpretation \\
    \midrule
    Decoder fine-tuning & lowers the step-1 error by $16\%$ ($34\%$ on Kolmogorov flow) and the step-16 error by $1\%$ & affects only the first steps \\
    Periodic handling (two seeds) & on the four doubly periodic families, $0.935$ / $0.932$ times the error without it at step 1 and $0.963$ / $0.961$ at step 16 & about $7\%$ lower error at step 1, more than twenty times the difference between seeds, and about $4\%$ lower at step 16 \\
    Sampling shift $8$ instead of $3$ & $0.986$ times the error with shift $3$ at step 1 and $0.959$ at step 16, and no family has a higher error & a test-time setting \\
    Random initialization, same network (first stage) & $1.15$ times the error (median over the splits), and it responds less to a change in $6$ of $6$ groups (\cref{tab:supp-init}) & video pretraining lowers the error and raises the response to a change \\
    $14$B video model with LoRA, same compute & $1.66$--$1.85$ times the error. With an equal number of training steps, it ties with the $1.3$B model at $6.9$ times the compute & no measurable gain from a larger video prior \\
    Regression instead of flow matching & at step 16, $1.2$--$1.9$ times the error on Kolmogorov flow and $2.19$ times on the extrapolation split of decaying turbulence. Over long rollouts, the generative model has a $13$--$35\%$ lower error & generative training helps \\
    No equation text (evaluation only) & median error $13\%$ higher at step 4 and $12\%$ at step 16. With the text of a wrong equation, $15\%$ and $19\%$. On decaying turbulence, $7$--$10$ times the error & the trained model uses the text, and no model was trained without it \\
    Walrus architecture trained from scratch & $1.02$ times the error of our model. Without its physics pretraining, Walrus has $1.70$ times its error. Without video pretraining, our model has $1.15$ times its error & under fixed conditions, a large video-pretrained transformer and a dedicated PDE architecture are on a par \\
    \bottomrule
  \end{tabular}
\end{table*}

Four further variants give no gain, and PhysDial does not include them.
Training on the model's own rollouts has a $7\%$ higher error at step 16, and augmentation with a doubled step hurts the original step.
A history of six frames is no better than two, and a projection onto physical constraints changes the error by at most $5\%$.

\subsection{Seeds and repeated runs}
\Cref{tab:supp-seeds} summarizes how much the results vary between seeds and between repeated runs.
For PhysDial and for Walrus, only the last training stage was repeated, so these numbers do not measure the variance of the whole training. The U-Net is retrained from scratch.

\begin{table}[t]
  \caption{\textbf{Variation between seeds and repeated runs.}}
  \label{tab:supp-seeds}
  \centering
  \scriptsize
  \setlength{\tabcolsep}{3pt}
  \begin{tabular}{@{}p{2.5cm}p{5.5cm}@{}}
    \toprule
    Repetition & Variation \\
    \midrule
    PhysDial, last stage with another seed & change of the error per split at step 4 / step 16 with a median of $0.4\%$ / $0.8\%$ and $90\%$ of the splits within $1.6\%$ / $3.0\%$. No comparison with Walrus changes. The shear-flow test split differs by $28\%$ at step 16 and is the only split that changes by more than $8\%$ \\
    Walrus, last stage repeated & change of the error per split at step 4 / step 16 with $90\%$ of the splits within $2.4\%$ / $3.7\%$. No split changes by more than $8\%$, and no comparison changes \\
    U-Net specialist, another seed & change of the step-16 error of at most $5\%$ on most Kolmogorov splits, unstable on decaying turbulence (\cref{tab:supp-unetseeds}) \\
    Switch training, two seeds (controlled study) & error ratios of the ten groups differ by at most $0.04$, and four of the five groups with repeated changes reproduce \\
    Anchors, two seeds & extrapolation cost at amplitude $0.5$ of $1.27$--$1.28$ in both, with the same sign of the side effect on the wavenumber split. No Kolmogorov split changes by more than $7.8\%$ \\
    \bottomrule
  \end{tabular}
\end{table}

\section{Limitations}
\label{sec:supp-limits}
\emph{Scope.} All families are two-dimensional on regular grids. Condition changes are trained and tested on Kolmogorov flow only, as abrupt changes of one condition at a time.
\emph{Autoencoder.} The video autoencoder sets a floor on the error of a single step.
\emph{Specialists.} Single-family specialists have the lower error on the training conditions of most families from The Well.
\emph{Runs.} Many results rest on a single run, including the switch training of PhysDial and all component studies on Walrus.
\emph{References.} The in-range reference model is not a clean upper bound.
Its starting point and its data are not exactly those of the tested model, and inside the training range it sometimes has the higher error of the two.
\emph{Walrus.} Its condition input was attached after pretraining, so the finding that the same data do not help it only suggests where our hypothesis stops.
\emph{A change that hurts.} When the viscosity drops to a value far below the training range, taking the new value makes the prediction of the switch-trained model worse.
\emph{Frame interval.} On the amplitude, drag and $\beta$ splits the $\Delta t$ input also leaves its training range, and our runs cannot separate this from the condition itself (\cref{sec:supp-results}).
\emph{Pretraining.} The comparison with a randomly initialized network is a single run and does not identify how pretraining raises the response to a change.
\emph{Missing ablations.} No experiment isolates the effect of restricting the loss to frames after the change. None tests whether the per-sample normalization of Walrus keeps it from using its condition input.

\section{Reproducibility}
\label{sec:supp-repro}
Every simulated trajectory has its own random seed and its own spin-up.
Training, evaluation and reference trajectories never share a seed.
The switch trajectories used for training also have seeds different from those of the evaluation groups.
All training and held-out values are those of \cref{tab:supp-overview,tab:supp-splits}, and all training settings are those of \cref{sec:supp-impl}.
The code for data generation, training and evaluation is available at \url{https://github.com/Dyloong1/PhysDial}.

\end{document}